\documentclass[prb,twocolumn,superscriptaddress]{revtex4-1}
\usepackage{latexsym}
\usepackage{amssymb}
\usepackage{amsmath}
\usepackage{amsbsy}
\usepackage{mathtools}
\usepackage{mathrsfs}
\usepackage{physics}
\usepackage[pdftex]{graphicx}
\usepackage{epstopdf}
\usepackage[usenames, dvipsnames]{color}
\usepackage{soul}
\usepackage[english]{babel}
\usepackage{xcolor}

\newcommand{\av}[1]{\bar{#1}}

\newcommand{\insa}{Universit\`e de Toulouse,
INSA-CNRS-UPS, LPCNO, 135 avenue de Rangueil,
31077 Toulouse, France}

\newcommand{\ioffe}{Ioffe Physical-Technical
Institute, 194021 St. Petersburg, Russia}

\newcommand{\uama}{\'Area de F\'isica Te\'orica y
Materia Condensada, Universidad Aut\'onoma Metropolitana
Azcapotzalco, Av. San Pablo 180, Col. Reynosa-Tamaulipas,
02200 Cuidad de M\'exico, M\'exico}

\begin{document}

\title{Polarization sensitive photodectector
based on GaAsN}

\author{V. G. Ibarra-Sierra}
\affiliation{\uama}
\author{J. C. Sandoval-Santana}
\affiliation{\uama}
\author{R. S. Joshya}
\author{H. Carr\`ere}
\affiliation{\insa}
\author{L.A. Bakaleinikov}
\author{V. K. Kalevich}
\author{E. L. Ivchenko}
\affiliation{\ioffe}
\author{X. Marie}
\author{T. Amand}
\author{A. Balocchi}
\affiliation{\insa}
\author{A. Kunold}
\affiliation{\uama}


\begin{abstract}
We propose and numerically simulate
an optoelectronic compact circular polarimeter.
It allows to electrically measure the 
degree of circular polarization and light
intensity at room temperature for a wide range of incidence angles
in a single shot.
The device, being based on GaAsN, is easy to integrate
into standard electronics and does not require
bulky movable parts nor extra detectors.
Its operation hinges mainly on two
phenomena: the spin dependent
capture of electrons
and the hyperfine interaction between bound electrons and
nuclei on Ga$^{2+}$ paramagnetic centers in GaAsN.
The first phenomenon confers the device with
sensitivity to the degree of circular polarization
and the latter allows to discriminate the handedness
of the incident light.
\end{abstract}

\maketitle

\section{Introduction}
Just as frequency and intensity, polarization
is one of the fundamental properties of light.
In particular circular polarimetry, the determination of the degree
of circular polarization,
plays a determinant role in many recent technological
developments. The capacity of circularly polarized light
to interact with chiral biological structures
has given rise to a wide range of applications
that go from remote sensing of
microbial organisms\cite{sparks2019classical}
to medical assessment techniques 
\cite{whittaker1994quantitative,
10.1117/12.2288761,10.1117/1.JBO.21.5.056002}.
The detection of circularly polarized photon states
is key to quantum information processing in the
observation of entanglement between photons
and electronic \cite{gao2012observation,bhaskar2020experimental} or
nuclear \cite{togan2010quantum,PhysRevB.92.081301} spins.

The detection of circularly polarized light conventionally requires
light to pass through a sequence of
various hefty optical elements \cite{Berry:s}
such as rotating polarizers and waveplates
hampering the integration of circular polarimeters
to standard electronics\cite{basiri2019nature}.
To circumvent the issues posed by standard circular polarimetry,
many circular polarizer
architectures have been proposed.
Although organic photovoltaics\cite{doi:10.1063/1.4868041}
has allowed to build
flexible and tunable full-Stokes polarimeters\cite{Roy:16,Yang:17}
from naturally abundant materials, it requires
the stacking of multiple components.
The distinct optical response of plasmonic chiral metamaterials
to left-handed (LCP) and right-handed (RCP) circularly polarized light
has been exploited to realize circular polarizers
\cite{Bai:s,Akbari:18,hu2017all,zhao2012twisted,basiri2019nature}.
Despite the relative ease with which these structures
can be merged with standard electronics
\cite{Bai:s},
they are limited by low circular
polarization extinction ratios and
low optical efficiency\cite{basiri2019nature}
due to the losses of the parasitic currents generated in the
nanoinclusions \cite{Lin:19}.
Regarding the compactness of the device, a major drawback of these
structures is the
requirement of off-chip detector to sense
the scattered light\cite{Lin:19}.
Similar approaches have been taken in the use of chiral
inclusions in dielectric metasurfaces\cite{hu2017all}
and transition metal dichalcogenides\cite{C9NR10768A}.
Even though the more recent Fabry-Pérot cavity based
polarimeters\cite{PjotrStoevelaar:20} can be, in principle,
extended to circularly polarized light
measurements, so far they have only been
demonstrated to have
low signal-to-noise ratios
under linearly polarized light.
The silicon-on-insulator
\cite{Lin:19,
FullyintegratedCMOScompatiblepolarizationanalyzer,
DONG2020125598}
is probably the most robust platform
for the integration of polarimeters with micro electronics
because of their compatibility with metal
oxide semiconductor fabrication process\cite{DONG2020125598}.
\begin{figure*}
    \centering
    \includegraphics[width=0.45\textwidth]{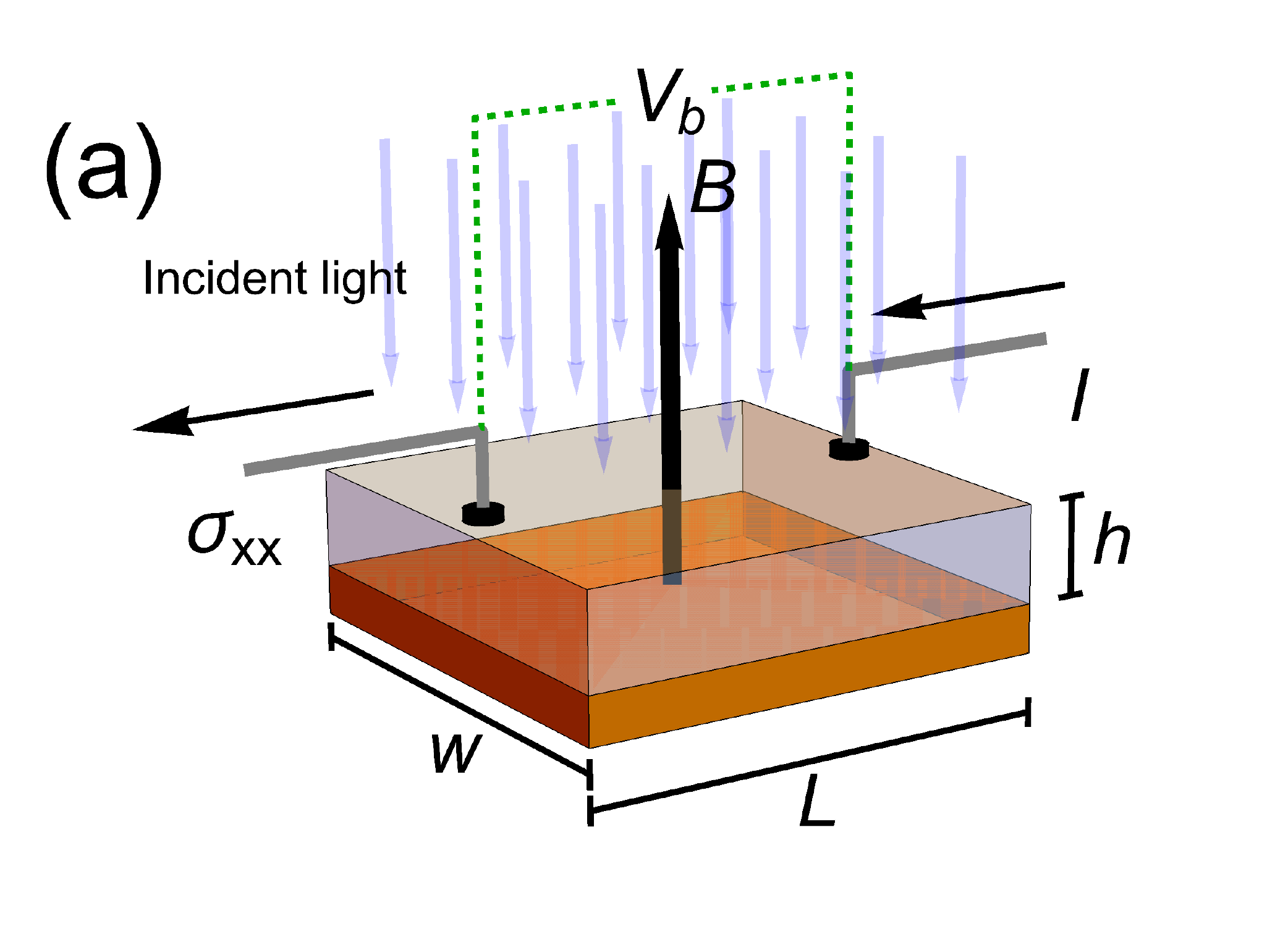}
    \includegraphics[width=0.45\textwidth]{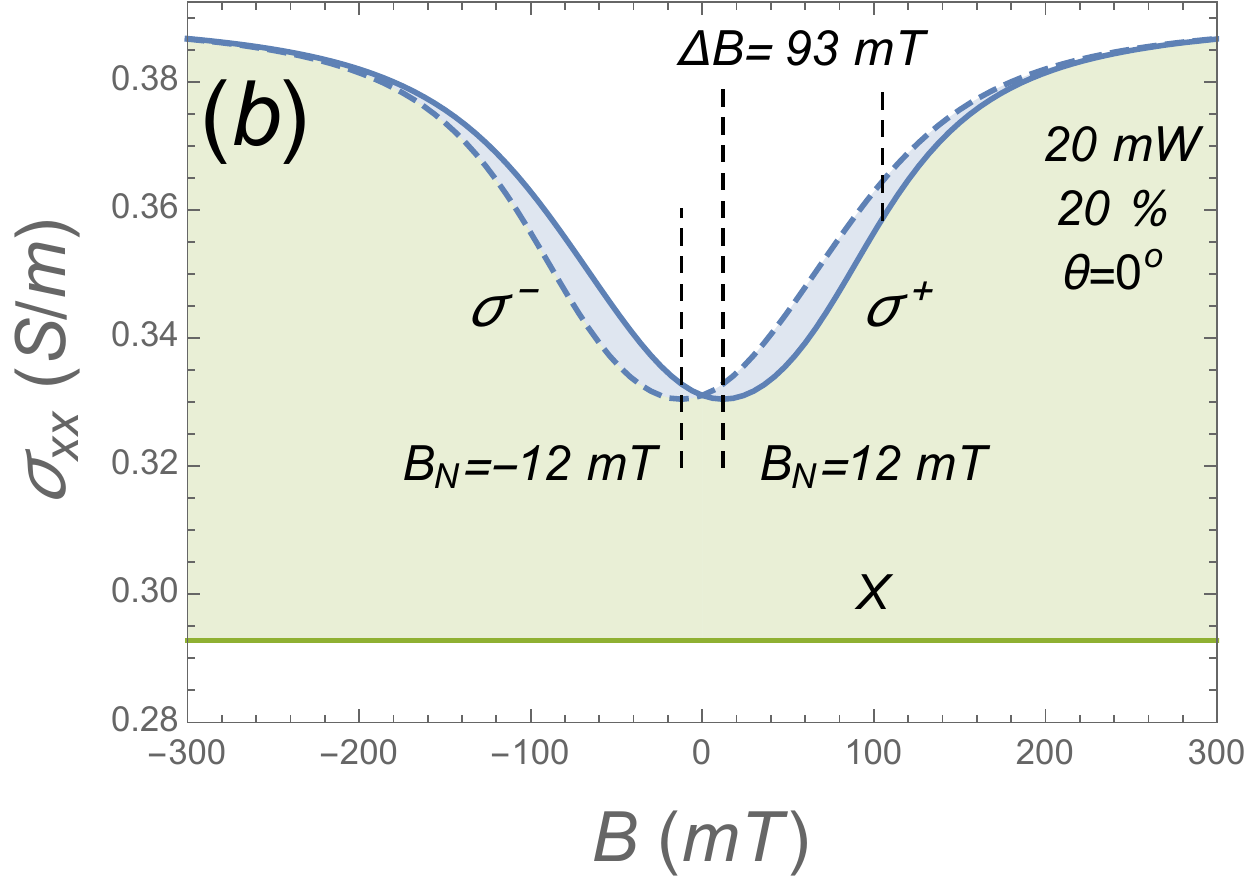}
    \includegraphics[width=0.45\textwidth]{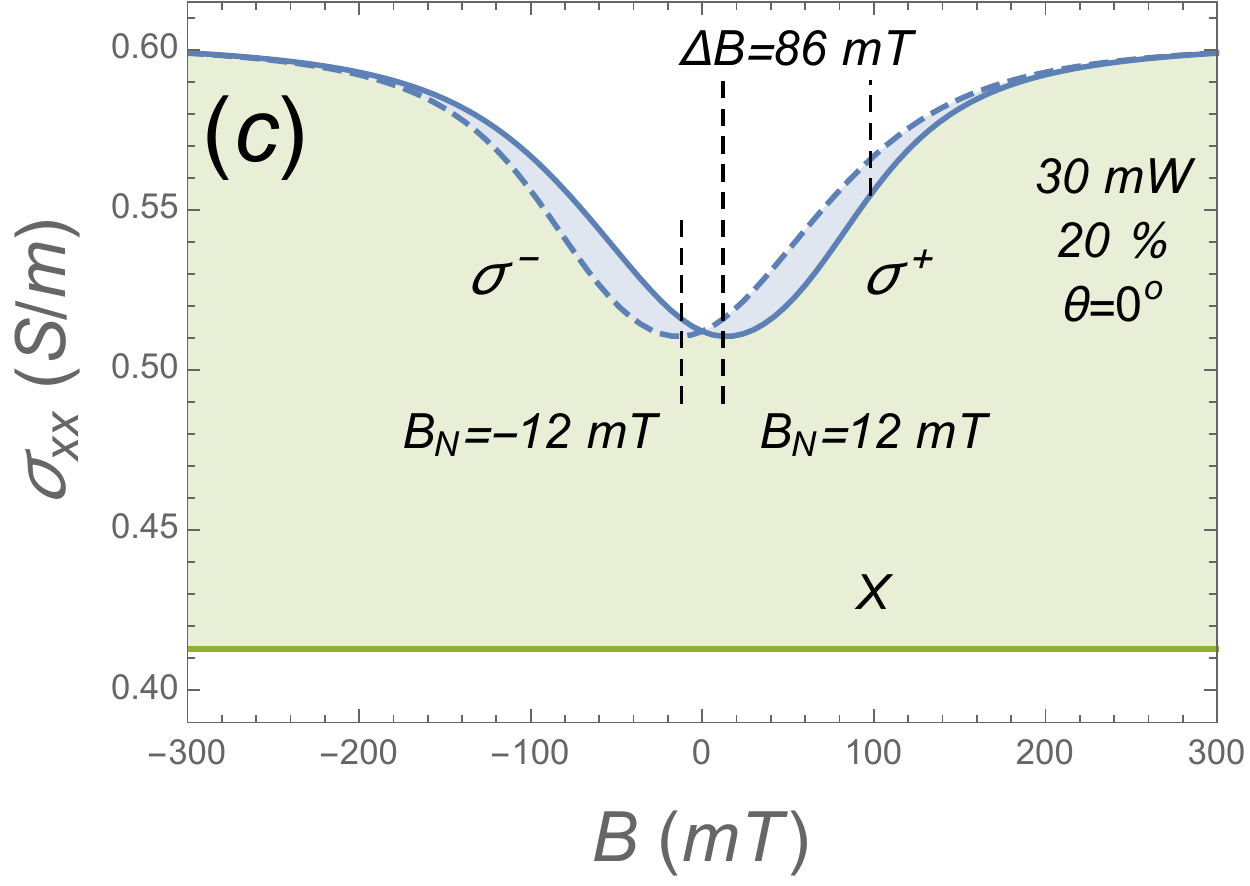}
    \includegraphics[width=0.45\textwidth]{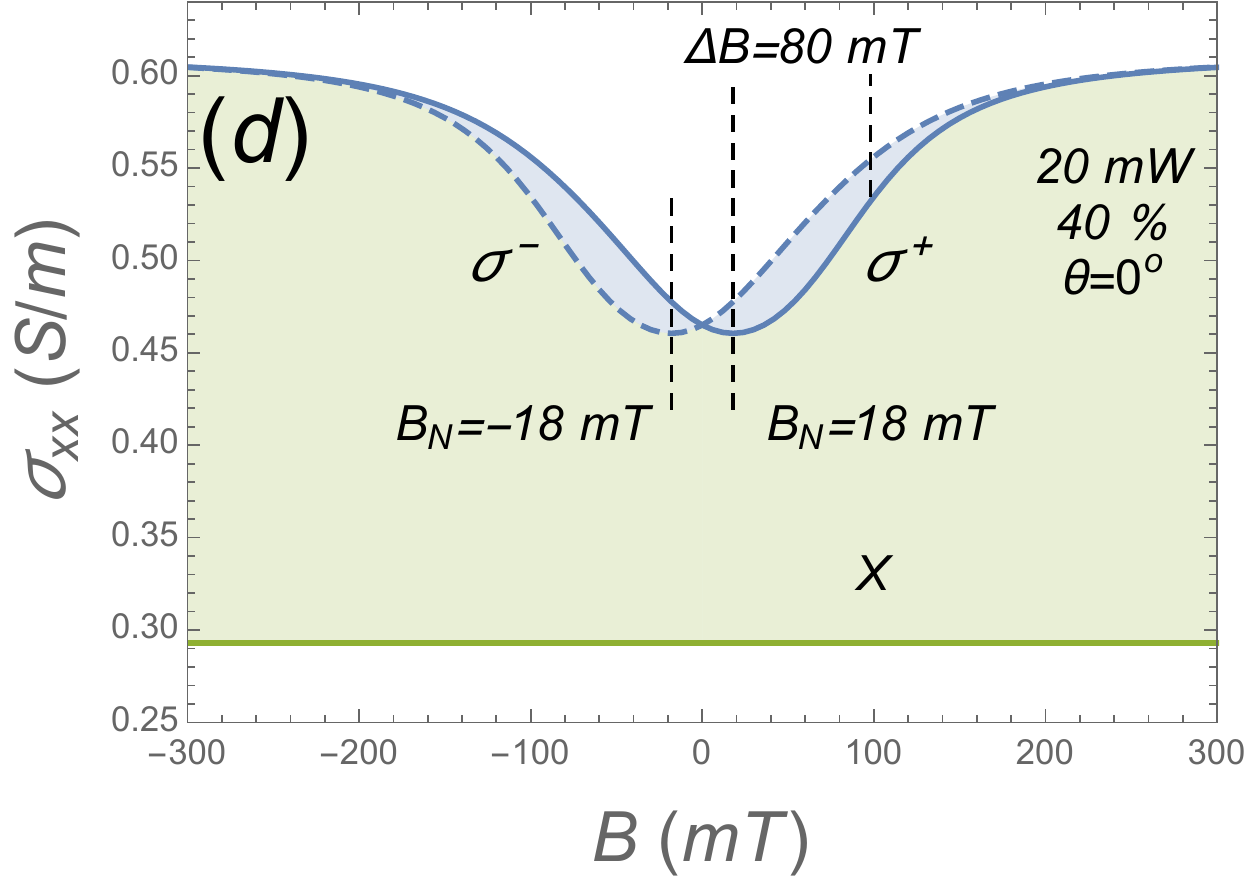}
    \caption{Longitudinal conductivity $\sigma_{xx}$ as a
       function of the magnetic field in Faraday configuration
       ($\theta=0^\circ$)
       for linearly ($X$), right circularly ($\sigma^+$)
       and left circularly ($\sigma^-$) polarized light.
       (a) Setup scketch of the GaAsN slab, magnetic field
       and incident light excitation. The magnetic field
       $B$ is in Faraday configuration and the conductivity
       $\sigma_{xx}$ is determined from the current $i$
       and the bias voltage $V_b$.
       The incident light power and the degree of circular 
       polarization are set to (b) $20$\,mW and $20\%$,
       (c) $30$\,mW and $20\%$ and (d) $20$\,mW and $40\%$.}
    \label{figure1}
\end{figure*}

Spin-optolectronic devices employ
the coupling of
photon angular momentum and
electron spin
\cite{doi:10.1063/1.3327809,doi:10.1063/1.4929326,ivchenko2017spindependent}.
These devices rely on the photogeneration
of spin-polarized conduction band electrons
due to the unique optical selection rules
of zinc-blende semiconductors under circularly polarized light.
Through the spin-orbit interaction the light-polarization
information can be translated first into an electron-spin polarization
and then into an electrical signal
via the inverse spin-Hall effect
\cite{doi:10.1063/1.3327809}.
Other similar alternatives convert the electron-spin polarization
into a photo-voltage through the asymmetries in the electron motion
due the built-in electric field induced
by the Shottky contacts of the device
\cite{doi:10.1063/1.4929326}.
Inverse spin-Hall effect devices operate
at room temperature\cite{doi:10.1063/1.2199473}
but allow only to detect the
degree of circular polarization of light.
In contrast, Shottky devices
can detect simultaneously
the degree of circular polarization and intensity
in a single shot but are restricted to cryogenic
temperatures.

In this paper we propose a spin-optoelectronic device
for the simultaneous detection of the intensity and
degree of circular polarization at room temperature.
The detector is based on a GaAsN epilayer grown
on GaAs.
The detection scheme relies mainly on two phenomena:
the generation of spin polarized conduction band (CB)
electrons through optically oriented pumping and
the spin dependent recombination (SDR)
\cite{PhysRevB.6.436,WEISBUCH1974141,PhysRevB.30.931,
Kalevich2005,doi:10.1063/1.2150252,Kalevich2007,
doi:10.1002/pssa.200673009,Zhao_2009,wang2009room,
KALEVICH20094929,
doi:10.1063/1.3186076,doi:10.1063/1.3273393,doi:10.1063/1.3275703,
doi:10.1063/1.3299015,Ivchenko_2010,PhysRevB.83.165202,
PhysRevB.85.035205,doi:10.1063/1.4816970,Kalevich2013,
puttisong2013efficient,PhysRevB.87.125202,PhysRevB.90.115205,
PhysRevB.91.205202,Ivchenko2016,PhysRevB.95.195204,
PhysRevB.97.155201,Sandoval-Santana2018,ibarra2018spin,
chen2018room}
that the CB electrons undergo through
Ga$^{2+}$ paramagnetic centers \cite{doi:10.1063/1.3275703,wang2009room,
doi:10.1063/1.4816970,ibarra2018spin}.
The orientation of the circularly polarized light
(RCP or LCP)
is discriminated from the sign
of the so called Overhauser-like effective
magnetic field \cite{Kalevich2013,PhysRevB.30.931,
PhysRevB.91.205202,Ivchenko2016}.

The paper is organized as follows.
Section \ref{sec:det} outlines how the
incidence power and the degree of circular polarization
affect the conductivity as a function of an applied
magnetic field. The device architecture
and the procedure to deconvolve
these quantities from the conductivity
are sketched in Sec. \ref{sec:dev}.
Appendix \ref{ap:model} presents a detailed
description of the model based on a master equation
for the density matrix that permits
the computation of the most relevant parameters
to obtain the conductivity of the sample.
In this same appendix we describe the method
followed to fit the model parameters through
measurements of the photoluminescence.

\section{Detection of the degree
of circular polarization through
the spin filtering effect}\label{sec:det}

SDR produces
an unbalance between the two CB electron spin polarization
populations under circularly polarized light excitation.
The unbalance is provoked by the different capture times
that hinge on
the relative spin orientation of CB electrons
and Ga$^{2+}$ center's bound electron
\cite{wang2009room,Zhao_2009,Ivchenko_2010}.
Whereas parallel-oriented spins block CB electrons
from recombining, antiparallel ones promote
recombination to the centers.
This sets in motion a spin filtering effect
in which electrons with a given spin-orientation
remain in the CB for long times, while
those with the opposite spin polarization
fastly recombine to the centers.
Spin polarizations close
to $100\%$ can be attained after
a few of these recombination cycles\cite{PhysRevB.91.205202}
at room temperature and over a wide wavelength excitation range
\cite{meier2012optical}.
The generated spin-polarized CB electron
excess population largely
enhances the photo-conductivity
under circularly polarized illumination
as compared to linearly polarized one
\cite{doi:10.1063/1.3273393,PhysRevB.83.165202}.
A similar argument applies to the holes
left behind by the photoexcited electrons.
The spin-dependent photo-conductivy thus provides
with the means to transduce the degree
of electron spin polarization into an
electrical signal.
\begin{figure*}
\includegraphics[width=0.45\textwidth,keepaspectratio=true]
{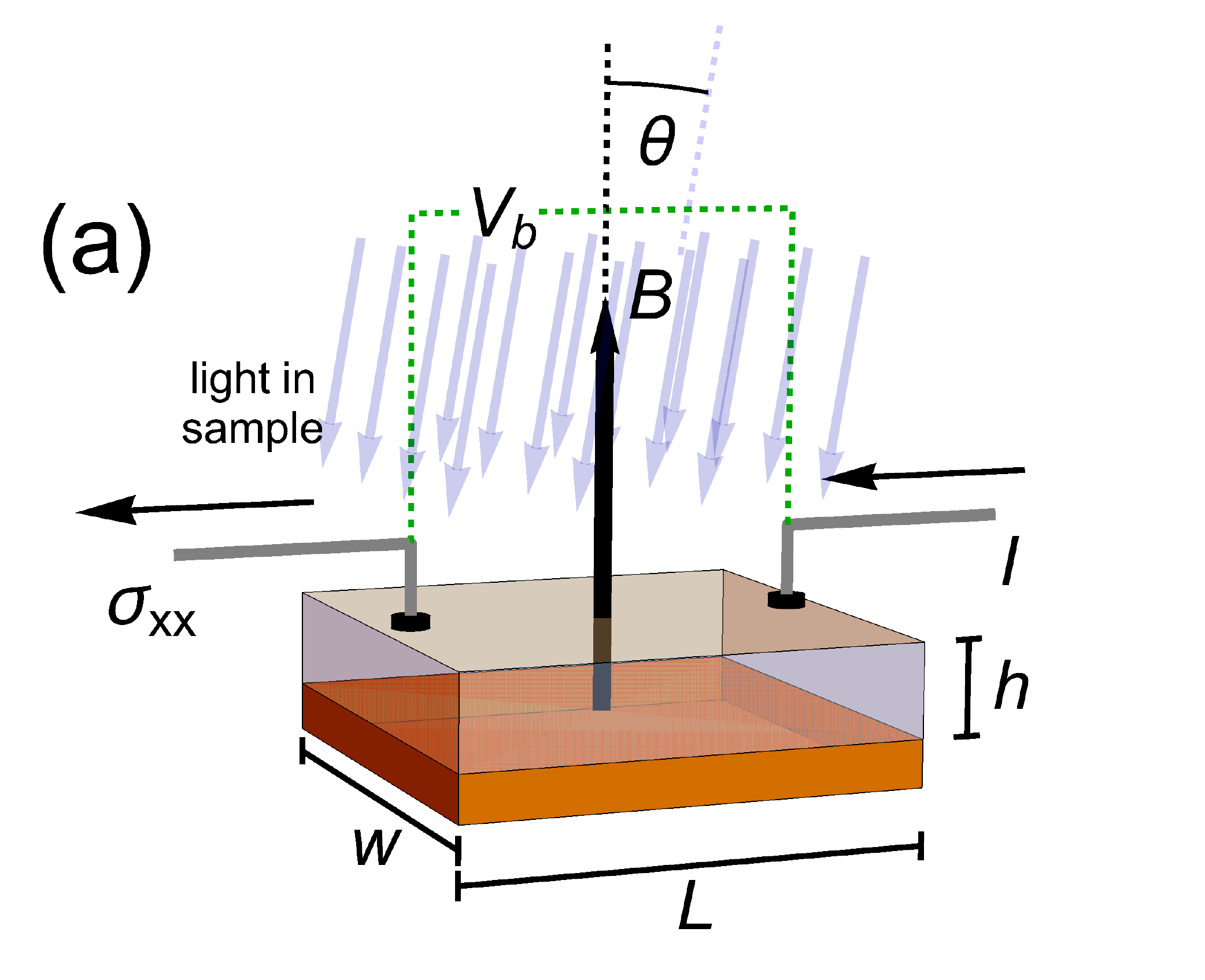}
\includegraphics[width=0.45\textwidth,keepaspectratio=true]
{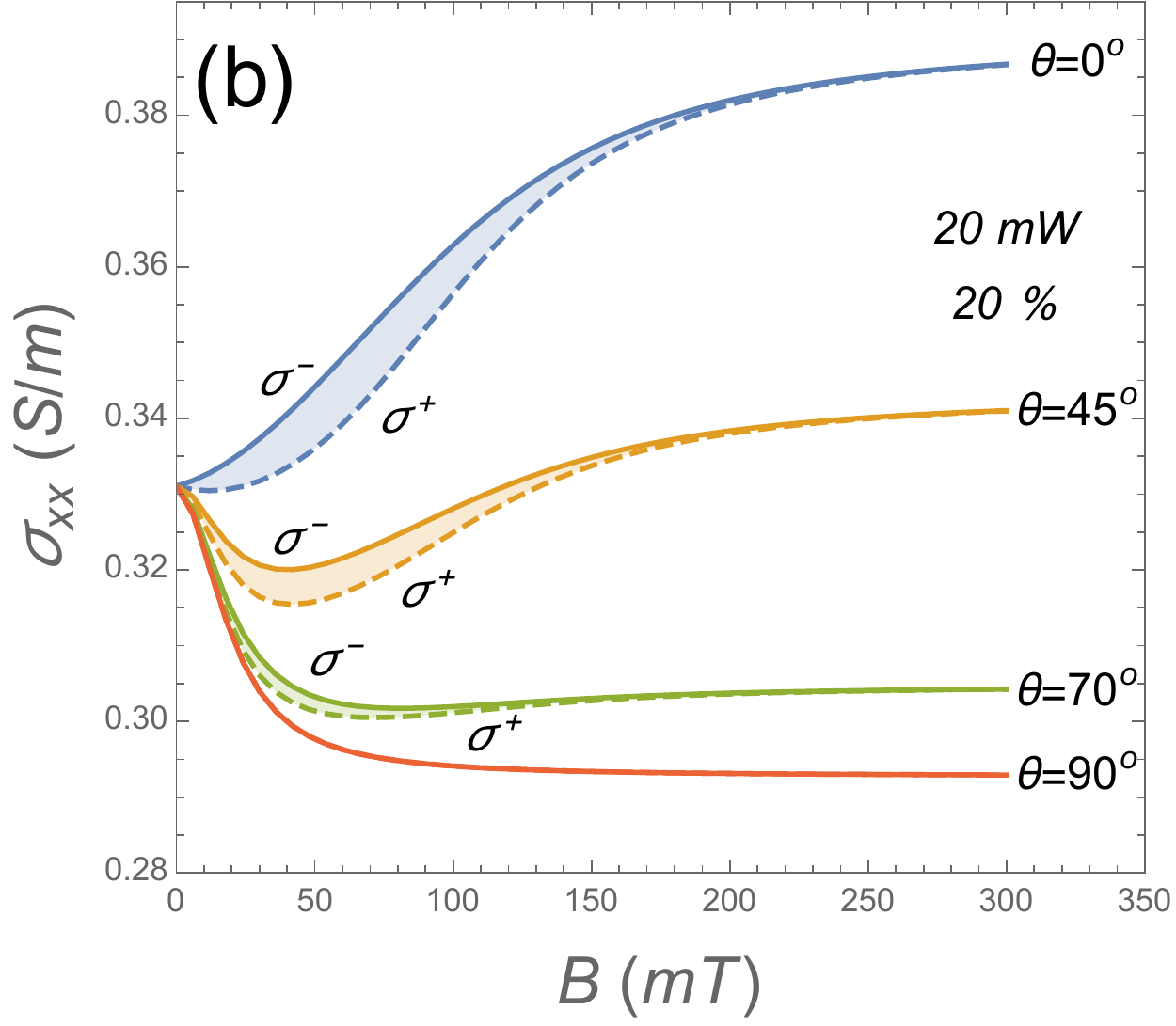}
\includegraphics[width=0.45\textwidth,keepaspectratio=true]
{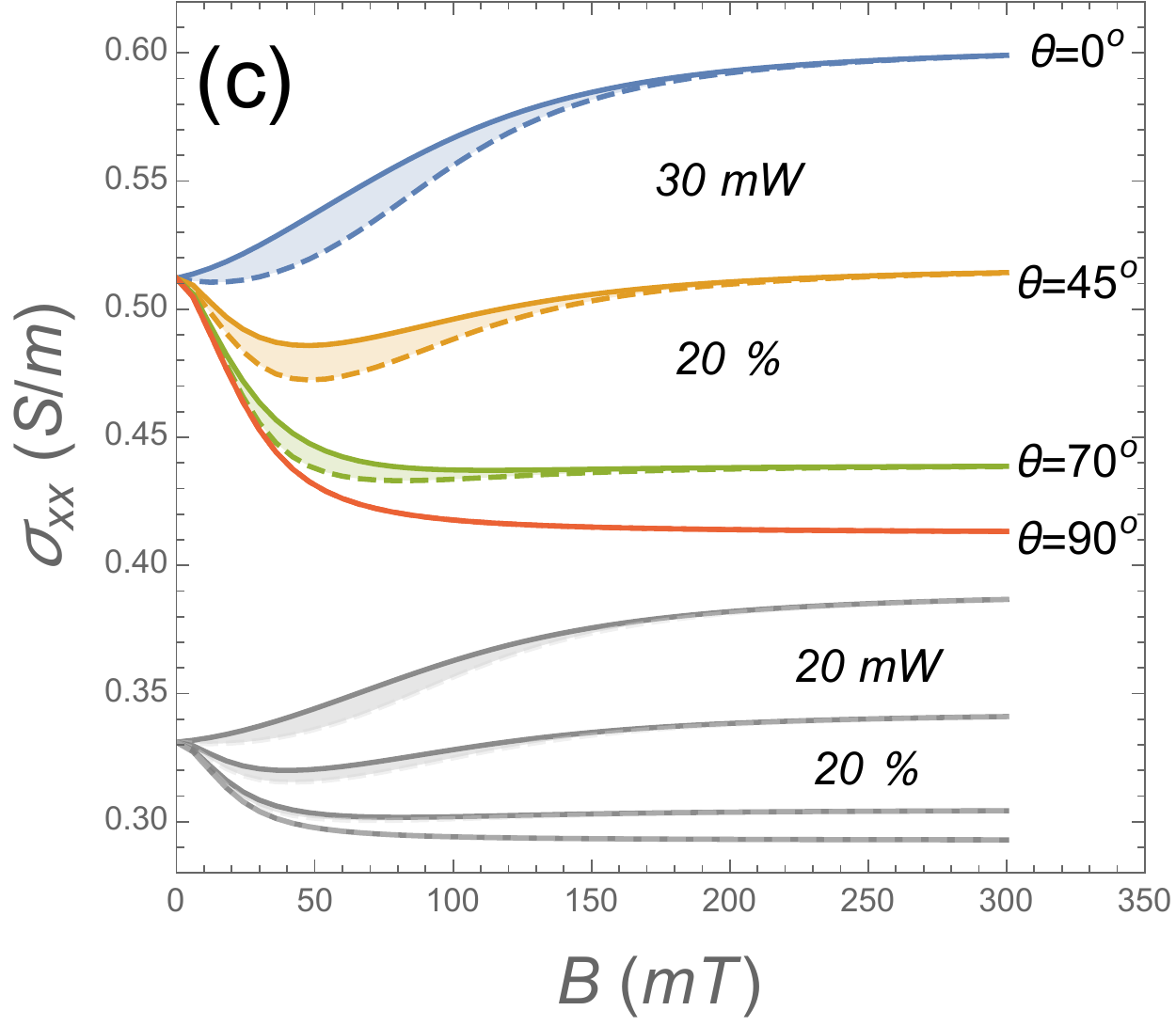}
\includegraphics[width=0.45\textwidth,keepaspectratio=true]
{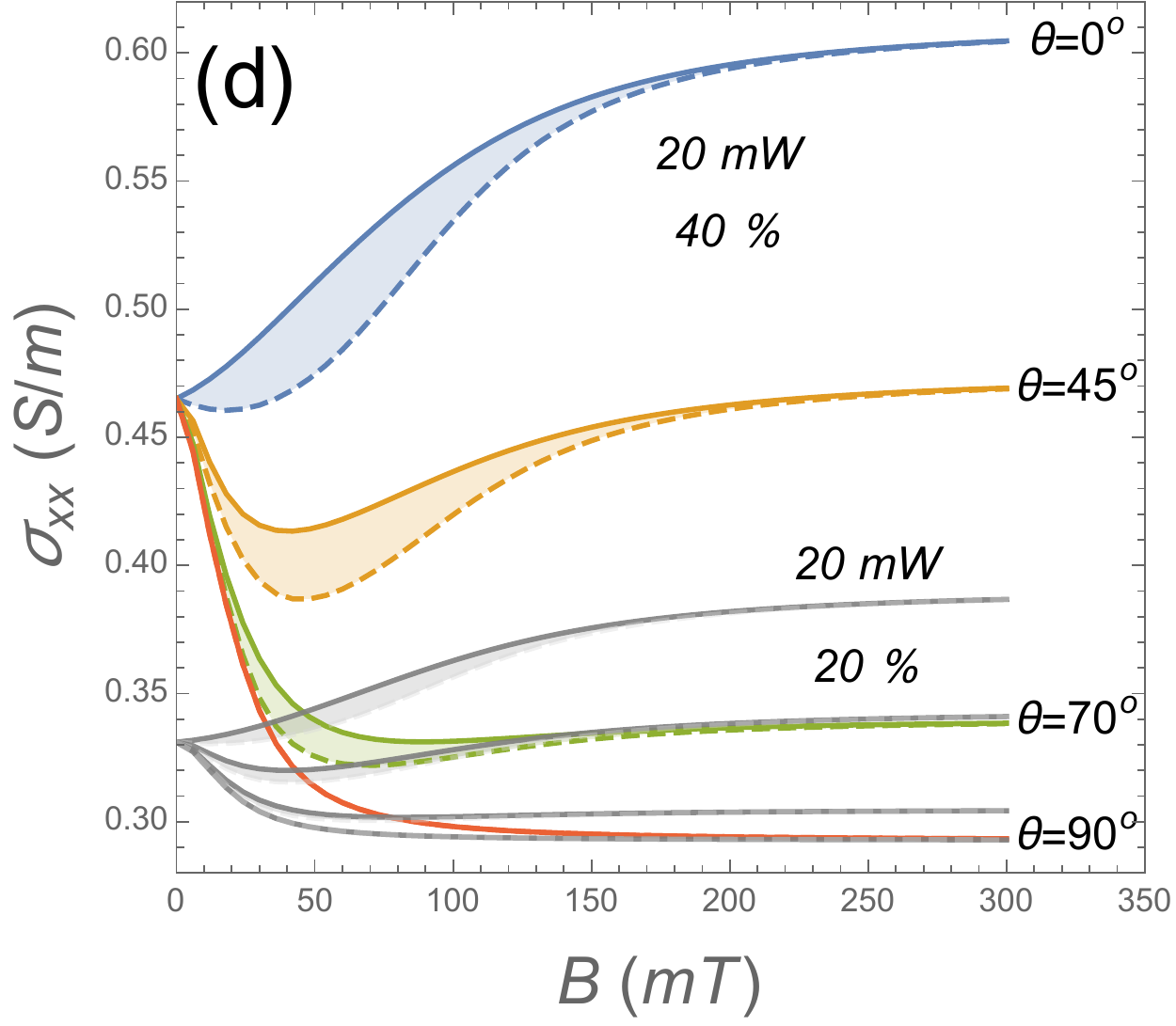}
\caption{Longitudinal conductivity $\sigma_{xx}$ as a
function of the magnetic field for different
relative angles, power intensities, degree
and orientation of circularly polarized light.
(a) Setup scketch of the GaAsN slab, magnetic field
and incident light excitation.
The relative angle $\theta$ between the
magnetic field and the propagation direction
of light inside the sample is marked above the
magnetic field.
The conductivity
$\sigma_{xx}$ is determined from the current $i$
and the bias voltage $V_b$.
Different incident light power and degree of
circular polarization
conditions are presented in the three panels:
(b) $20$\,mW and 40\% right circular polarization,
(d) $30$\,mW and 40\% right circular polarization 
and
(d) $20$\,mW and 80\% right circular polarization.
Panels (b) and (c) show the plots of $20$ mW
and 40\% right circular polarization for
comparison. The solid and dashed lines correspond to
$\sigma^-$ (LCP) and $\sigma^+$ (RCP), respectively.} 
\label{figure2}
\end{figure*}

Even though it can be modulated
by the degree of spin polarization,
the conductivity itself
does not allow to discriminate between
RCP and LCP light; it is insensitive
to the spin direction of the
excess CB electrons and, consequently,
the  circular polarization handedness of
the incident light.
However, the two handedness 
of circularly polarized light can be made
to have a different effect on the Ga$^{2+}$ centers
by applying a relatively small magnetic field
in Faraday configuration.
This
has primarily two effects:
first, it amplifies the spin filtering effect
\cite{PhysRevB.85.035205,Kalevich2013,PhysRevB.87.125202,
puttisong2013efficient} and
second, it generates an Overhauser-like magnetic
field.
The main cause of these
two phenomena has been identified
as the hyperfine interaction between
the outer shell bound electron and the
nucleus of the Ga$^{2+}$ center
\cite{PhysRevB.87.125202,puttisong2013efficient,
PhysRevB.90.115205,PhysRevB.91.205202,PhysRevB.95.195204}.
The amplification of the spin filtering effect
gives the photoluminescence intensity as a function of
the magnetic field the shape of an inverted
Lorentzian function whose minimum
is located at the Overhauser-like
magnetic field \cite{Kalevich2013,PhysRevB.30.931,
PhysRevB.91.205202,Ivchenko2016}.
The position of this minimum, or more precisely, the
sign of the Overhauser-like magnetic field are key
to differentiating between the two orientations
of circularly polarized light.
The Overhauser-like magnetic field is positive
for RCP and negative for LCP.

Figure \ref{figure1} shows plots
of the calculated longitudinal conductivity
$\sigma_{xx}$ as a function
of the magnetic field for linearly
polarized light ($X$, green), right circularly polarized
light ($\sigma^+$, solid blue line)
and left circularly polarized
light ($\sigma^-$, dashed blue line).
The electrical current is calculated through
the Drude model. The setup is sketched in Fig.
\ref{figure1}(a). We consider rectangular GaAsN slabs
of length $L$, width $w$ and height $h$ subject to a perpendicular
magnetic field $B$ and a bias voltage $V_b$ applied along
the length $L$ [see Fig. \ref{figure1}(a)].
The longitudinal component of
the conductivity is given by the sum of the
electron, light hole and heavy hole contributions
\begin{multline}
  \sigma_{xx}=e\Bigg[\frac{\av{n}\mu_e}{1+\mu_e^2B^2}
      +\frac{\av{p}\mu_{lh}/2}{1+\mu_{lh}^2B^2}
      +\frac{\av{p}\mu_{hh}/2}{1+\mu_{hh}^2B^2}\\
      +\frac{\big(
         \frac{\av{n}\mu_e^2B}{1+\mu_e^2B^2}
        -\frac{\av{p}\mu_{lh}^2B/2}{1+\mu_{lh}^2B^2}
        -\frac{\av{p}\mu_{hh}^2B/2}{1+\mu_{hh}^2B^2}
        \big)^2
      }{\frac{\mu_e \av{n}}{1+\mu_e^2B^2}
      +\frac{\mu_{lh} \av{p}/2}{1+\mu_{lh}^2B^2}
      +\frac{\mu_{hh} \av{p}/2}{1+\mu_{hh}^2B^2}}
      \Bigg],
       \label{eq:drude01}
\end{multline}
where $e$ is the electron charge,
$\mu_e=300$\,cm$^2/$Vs,
$\mu_{lh}=50$\,cm$^2/$Vs,
and $\mu_{hh}=50$\,cm$^2/$Vs \,\,
\cite{doi:10.1063/1.2798629,Dhar_2007,
doi:10.1063/1.2927387,5411156,
https://doi.org/10.1002/pssc.201200383,Patan_2009}
are the electron, light hole and heavy hole mobilities.
Naturally, the current intensity through a slab
is given by
\begin{equation}
    I=\sigma_{xx} (V_b/L)wh.
\end{equation}
The electron and hole densities $\av{n}$ and $\av{p}$,
needed in the expressions above,
are obtained from the quantum statistical 
averages of the density matrix
\eqref{eq:nav01} and \eqref{eq:pav01}.
The density matrix is worked out from the numerical
solution of a master equation previously developed
by us \cite{PhysRevB.101.075201}.
In Appendix \ref{ap:model} we present
a detailed description of this method
and list the input parameters.
The second line of Eq. \eqref{eq:drude01}
comes from the transverse Hall field.
These terms also introduce
a magnetic field dependence in $\sigma_{xx}$, but
they are far exceeded by the one
introduced by $\av{n}$ and $\av{p}$.
Indeed, the inverted Lorentzian shape of this
curve comes entirely from $\av{n}$ and $\av{p}$ as functions
of the magnetic field.
These curves strongly resemble the inverted
Lorentzian obtained for the photoluminescence 
\cite{PhysRevB.85.035205},
a trait of the amplification of the spin filtering effect
under a Faraday configuration magnetic field.
This particular behaviour enters
the conductivity in Eq. (\refeq{eq:drude01})
through the electron and hole densities.
We observe that the
conductivity is enhanced
when the illumination is switched to
$\sigma^+$ or $\sigma^-$.
This difference (green shaded region)
permits to distinguish between
linearly and circularly polarized incident light.
One of the most significant characteristics of these
plots is the different displacements created by
$\sigma^+$ and $\sigma^-$ incident light.
While the incident RCP light shifts the Lorentzian curve
$12$\, mT to the right, the LCP light shifts it
to the left by the same amount. The contrast
between these two curves (blue shaded region)
allows to discriminate between RCP and LCP light.
Comparing panel (b) with panels (c) and (d) we notice that
either increasing the illumination power $P_{\mathrm{exc}}$
(from $20$\,mW to $30$\,mW),
or its degree of circular polarization $P_{\mathrm{c}}$
(from $20\%$ to $40\%$)
appreciably changes some of the
$\sigma_{xx}(B)$ features.
Particular attention should be payed to
the Overhauser-like magnetic field $B_{N}$
and the Lorentzian width $\Delta B$.
These two quantities are known to be constant functions
of the power intensity after they saturate
\cite{Kalevich2013,PhysRevB.101.075201}
at a threshold power which, in this case, is roughly $10$\,mW.
Further on, this property of $B_{N}$ and $\Delta B$
will be important to select the relevant
parameters used to characterize some of the properties of
light as incident power, degree of circular polarization
and relative angle.

\begin{figure*}
\includegraphics[width=0.4\textwidth,keepaspectratio=true]
{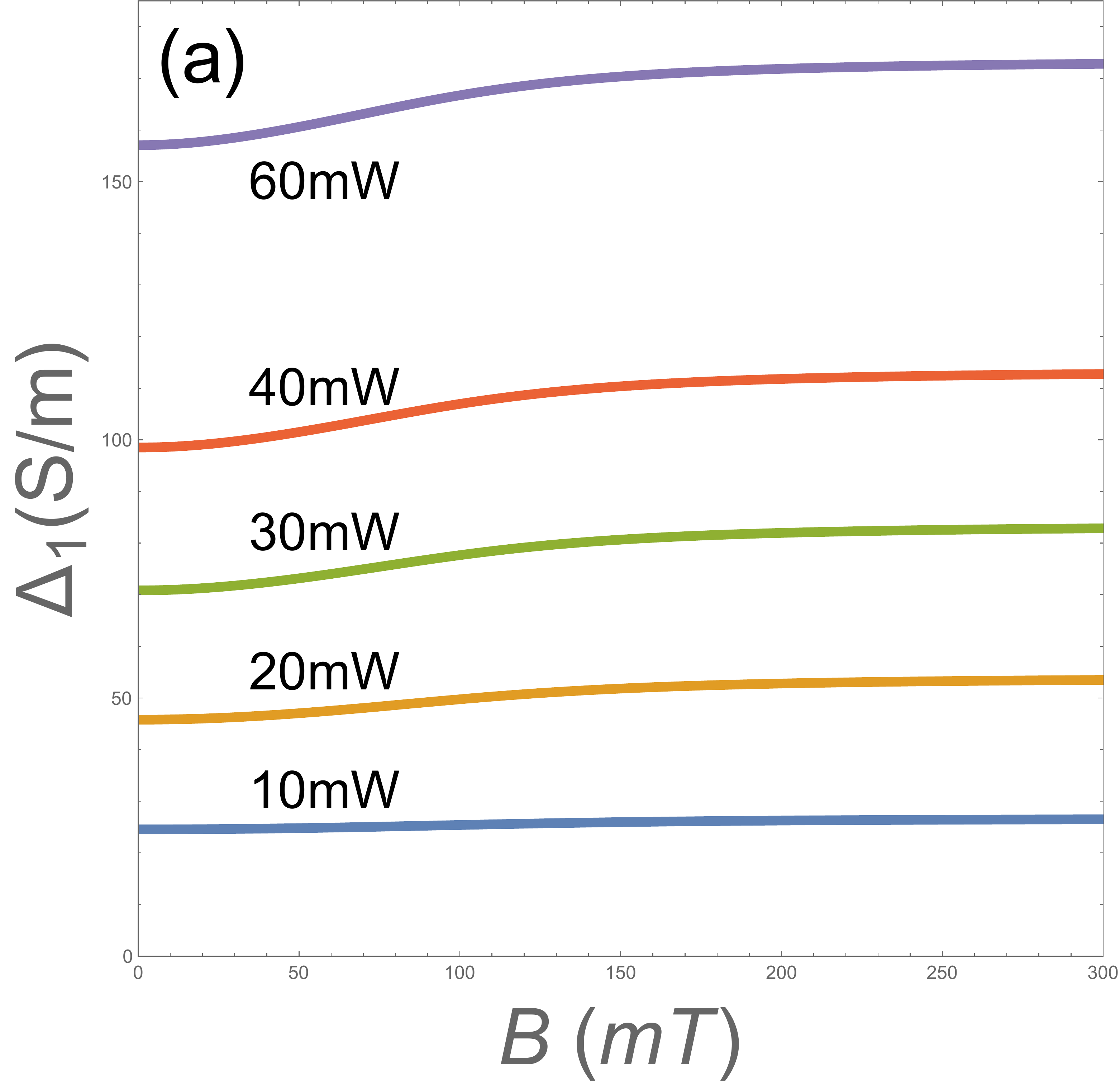}
\includegraphics[width=0.4\textwidth,keepaspectratio=true]
{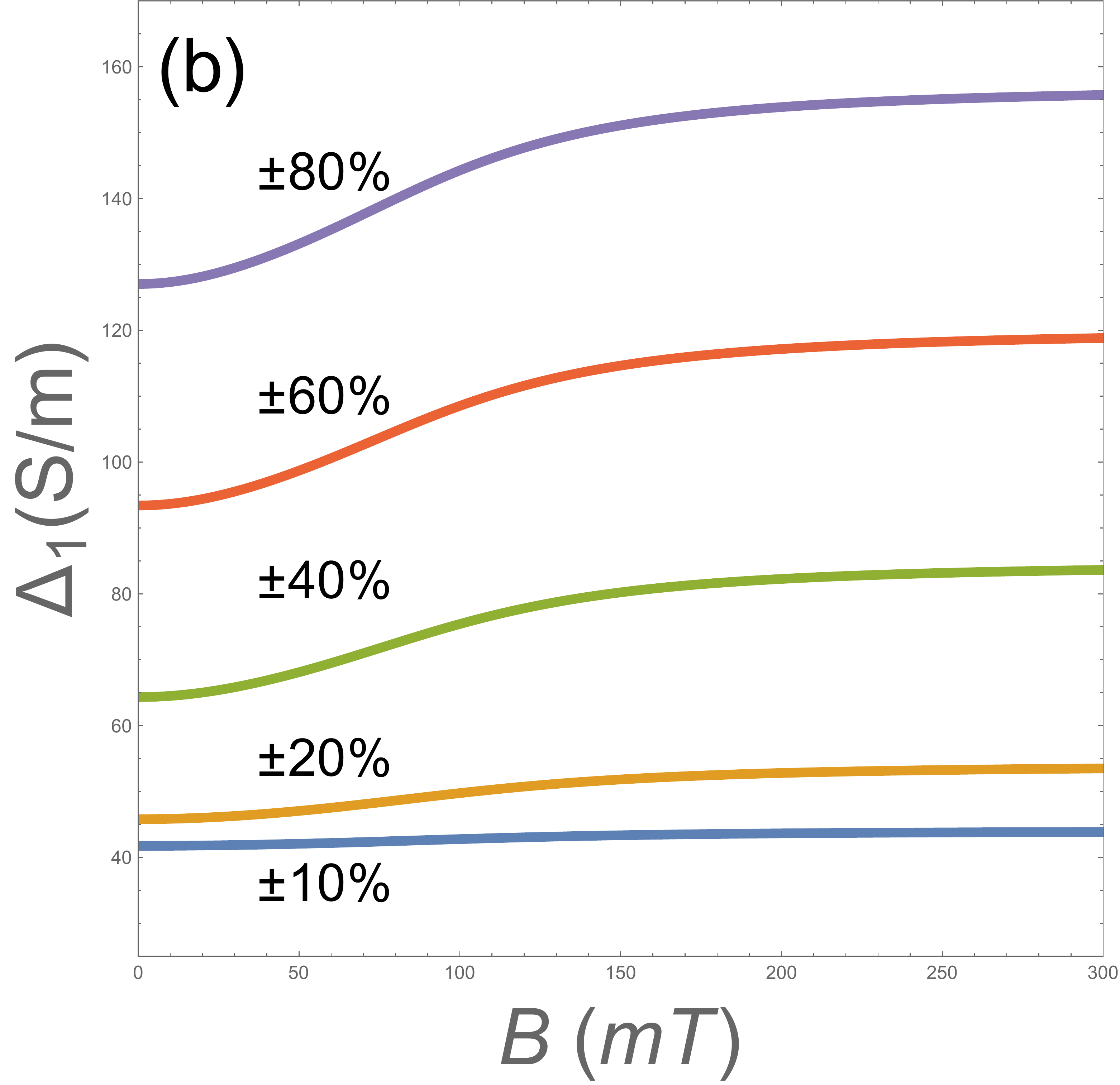}
\includegraphics[width=0.4\textwidth,keepaspectratio=true]
{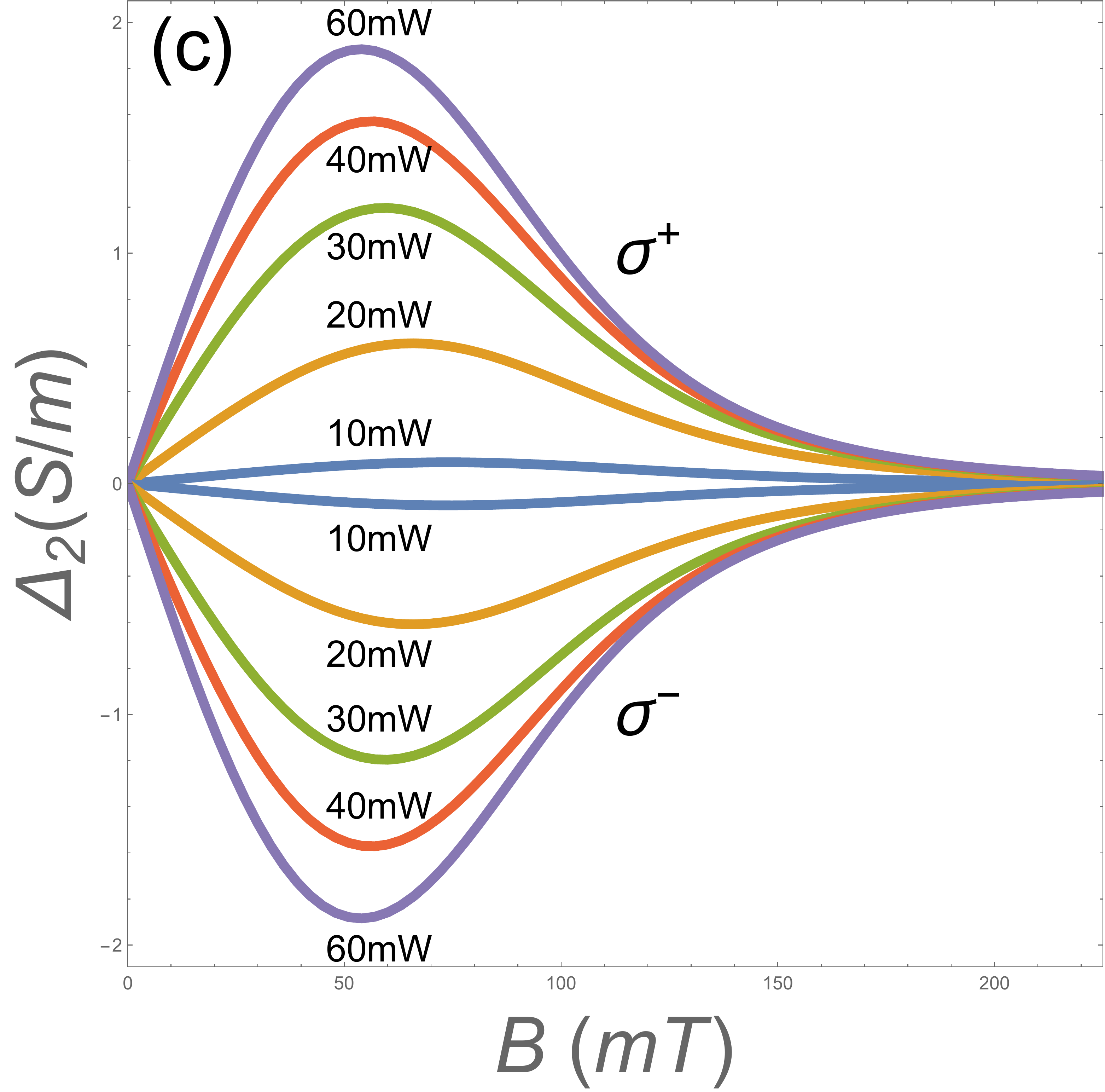}
\includegraphics[width=0.4\textwidth,keepaspectratio=true]
{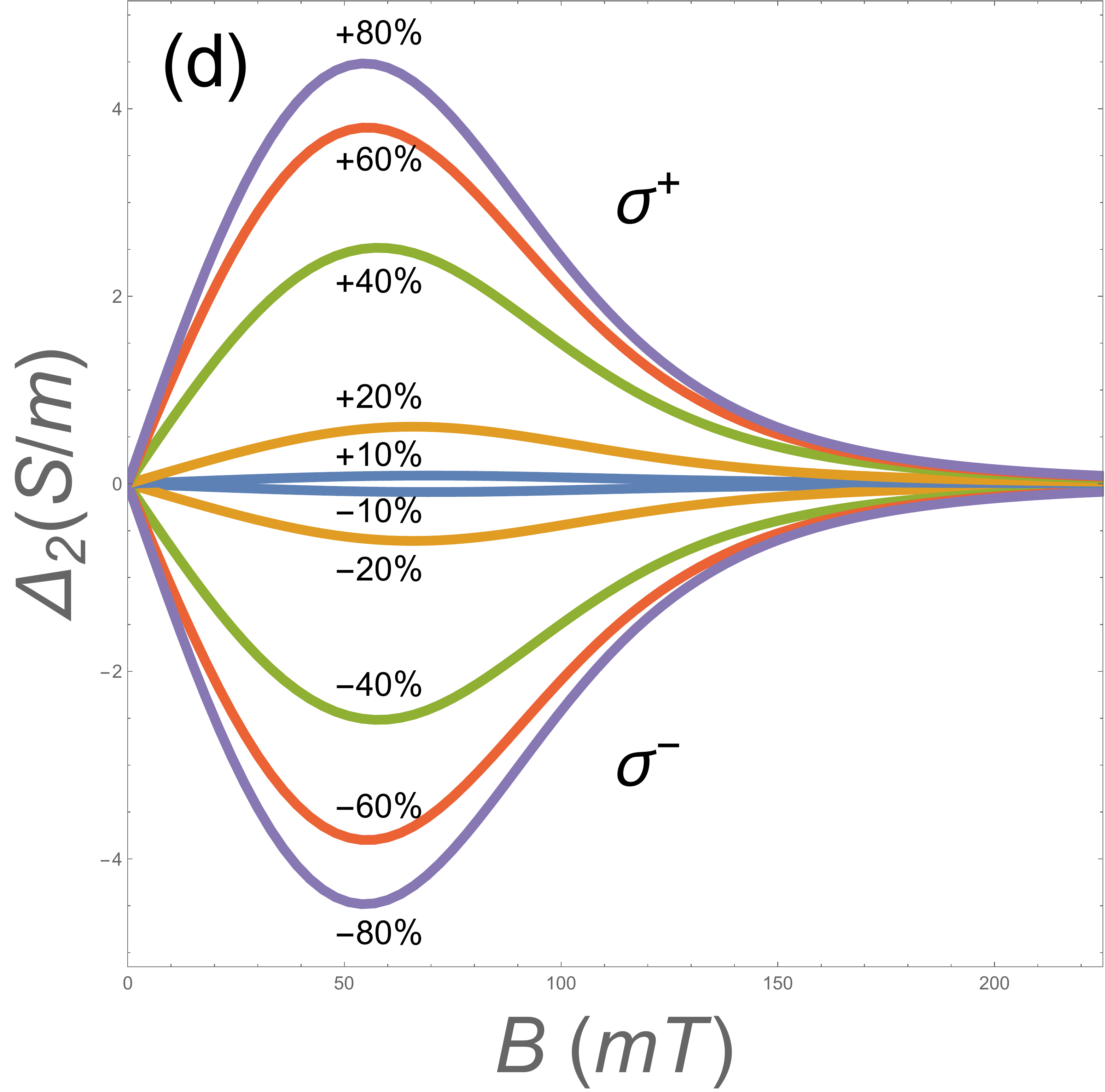}
\caption{Parameters (a,b) $\Delta_1(B)$ and (c,d) $\Delta_2(B)$
as a function of the magnetic field for
different incident powers (a,c) and 
degree of circular polarizations (b,d) for $P_{exc}=20$ mW.}
\label{figure3}
\end{figure*}

The incidence angle of circularly polarized light
also has a strong impact on
the conductivity of the sample.
A complete analysis of its multiple effects
outside normal incidence
would require to consider the variations
of the intensity,  propagation direction and
the polarization predicted
by the Fresnel equations and the Snell law.
These depend on the
widths of the GaAs cap, present in some
architectures, and the GaAsN  slab. This would, however,
yield sample geometry dependent conductivities.
So as to obtain data of a more general
nature it is convenient to express our results
in terms of the relative angle $theta$ between
the applied magnetic field and the propagation
line of light inside the sample.
Given its isotropy, the model
(presented in Appendix \ref{ap:model}),
is sensitive only to the relative orientation
of the applied magnetic field and the
propagation direction of light inside the sample.
This reduces the number of used variables and
considerably simplifies the discussion.
Even though
the angle $\theta$ has
a noticeable effect on the conductivity
it is still possible to carry out undistorted intensity and
degree of circular polarization measurements.
In Fig. \ref{figure2} we observe the conductivity
as a function of the magnetic field for
angles between $0^\circ$ (Faraday configuration)
to $90^\circ$ (Voigt configuration).
A sketch of the setup is presented in Fig. \ref{figure2} (a)
We observe that at Faraday geometry ($\theta=0$)
the $\sigma_{xx}(B)$ adopts the features
of the amplification of the spin filtering effect,
namely, it takes the form
of a downward Lorentzian-like curve shifted by $B_{N}$.
Meanwhile, at Voigt geometry ($\theta=90^{\circ}$),
$\sigma_{xx}(B)$ takes the shape of the upward
centered Lorentzian that characterizes
the Hanle effect \cite{KALEVICH20094929}.
At oblique angles the conductivity behaves
as a combination of both downward and upward Lorentzians
\cite{Ivchenko2016}.
In panel (b) we use as reference $\sigma_{xx}(B)$ under
illumination with an
incident power of $20$\, mW and a degree of circular polarization
of $20\%$.
The conductivity in panel (b) is contrasted 
with the conductivity calculated for an incident power
of $30$\, mW in panel (c) and $40\%$ degree of circular polarization
in panel (d).
We notice that even for very wide relative angles
($\theta\le 70^{\circ}$) there is a clear
distinction between $\sigma^+$ and $\sigma^-$.
The most efficient detection configuration
is at normal incidence where this difference
is maximized.
Further on we show that even at wide relative angles
it is possible to measure the intensity
and the degree of circular polarization.

From Fig. \ref{figure1}
it can then
be said that in general the intensity,
the degree of circular polarization
and its handedness are encoded in $\sigma_{xx}(B)$.
It still remains to find a way to deconvolve
$P_{\mathrm{exc}}$, $P_{\mathrm{c}}$ and $\theta$
from the conductivity.
For this purpose we have defined the following two parameters 
\begin{eqnarray}
\Delta_1(B) &=& \sigma_{xx}(-B)+\sigma_{xx}(B), \label{eq:delta1}\\
\Delta_2(B) &=& \sigma_{xx}(-B)-\sigma_{xx}(B),
\label{eq:delta2}
\label{eq:delta3}
\end{eqnarray}
which are plotted in Fig. \ref{figure3} 
as a function of the magnetic
field.
The parameter $\Delta_1(B)$ depends both on the incident
power [panel (a)] and degree of circular polarization
[panel (b)]
It is however insensitive to the orientation of the
circular polarization.
Additionally it is a fairly smooth function of the magnetic field
and can therefore be expected to provide information
on the power and the degree of circular polarization regardless
of the applied magnetic field.
Instead, $\Delta_2(B)$ has a pronounced sensitivity to
power [panel (c)] and both degree of circular polarization
[panel (c)] and its orientation [panels (c) and (d)]
at approximately $B=50$\,mT. This value depends on
$B_{N}$ and $\Delta B$ which, as mentioned earlier,
remain constant above a certain power threshold.
Due to this, the maximum sensitivity of $\Delta_2(B)$ to power
and degree of circular polarization always occurs at
the same magnetic field value.
The fact that both parameters
are very sensitive at fixed values
of the magnetic field
is central to the design of the
polarimeter architecture; not only no variable magnetic
field is needed to characterize light,
but its magnitude is small enough
that it can be generated by small permanent
magnets.

The previous results give us an inkling on
what parameters may encode all the information
necessary to characterize light's
degree of circular
polarization and power.
We know that at $B=50$\,mT $\Delta_1(B)$
and $\Delta_2(B)$ are strongly responsive to
changes in power and degree of circular polarization.
Moreover, the sign of $\Delta_2(50\,\mathrm{mT})$,
being a measure of the asymmetry, provides information
on the handedness of the degree of polarization.
Thereby, a suitable set of parameters
to fully characterize an incident
beam of light is
$\Gamma_1=\Delta_1(50\,\mathrm{mT})$ and $\Gamma_2=\Delta_2(50\,\mathrm{mT})$.
In the next section we propose a device architecture
that allows to measure these three parameters
and explain how to extract from them
the incidence power
$P_{\mathrm{exc}}$,
the degree of circular polarization $P_{\mathrm{c}}$
and its handedness.
\begin{figure}
\includegraphics[width=0.48\textwidth,keepaspectratio=true]
{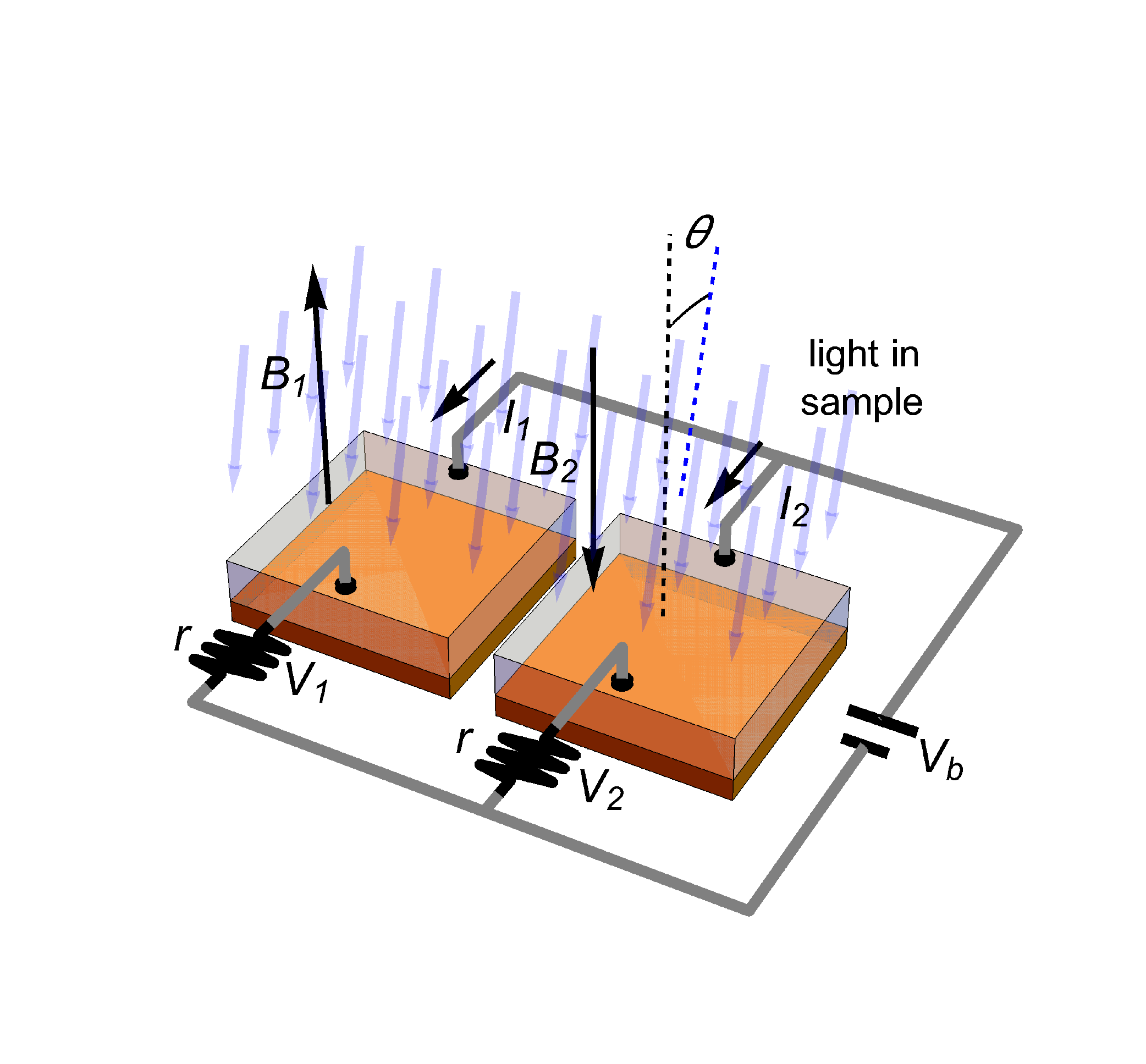}
\caption{Schematics of the polarimeter. Both slabs are connected
in parallel with a bias voltage source $V_b$. The voltages in
the resistances $r$, $V_1$ and $V_2$, are
used to determine the current intensities
$I_1$ and $I_2$ that travel through
each GaAsN epilayer slab.
The magnetic fields $B_1$ and $B_2$ are produced by the permanent
magnets (dark blue) located on the base of each slab.
The relative angle $\theta$ between the magnetic field and
the propagation direction of light inside the sample is indicated.}
\label{figure4}
\end{figure}

\section{Device architecture and deconvolution
of the incident power, degree of polarization
and its orientation}\label{sec:dev}
\begin{figure*}
\includegraphics[width=0.31\textwidth,keepaspectratio=true]
{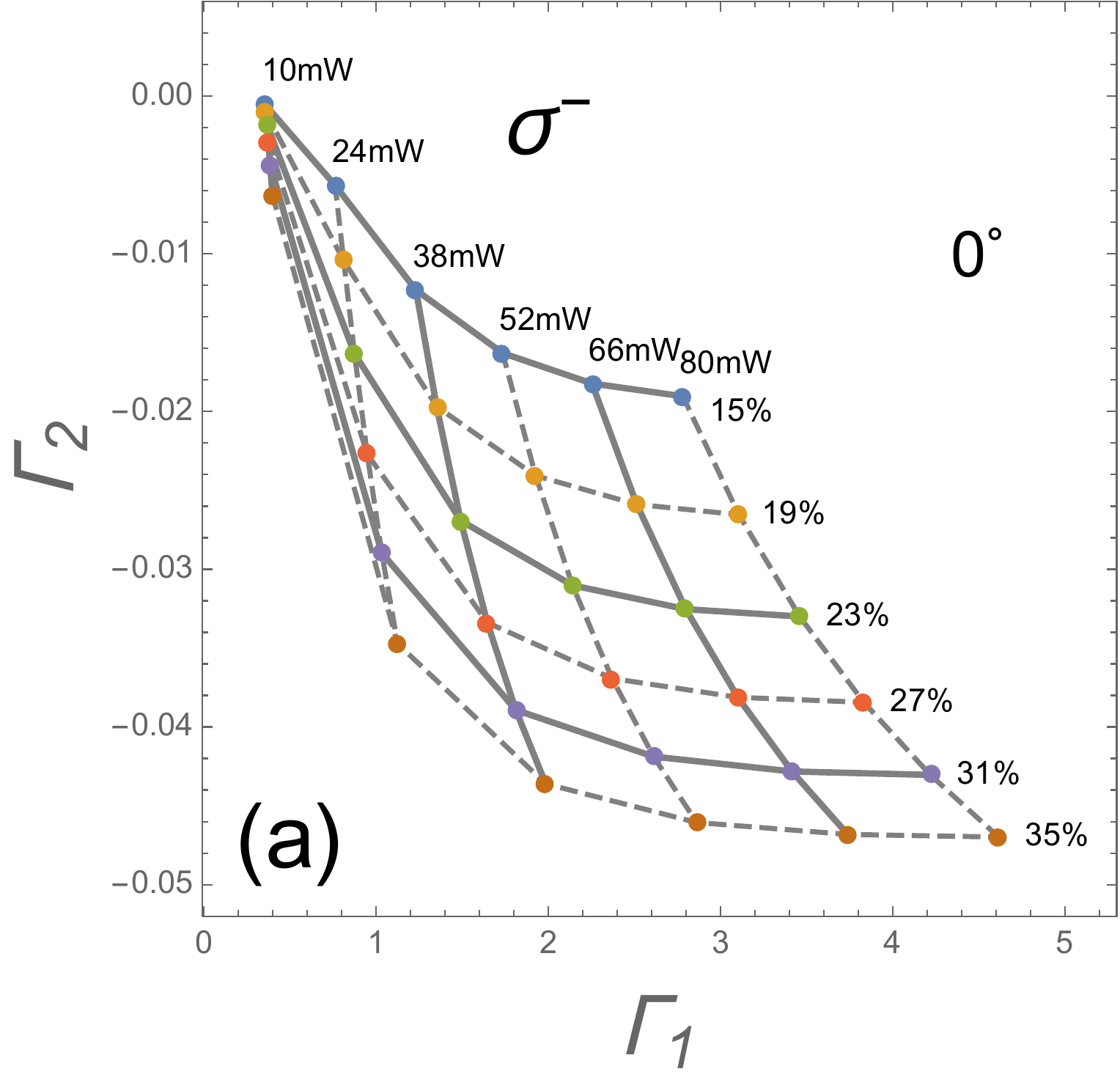}
\includegraphics[width=0.31\textwidth,keepaspectratio=true]
{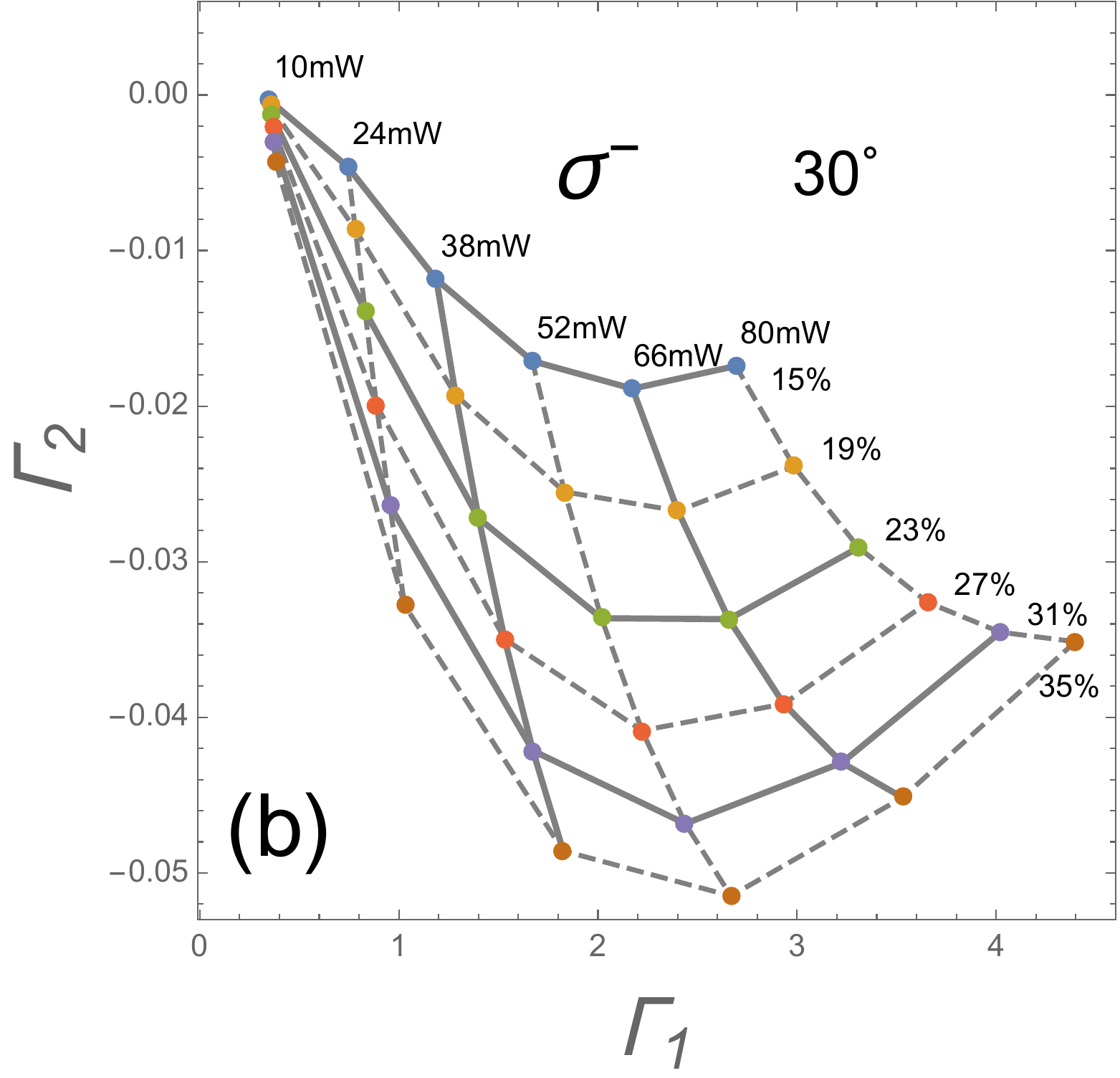}
\includegraphics[width=0.31\textwidth,keepaspectratio=true]
{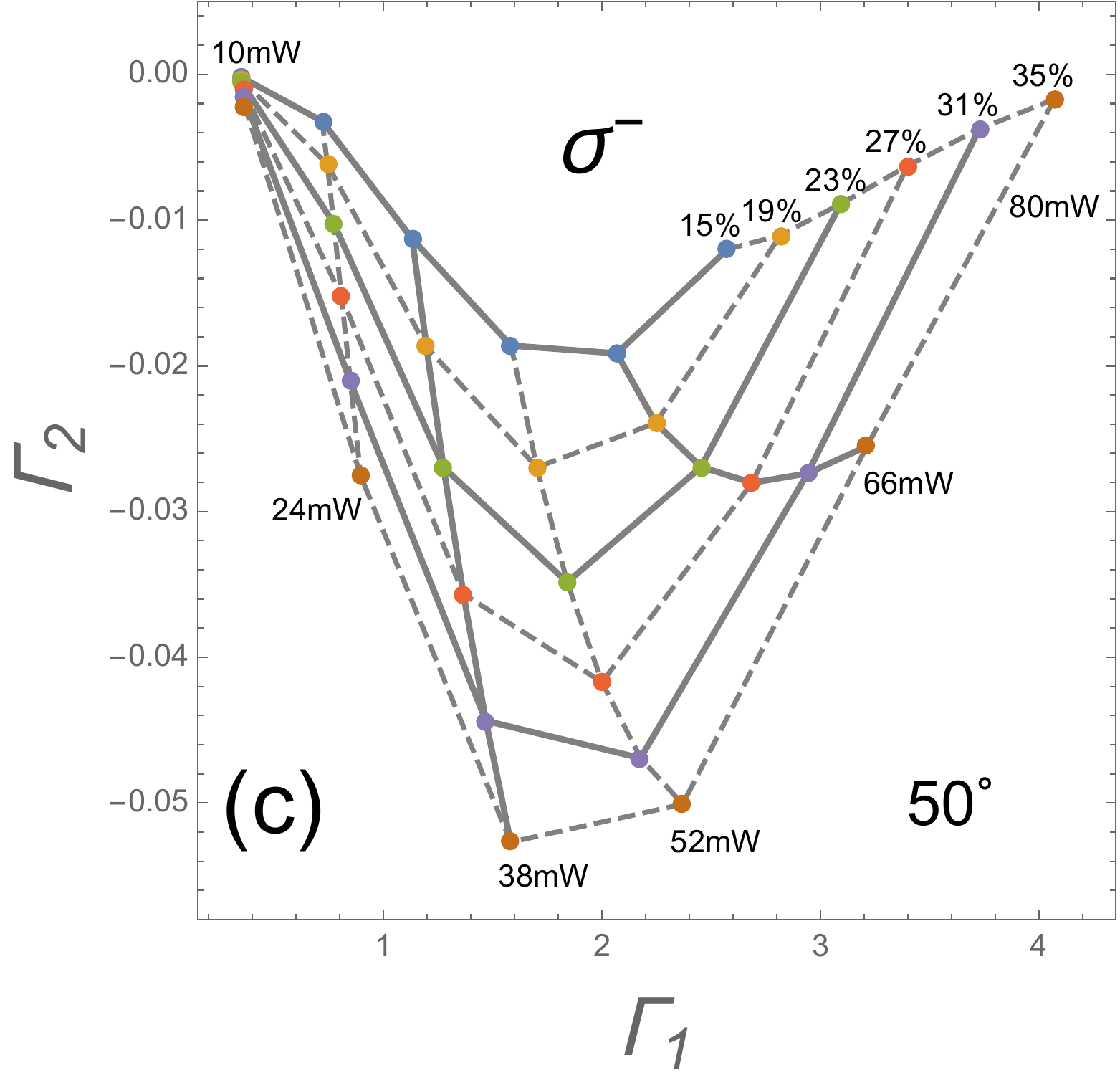}
\caption{
Power and degree of circular polarization
isolines as a function of $\Gamma_1 = \Delta_1(50\,\mathrm{mT})$
and $\Gamma_2=\Delta_2(50\,\mathrm{mT})$
under $\sigma^-$ illumination at a relative
angle of
(a) $0^{\circ}$, (b) $30^{\circ}$ and (c) $50^{\circ}$.}
\label{figure5}
\end{figure*}

The proposed device consists of interconnected slabs
of GaAs$_{1-x}$N$_x$ as shown in Fig. \ref{figure4}.
Each slab consists of a GaAs$_{1-x}$N$_x$ epilayer
on a (100)-oriented semi-insulating GaAs substrate
\cite{doi:10.1063/1.2150252,doi:10.1002/pssa.200673009}.
This whole heterostructure is depicted as a transparent
parallelepiped. The magnetic fields
$\boldsymbol{B}_1$ and $\boldsymbol{B}_2$
are produced by permanent magnets beneath the GaAsN/GaAs
heterostructures.
The voltage source $V_b$ is used to drive the currents
$I_1$ and $I_2$ through the GaAsN slabs.
The voltages $V_1$ and $V_2$ in the resistances $r$
allow to detect the currents $I_1$ and $I_2$.
The permanent magnets sizes should be chosen
as to generate the magnetic fields
$B_1=-B_2=50$\,mT providing with enough information to
obtain $\Gamma_1$ and $\Gamma_2$. According to
Eqs. (\ref{eq:delta1}) and (\ref{eq:delta3})
it would seem that three values of the conductivity
[$\sigma_{xx}(0)$, $\sigma_{xx}(50\,\mathrm{mT})$ and
$\sigma_{xx}(-50\,\mathrm{mT})$]
are needed to obtain $\Gamma_1$ and $\Gamma_2$.
This would imply the need for an extra third slab of GaAsN.
However, $\sigma_{xx}(0)$ can be estimated as
$(\sigma_{xx}(50\,\mathrm{mT})+\sigma_{xx}(-50\,\mathrm{mT}))/2$
restricting the number of slabs to two.
An equivalent set-up would consist of
a single slab of GaAsN where a microcoil
would generate an alternating magnetic field.

\begin{figure*}
\includegraphics[width=0.32\textwidth,keepaspectratio=true]
{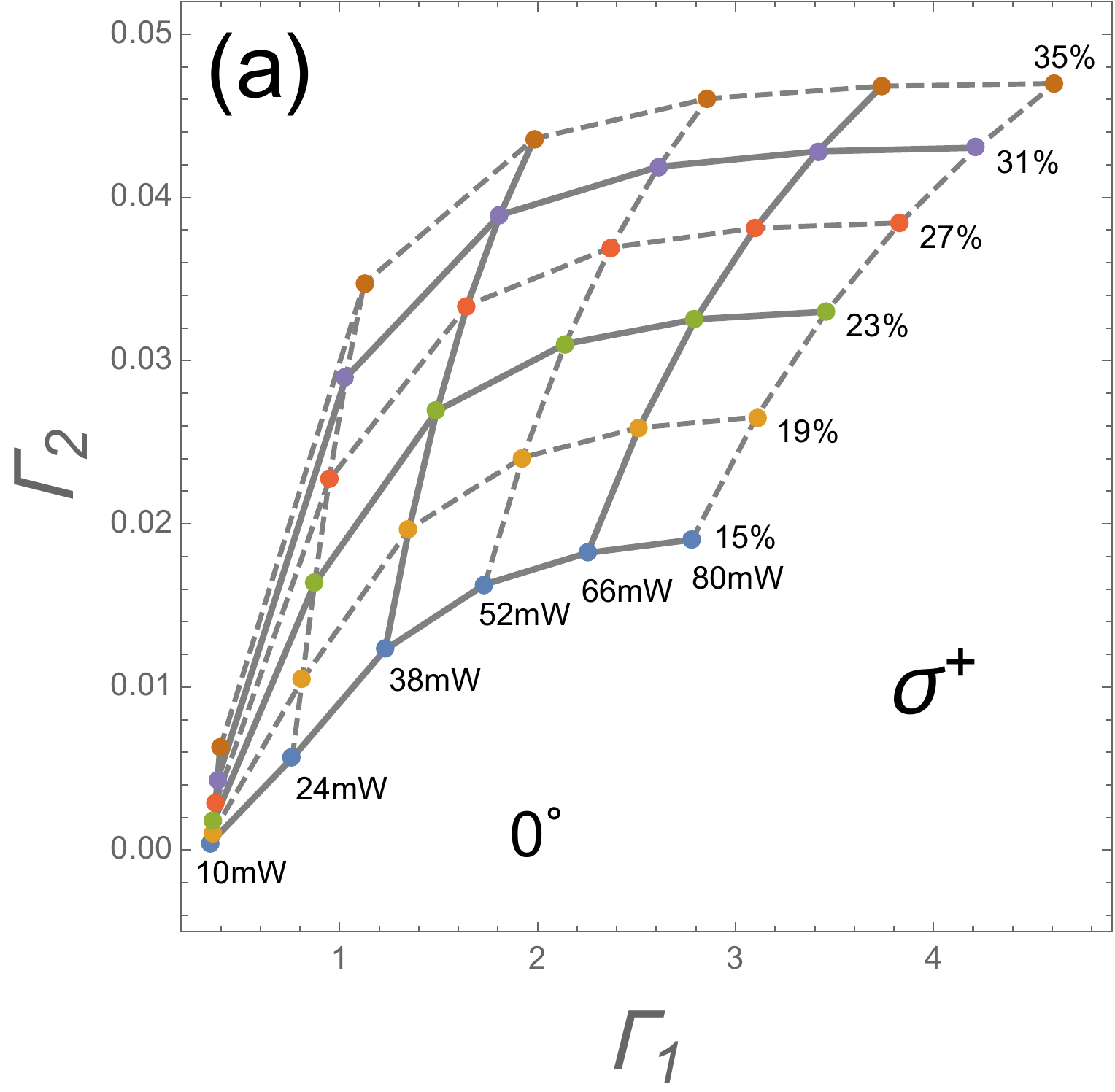}
\includegraphics[width=0.32\textwidth,keepaspectratio=true]
{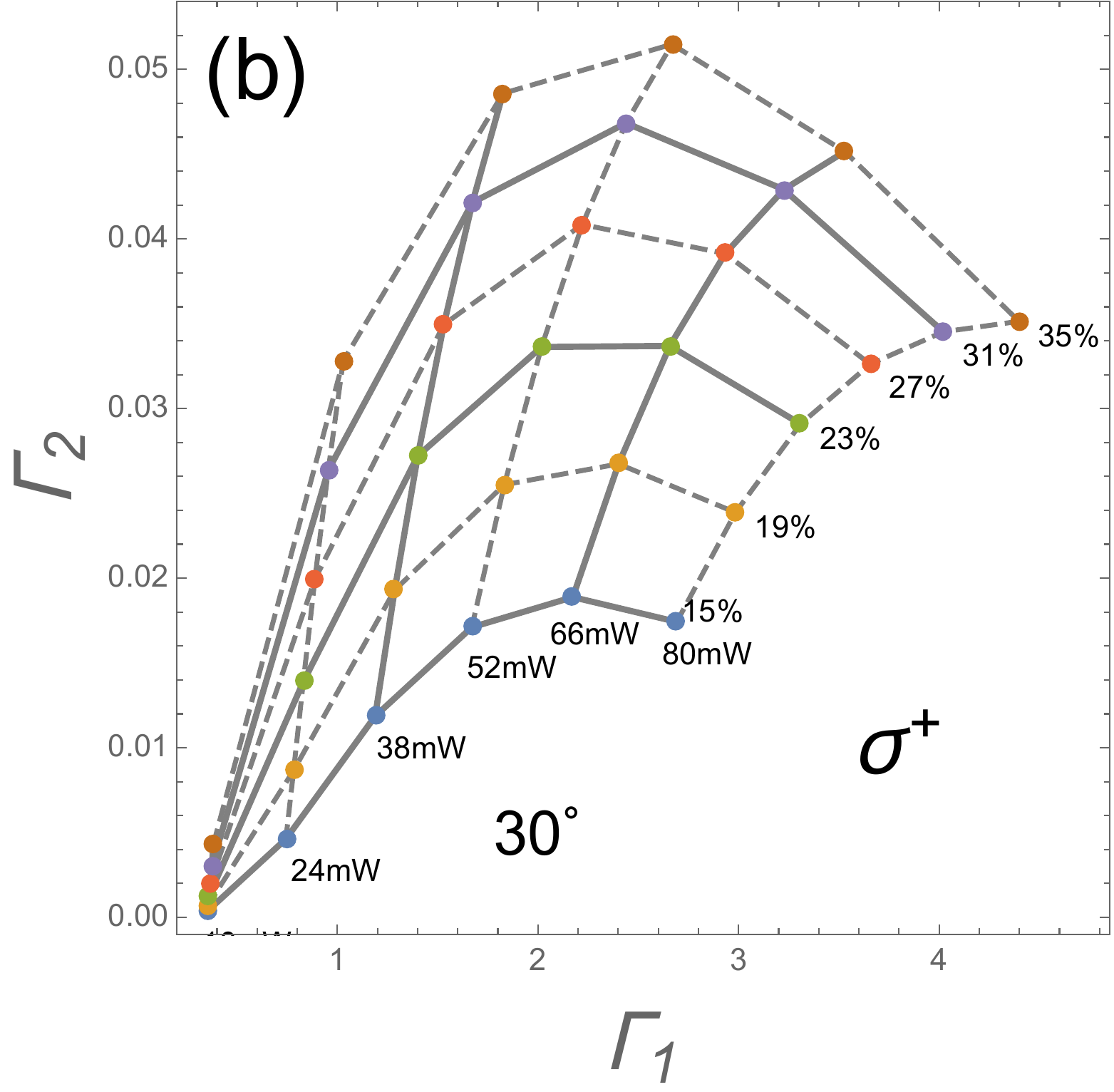}
\includegraphics[width=0.32\textwidth,keepaspectratio=true]
{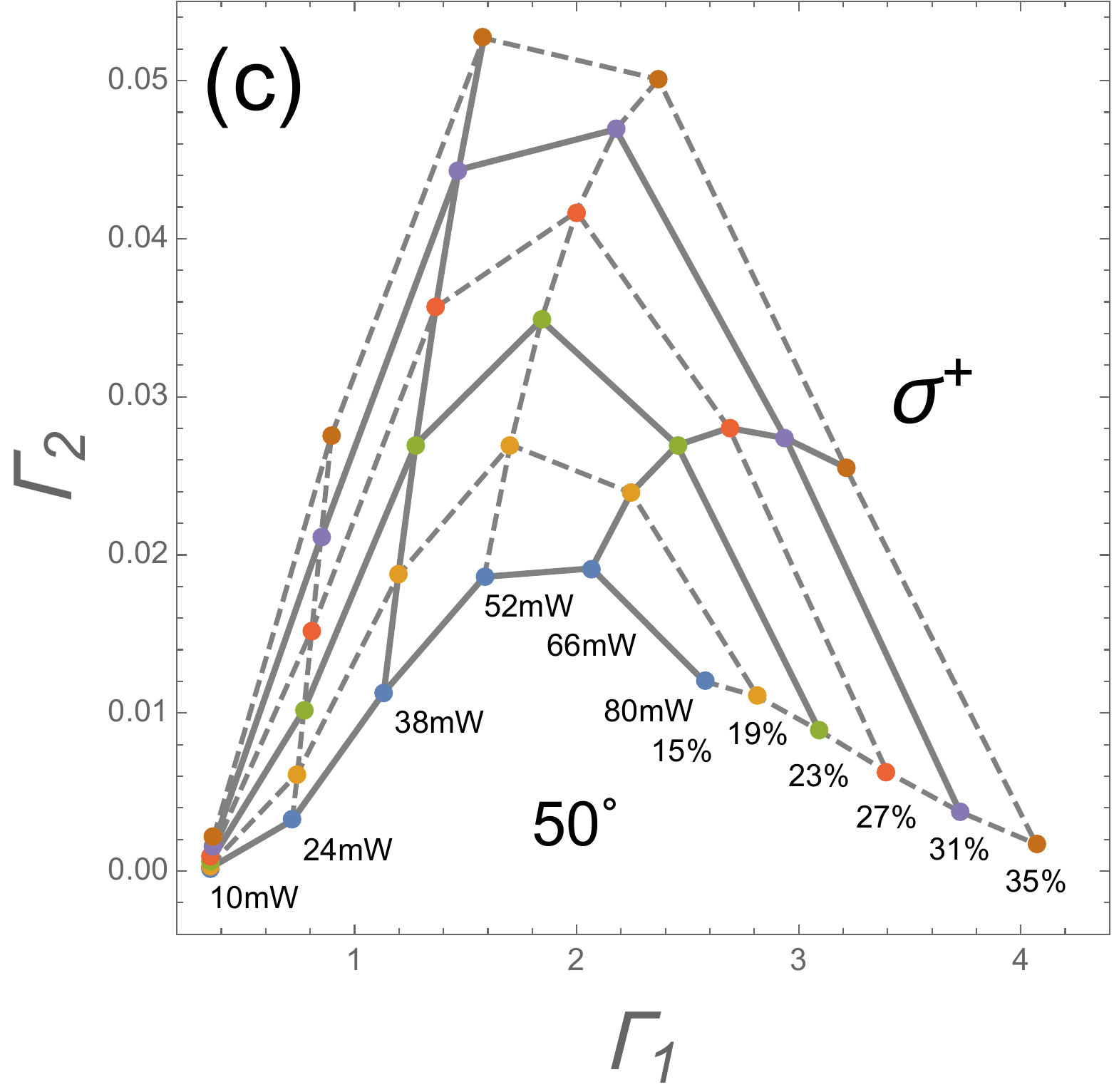}
\caption{
Power and degree of circular polarization
isolines as a function of $\Gamma_1$ and $\Gamma_2$
under $\sigma^+$ illumination at a relative
angle of
(a) $0^{\circ}$, (b) $30^{\circ}$ and (b) $50^{\circ}$. }
\label{figure6}
\end{figure*}

Figures \ref{figure5} and \ref{figure6} show
the power and degree of polarization
isolines on the $\Gamma_1-\Gamma_2$ plane
under $\sigma^-$ and $\sigma^+$ illumination.
Under normal incidence ($\theta=0$)
the degree of circular polarization
$P_{\mathrm{c}}$ and the power $P_{\mathrm{exc}}$
can be determined by interpolating
between the isolines in Figs. \ref{figure5}a or
\ref{figure6}a depending on if
$\Gamma_2<0$ ($\sigma^-$) or $\Gamma_2>0$ ($\sigma^+$).
The progression of
Figs. \ref{figure5}a, b, c and
Figs. \ref{figure6}a, b, c, shows that despite the
slight deformation due to the variation
in the incidence angle, the isolines preserve
their main topological properties.
Moreover, it can be readily proven from the Fresnel equations,
that the refracted light remains elliptically polarized
even for incidence angles far from $\theta=0^\circ$.
Hence, the values of $P_{\mathrm{c}}$ and $P_{\mathrm{exc}}$
can be determined for a very wide range
of incidence angles that approximately goes
from $\theta=0^\circ$ to $\theta=50^\circ$.

\section{Conclusions}\label{sec:conclusions}
In summary, we have developed the concept of
a spin-optoelectronic detector for the simultaneous
measurement of 
the degree of circular polarization, its handedness and
the intensity of a light excitation.
We report on a theoretical analysis and numerical simulations
based on a master equation approach
that has been previously shown to yield very
good quantitative agreement with experimental results
shown in Fig. \ref{figure7}.
This all-electronic compact device
would operate at room temperature
requiring no additional bulky movable parts.
It relies on
the unique spin selection rules of GaAs and 
spin-dependent capture of CB electrons that takes
place through the Ga$^{2+}$ centers in GaAsN.
The hyperfine interaction that couples bound
electrons and nuclei in the centers induces
an asymmetry in the electron and hole population
that is sensitive to the circular polarization orientation
of the incident light.
The device consists of two independent GaAsN based
independent detectors each subject to different
magnetic fields generated by small permanent magnets.
The values of the magnetic fields are tuned to
enhance the sensitivity of each detector
to a given property of light. 
We expect that
our proposed circular polarimeter
will allow for the possibility of
simultaneously detecting the intensity and
the degree of circular polarization
in an integrated optoelectronic platform.

\section{Acknowledgements}
We acknowledge funding from LIA CNRS-Ioffe RAS ILNACS.
A.K., J.C.S.S. and V.G.I.S. gratefully 
appreciate the financial support of Departamento de Ciencias B\'asicas 
UAM-A grant numbers 2232214 and 2232215.
X.M. also thanks Institut
Universitaire de France.
This work was supported by Programme Investissements d'Avenir
under the program ANR-11-IDEX-0002-02, reference ANR-10-LABEX-0037-NEXT.

\appendix

\section{Model}\label{ap:model}
In this appendix we present a summary of the
method used for the computation of
the quantum statistical averages of the
electron and hole populations $\av{n}$ and
$\av{p}$, necessary to obtain the conductivity
$\sigma_{xx}$.
For a detailed presentation of this
method see Ref. [\onlinecite{PhysRevB.101.075201}].

Any quantum statistical average may be obtained
from the trace of density matrix multiplied by
the appropriate operator.
The density
matrix $\rho$, in turn, is ensued from the master
equation\cite{PhysRevB.101.075201}
\begin{equation}
\frac{d\rho}{dt}=\frac{i}{\hbar}\left[\rho,H\right]
  +\mathcal{D}\left(\rho\right),
\label{eq:master}
\end{equation}
where $H$ is
the Hamiltonian and $\mathcal{D}$
is the dissipator.

The density matrix may be expressed
as the direct sum
\begin{equation}
\rho=\rho_v \oplus
 \rho_c\oplus
 \rho_1 \oplus
 \rho_2 ,
\end{equation}
where $\rho_v$ and $\rho_c$ are the
density submatrices for valence band and conduction
band electrons respectively. 
The density submatrices $\rho_1$ and $\rho_2$
correspond to singly and double occupied
Ga$^{2+}$ centers.
The density matrix, as any other matrix
relevant to \eqref{eq:master}, can be conveniently
expanded in terms of the elements
of an internal space of Hermitian matrices
\cite{PhysRevB.101.075201}
\begin{multline}
\Lambda=\left\{p,S_k,U_{k,j,i},V_{j,i},
\right\}
=\left\{\lambda_1,\lambda_2,\dots,\lambda_{d}\right\},\\
\,\,\,\, i,j,k=0,1,2,3\,\, , \,\,\, d=85 ,\label{base}
\end{multline}
where
\begin{eqnarray}
p&=&1_{1\times 1}\oplus 
  0_{2\times 2}
  \oplus 0_{8\times 8} 
  \oplus 0_{4\times 4},
\\
S_k&=& 0_{1\times 1}
  \oplus \left(s_k\right)
  \oplus 0_{8\times 8} 
  \oplus 0_{4\times 4},
\\
U_{k,j,i} &=& 0_{1\times 1}
  \oplus 0_{2\times 2}\oplus
  \left(s_k\otimes s_j\otimes s_i\right)
  \oplus 0_{4\times 4},
\\
V_{j,i} &=& 0_{1\times 1} 
  \oplus 0_{2\times 2}
  \oplus 0_{8\times 8}
  \oplus\left(s_j\otimes s_i\right),
\end{eqnarray}
with $i,j,k=0,1,2,3$.
Here, $s_0=1_{2\times 2}/2$ where $1_{2\times 2}$ 
is the $2\times 2$ identity matrix and
$s_{1,2,3}$ are the spin matrices
following the standard commutation rules
\begin{equation}
\left[s_i,s_j\right]=i\hbar \sum_k\epsilon_{i,j,k}s_k,
\,\,\,\, i,j,k=1,2,3.
\end{equation}
This particular choice of a matrix basis
is advantageous for a number of reasons.
First, any operator
can be expressed as a linear combination
of the elements of $\Lambda$. Second,
its elements are orthogonal with respect
to the trace,
\begin{equation}
    \mathrm{tr}\left[\lambda_q^{\dagger}\lambda_{q^\prime}\right]=
    \mathrm{tr}\left[\lambda_q^2\right]\delta_{q,q^\prime},
\end{equation}
which is related to the quantum statistical average
of physical observables.
Of particular importance are the quantum statistical averages
of the elements of $\Lambda$,
\begin{equation}
\av{\lambda}_q=\Tr\left[\rho \lambda_q\right],
\end{equation}
whereby the density matrix can be expanded as
\begin{multline}
\rho=
\sum_{q=1}^d
\frac{\av{\lambda_q}\lambda_q}{\Tr\left[\lambda_q^2\right]}
  =\frac{\av{p}p}{\Tr\left[p^2\right]}
  +\sum_{k=0}^3\frac{\av{S}_kS_k}{\Tr\left[S_k^2\right]}
  \\
  +\sum_{k,j,i=0}^3\frac{\av{U}_{k,j,i}U_{k,j,i}}
  {\Tr\left[U_{k,j,i}^2\right]}
  +\sum_{j,i=0}^3\frac{\av{V}_{j,i}V_{j,i}}
  {\Tr\left[V_{j,i}^2\right]}\,\,\, .
\label{eq:densmat01}
\end{multline}
In general, any operator $O$ can be expanded in this
basis as
\begin{multline}
    O=\sum_{q=1}^d
    \frac{\mathrm{Tr}\left[O\lambda_q\right]}
    {\mathrm{Tr}\left[\lambda_q^2\right]}\lambda_q
    =\frac{\Tr\left[Op\right]p}{\Tr\left[p^2\right]}
  +\sum_{k=0}^3\frac{\Tr\left[OS_k\right]S_k}{\Tr\left[S_k^2\right]}
  \\
  +\sum_{k,j,i=0}^3\frac{\Tr\left[OU_{k,j,i}\right]U_{k,j,i}}
  {\Tr\left[U_{k,j,i}^2\right]}
  +\sum_{j,i=0}^3\frac{\Tr\left[OV_{j,i}\right]V_{j,i}}
  {\Tr\left[V_{j,i}^2\right]}\,\,\, ,
    \label{eq:decompose}
\end{multline}
and its corresponding quantum statistical average is given by
\begin{equation}
\av{O}=\Tr\left[O\rho\right]=
\sum_{q=1}^d\frac{\Tr\left[O\lambda_q\right]}
{\Tr\left[\lambda_q^2\right]}
\av{\lambda}_q\,\,\, .\label{eq:averagedecomp}
\end{equation}
In this way,
the density of VB holes is represented by the matrix $p$,
the density of CB electrons by $n=2S_0$,
the spin components of CB electrons by $S_k$ ($k=1,2,3$),
the concentration of singly occupied traps by $N_1=8U_{0,0,0}$,
the spin components of bound electrons in Ga$^{2+}$ centers
by $S_{ck} = 4U_{k,0,0}$ ($k=1,2,3$)
and the concentration of doubly occupied traps by
$N_2=4V_{0,0}$.
Using \eqref{eq:decompose}
we can also decompose more complicated operators.
For instance, the components
of the nuclear spin operators of
singly charged centers $\boldsymbol{I}_1$
and doubly charged centers $\boldsymbol{I}_2$
can readily be expressed
as a superposition of the elements of (\ref{base}) by
\begin{eqnarray}
I_{1,k} &=& \sum_{j,i=0}^3 
  \frac{ \Tr\left[I_{1,k}U_{0,j,i}\right]}
  {\Tr\left[U_{0,j,i}U_{0,j,i}\right]}U_{0,j,i}\,\, ,\\
I_{2,k} &=& \sum_{j,i=0}^3 
  \frac{ \Tr\left[I_{2,k}V_{j,i}\right]}
  {\Tr\left[V_{j,i}V_{j,i}\right]}V_{j,i}\,\,.
\end{eqnarray}
Note that in the previous expressions only
a few elements of the basis are needed.

The Hamiltonian in \eqref{eq:master}
\begin{equation}
H=\hbar\boldsymbol{\omega}\cdot\boldsymbol{S}
+\hbar\boldsymbol{\Omega}\cdot\boldsymbol{S}_c
+A\boldsymbol{I}_1\cdot \boldsymbol{S}_c,
\end{equation}
accounts for the Zeeman and hyperfine interactions.
The first two terms correspond to
the Zeeman interaction of an external magnetic field
$\boldsymbol{B}$ with the CB electrons spin $\boldsymbol{S}$
and the centers bound electrons spin $\boldsymbol{S}_c$.
In these expressions
$\boldsymbol{\omega}=g\mu_B\boldsymbol{B}/\hbar$,
$\boldsymbol{\Omega}=g_c\mu_B\boldsymbol{B}/\hbar$,
where $\mu_B$ is the Bohr
magneton, $g$ is the gyromagnetic factor for
CB electrons and $g_c=2$\, \cite{PhysRevB.95.195204}
is the gyromagnetic factor for bound electrons in
singly occupied centers.
The third term is responsible for the hyperfine
interaction that takes place in singly occupied centers
between the bound electron spin $\boldsymbol{S}_c$
and the nuclear spin $\boldsymbol{I}_1$.
The hyperfine parameter is $A$.
In doubly occupied traps electrons form
a singlet state that does not interact
with the nuclear spin $\boldsymbol{I}_2$.

The dissipator
\begin{equation}
\mathcal{D}\left(\rho\right)=\mathcal{G}
   +\mathcal{D}_S
   +\mathcal{D}_{SC}
   +\mathcal{D}_1+\mathcal{D}_2
   +\mathcal{D}_{SDR}+\mathcal{D}_{P},
\end{equation}
is primarily composed of
generation ($\mathcal{G}$),
spin relaxation
($\mathcal{D}_S$ and $\mathcal{D}_{SC}$,
$\mathcal{D}_{1}$ and $\mathcal{D}_2$)
and recombination terms ($\mathcal{D}_{SDR}$ and $\mathcal{D}_{P}$)
The electron-hole pair generating term is given by
\begin{equation}
\mathcal{G}=\left(G_{+}+G_{-}\right)\left(p+n\right)
+2\left(G_{+}-G_{-}\right)\boldsymbol{e}\cdot \boldsymbol{S},
\end{equation}
where $\boldsymbol{e}$ is a unitary vector in
the excitation direction.
Spin-up and spin-down CB electron generation rates
are given by the smooth step function
\begin{equation}
G(t)_\pm =\frac{G_0 P_{exc}}{2}
\frac{1\pm P_{e}}{2}\left[
1+\tanh\left(\frac{t-t_0}{\Delta t}\right)
\right],
\end{equation}
where $P_{exc}$ is the excitation power,
$G_0$ is the
power to generation factor,
$\Delta t$ is the width of the step function
and the spin polarization degree is parametrized
by $P_{e}\in\left[-1,1\right]$.

The electronic and nuclear spin relaxation
dissipators are in general derived from
Wangsness - Bloch - Redfield theory
\cite{PhysRev.89.728,REDFIELD19651,
PhysRev.142.179,PhysRevB.95.195204}.
The dissipator
for CB electrons straightforwardly gives
\begin{equation}
\mathcal{D}_S=-\frac{1}{2\tau_s}
\sum_{k=1}^{3}\left[S_k,\left[S_k,\rho\right]
\right].
\end{equation}
The spin relaxation dissipators in 
centers require closer attention.
Firstly, in singly occupied centers
it is important to distinguish
two different electronic
and nuclear relaxation times:
one belonging to the electronic and nuclear spins
themselves
and the other to the correlation
between electronic and nuclear spins
\cite{PhysRevB.101.075201}.
These two times are quite different in magnitude.
Secondly, in singly and doubly occupied traps,
we assume that
dipole-dipole interaction between
centers and neighbouring Ga atoms is the
leading mechanism of nuclear spin relaxation
\cite{PhysRevB.95.195204}.
The dissipator for the bound electron spin
can thus be expanded in terms
of the $\Lambda$ subbasis elements
corresponding to the singly occupied traps
\begin{multline}
\mathcal{D}_{SC}=-\frac{1}{2\tau_{sc}} \sum_{k=1}^3
\frac{\mathrm{Tr}\left[\Gamma_S U_{k,0,0}\right]}
 {\mathrm{Tr}\left[U_{k,0,0}U_{k,0,0}\right]}
 U_{k,0,0}\\
 -\frac{1}{2\tau_{sco}} \sum_{k,j,i=0}^3
\mu_{k,j,i}
\frac{\mathrm{Tr}\left[\Gamma_S U_{k,j,i}\right]}
 {\mathrm{Tr}\left[U_{k,j,i}U_{k,j,i}\right]}
 U_{k,j,i} ,
 \label{eq:dissipatordsc}
\end{multline}
where
\begin{equation}
    \Gamma_S 
    = \sum_{k=1}^{3}\left[S_{ck},\left[S_{ck},\rho\right]
     \right]
\end{equation}
is the usual commutator arising from
Wangsness - Bloch - Redfield theory for dipole-dipole
interaction.
\begin{figure}
\includegraphics[width=0.45\textwidth,keepaspectratio=true]
{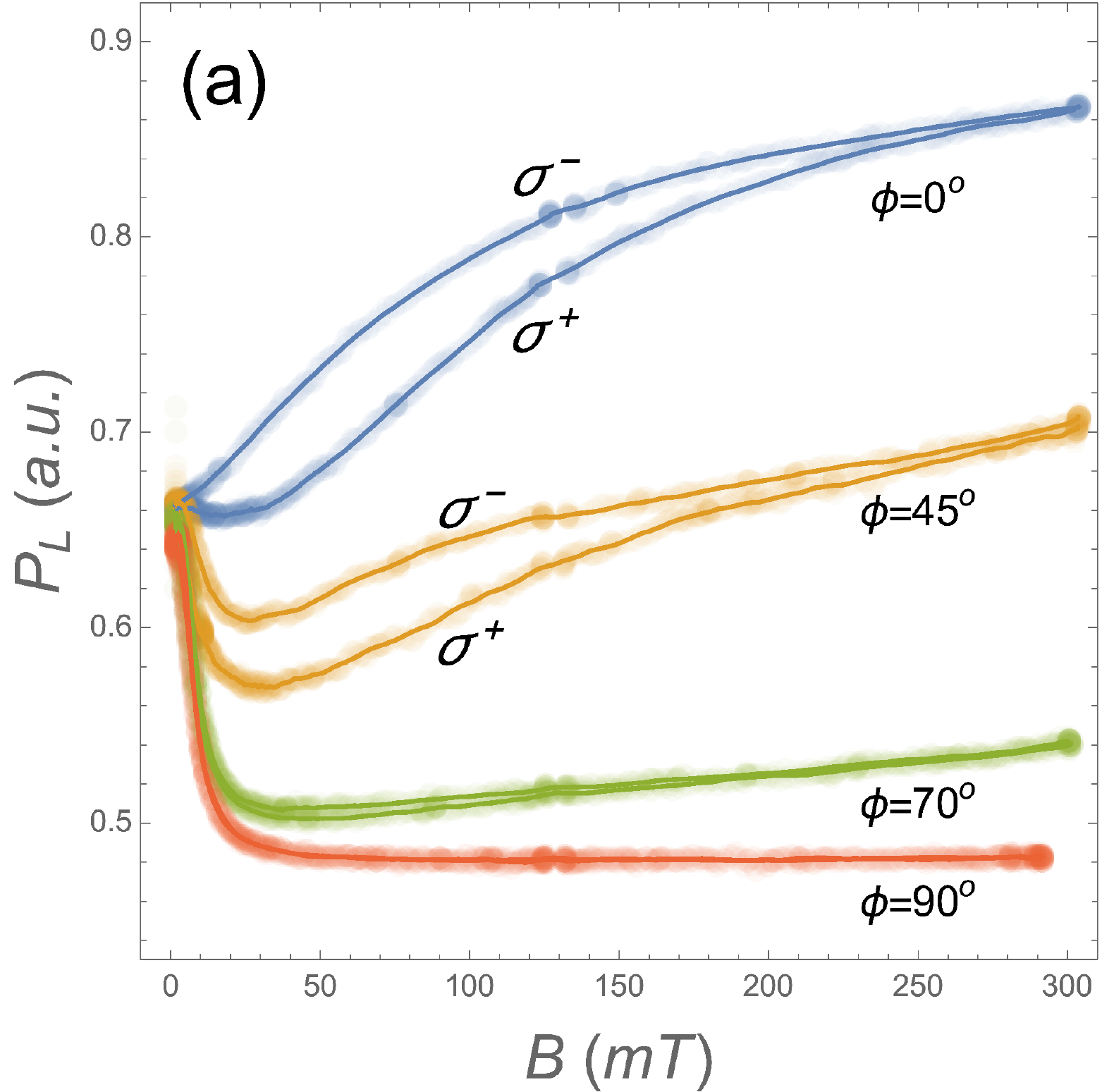}
\includegraphics[width=0.45\textwidth,keepaspectratio=true]
{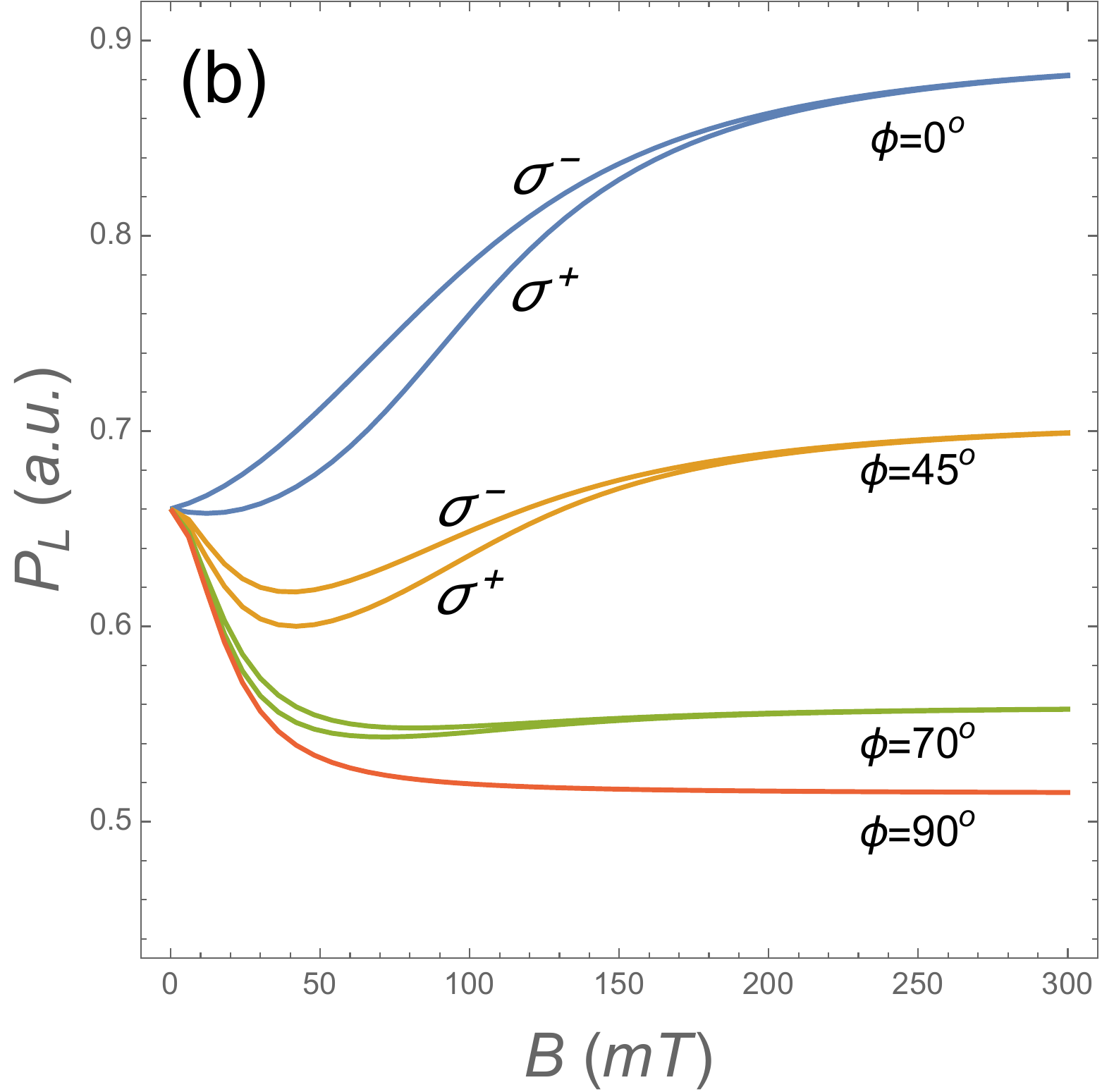}
\caption{(a) Experimental and (b) theoretical
results for the photoluminescence $P_L$  as a function
of an oblique magnetic field at an incident power
of $20$mW. The magnetic field
is rotated from Faraday ($\varphi=0^{\circ}$)
to Voight ($\varphi=90^{\circ}$) configuration.
In panel (a) the dots correspond to the experimental
results and the solid lines are a guide to the eye.
}
\label{figure7}
\end{figure}
The first line of \eqref{eq:dissipatordsc}
is the projection of $\Gamma_S$ on to
the electron spin subbasis of $\Lambda$, namely,
the spin components of the bound electron $S_{ck}=4U_{k,0,0}$ for
$k=1,2,3$.
Accordingly, the second line of \eqref{eq:dissipatordsc}
is the projection of
$\Gamma_S$ on to the subbasis of electron-nucleus
correlations which are selected by
\begin{equation}
\mu_{k,j,i}=\begin{cases}
0,\,\,\,\,  k=0 \,\, \lor  \,\, (k\ne 0 \land j=0\land i=0) \,\, \\
1,\,\,\,\,  k\ne 0 \land (j\ne 0 \lor i\ne 0)\,\,  \\
\end{cases}.
\end{equation}
Each one of these projections is governed by a different
relaxation time:  $\tau_{sc}$ is
the centers' electronic spin relaxation time and
$\tau_{sco}$ is the relaxation time of correlations \cite{PhysRevB.101.075201}.
Similarly, the nuclear spin relaxation
dissipator is expanded as
\begin{multline}
\mathcal{D}_{1}=-\frac{1}{3\tau_{n1}} \sum_{k=1}^3
\frac{\mathrm{Tr}\left[\Gamma_I U_{k,0,0}\right]}
 {\mathrm{Tr}\left[U_{k,0,0}U_{k,0,0}\right]}
 U_{k,0,0}\\
 -\frac{1}{2\tau_{n1co}} \sum_{k,j,i=0}^3
\mu_{k,j,i}
\frac{\mathrm{Tr}\left[\Gamma_I U_{k,j,i}\right]}
 {\mathrm{Tr}\left[U_{k,j,i}U_{k,j,i}\right]}
 U_{k,j,i},
\end{multline}
where
\begin{equation}
    \Gamma_I 
    = \sum_{k=1}^{3}\left[I_{1k},\left[I_{1k},\rho\right]
     \right] 
\end{equation}
comes out of the dipole-dipole interaction
between the centers nuclear spin and the neighbouring
Ga nuclei.
The dissipator of doubly occupied centers
reads
\begin{equation}
    \mathcal{D}_2=-\frac{1}{3\tau_{n2}}
\sum_{k=1}^{3}\left[I_{2k},\left[I_{2k},\rho\right]
\right].
\end{equation}
Since in doubly occupied centers both electrons residing
in the 4s orbital form a singlet state
which is not correlated to the
nuclear spin,
$\mathcal{D}_2 $ needs not be separated 
into the spin and correlation parts.

The spin selective capture of electrons
into singly occupied centers is
accounted for by the dissipator\cite{PhysRevB.101.075201}
\begin{multline}
\mathcal{D}_{SDR}=
  -2\sum_{k=0}^3
  \left(4c_n\sum_{k^{\prime},k^{\prime\prime}=0}^3
  \av{S}_{k^{\prime}}
  Q^{\top}_{k,k^{\prime},k^{\prime\prime}}
  \av{U}_{k^{\prime\prime},j,i}\right)
  S_{k}\\
 -8\sum_{k,j,i=0}^3\mu_{k,j,i}
 \left(c_n\sum_{k^{\prime},k^{\prime\prime}=0}^3
  \av{S}_{k^{\prime}}
  Q_{k,k^{\prime},k^{\prime\prime}}
  \av{U}_{k^{\prime\prime},j,i}\right)
  U_{k,j,i}\\
 +4\sum_{j,i=0}^3
  \left(2c_n\sum_{k^{\prime},k^{\prime\prime}=0}^3
  \av{S}_{k^{\prime}}
  Q^{\top}_{0,k^{\prime},k^{\prime\prime}}
  \av{U}_{k^{\prime\prime},j,i}\right)
  V_{j,i}\,\,\, ,\label{eq:dissdrnew}
\end{multline}
where the capture coefficient of CB electrons $c_n=1/N_0\tau^*$,
$\tau^*$ is the electron recombination time in the
low excitation power regime and
$N_0$ is the total number of centers in the sample.
The matrices $Q_{k,k^{\prime},k^{\prime\prime}}
=\left(\mathbb{Q}_k\right)_{k^{\prime},k^{\prime\prime}}$,
related to the space rotations generators \cite{book:729155},
are responsible for the spin dependent capture. These
are defined by
\begin{align}
&\mathbb{Q}_0 =    \left(
\begin{array}{cccc}
 1 & 0 & 0 & 0 \\
 0 & -1 & 0 & 0 \\
 0 & 0 & -1 & 0 \\
 0 & 0 & 0 & -1 \\
\end{array}
\right),&
\mathbb{Q}_1 = \left(
\begin{array}{cccc}
 0 & 1 & 0 & 0 \\
 -1 & 0 & 0 & 0 \\
 0 & 0 & 0 & 0 \\
 0 & 0 & 0 & 0 \\
\end{array}
\right),& \nonumber \\
& \mathbb{Q}_2 =\left(
\begin{array}{cccc}
 0 & 0 & 1 & 0 \\
 0 & 0 & 0 & 0 \\
 -1 & 0 & 0 & 0 \\
 0 & 0 & 0 & 0 \\
\end{array}
\right), &
\mathbb{Q}_3 =\left(
\begin{array}{cccc}
 0 & 0 & 0 & 1 \\
 0 & 0 & 0 & 0 \\
 0 & 0 & 0 & 0 \\
 -1 & 0 & 0 & 0 \\
\end{array}
\right).
\end{align}
Note that while $\mathbb{Q}_0$ remains invariant under
rotations,
$\mathbb{Q}_1$, $\mathbb{Q}_2$ and $\mathbb{Q}_3$
rotate as the components of a three-dimensional vector.

The recombination of VB holes into
doubly occupied centers is governed
by the dissipator\cite{PhysRevB.101.075201}
\begin{multline}
\mathcal{D}_P= -\left(4 c_p\, \av{p}\,\av{V}_{0,0} \right)p
  +\frac{1}{8}\left(\frac{1}{2}
  \sum_{j,i=0}^3\av{p}\,\av{U}_{0,j,i}\right)U_{0,j,i}\\
-\frac{1}{4}\left(\sum_{j,i=0}^3\av{p}\,\av{V}_{j,i}\right)V_{j,i}
\,\,\, ,
\end{multline}
where $c_p=1/N_0\tau_h$ is the capture coefficient for holes
and $\tau_h$ is the hole recombination time in the
high excitation power regime.

Having established the components of
the master equation we now turn our
attention to
the determination of the time dependence
of observables such as $\av{n}$ and $\av{p}$.
Instead of directly solving the system of ordinary
differential equations arising from \refeq{eq:master} for
the matrix elements of $\rho$,
we work out the system of ordinary differential equations
for the statistical averages of the elements of
$\Lambda$. This procedure yields $d=85$ differential
equations of the form
\begin{eqnarray}
\frac{d\av{\lambda}_q}{dt}
  =\frac{i}{\hbar}\Tr\left[\left[H,\lambda_q\right]
  \rho\right]
  +\Tr\left[\mathcal{D}\left(\rho\right) \lambda_q\right].
  \label{eq:solmodel01}
\end{eqnarray}
These are solved numerically
over a sufficiently long time ($\approx 400$ns) to allow
steady state conditions to be reached.
Finally, the quantum average of any operator
can be computed from the decomposition
\eqref{eq:averagedecomp}.
In particular, the CB electron and VB hole densities
are
\begin{eqnarray}
  \av{n}=\sum_{q=1}^d\frac{\Tr\left[n\lambda_q\right]}
{\Tr\left[\lambda_q^2\right]}
\av{\lambda}_q,\label{eq:nav01}\\
  \av{p}=\sum_{q=1}^d\frac{\Tr\left[p\lambda_q\right]}
{\Tr\left[\lambda_q^2\right]}
\av{\lambda}_q.\label{eq:pav01}
\end{eqnarray}

In order for the model to yield quantitatively
realistic results, we have
fitted its parameters using experimental measurements of the
photoluminescence for various incident powers
and magnetic field orientations.
Varying the magnetic field orientation provides
a complete set of experimental points,
especially useful to determine the electronic
and nuclear spin relaxation times
$\tau_h$, $\tau_r$, $\tau^*$, $\tau_s$,
$\tau_{sc}$, $\tau_{sco}$, $\tau_{n1}$,
$\tau_{n1co}$ and $\tau_{n2}$.
The photoluminescence, on the other hand,
yields information on the power to generation
factor $G_0$ and the overall number of
paramagnetic centers $N_c=N_1+N_2$ in the sample.
The photoluminescence measurements were
performed on a
$100$ nm thick GaAs$_{1-x}$N$_x$ epilayer ($x=0.021$)
grown by molecular beam epitaxy on a (001) semi-insulating
GaAs substrate and capped with $10$\,nm GaAs.
The room temperature epilayer gap is $1080$\,nm.
The excitation light was provided by a
$850$\,nm laser diode.
Figure \ref{figure7} shows a comparison between
the measured and the theoretical photoluminescence
at an incident power of $20$\,mW. The model was also
fitted using different incident powers which are not
shown in this figure.
The parameters that best fit experimental results
on photoluminescence are
$G_0 = 3 \times 10^{23}$\,cm$^{-3}$\,mW$^{-1}$,
$\Delta t=10$\,ps,
$N_c=N_1+N_2=2.2\times 10^{15}$\,cm$^{-3}$,
$\tau_h=20$\,ps,
$\tau_r=1.4$\,ns,
$\tau^*=10$\,ps,
$\tau_s=120$\,ps,
$\tau_{sc}=2$\,ns,
$\tau_{sco} = 18$\,ns,
$\tau_{n1}=2$\,ns,
$\tau_{n1co}=8$\,ns,
$\tau_{n2}=100$\,ps,
$A=0.069$\,cm$^{-1}$ \, \cite{puttisong2013efficient},
$g=1$ and  $g_c = 2$\,\,
\cite{PhysRevB.95.195204,PhysRevB.101.075201}.


\begin{thebibliography}{67}%
\makeatletter
\providecommand \@ifxundefined [1]{%
 \@ifx{#1\undefined}
}%
\providecommand \@ifnum [1]{%
 \ifnum #1\expandafter \@firstoftwo
 \else \expandafter \@secondoftwo
 \fi
}%
\providecommand \@ifx [1]{%
 \ifx #1\expandafter \@firstoftwo
 \else \expandafter \@secondoftwo
 \fi
}%
\providecommand \natexlab [1]{#1}%
\providecommand \enquote  [1]{``#1''}%
\providecommand \bibnamefont  [1]{#1}%
\providecommand \bibfnamefont [1]{#1}%
\providecommand \citenamefont [1]{#1}%
\providecommand \href@noop [0]{\@secondoftwo}%
\providecommand \href [0]{\begingroup \@sanitize@url \@href}%
\providecommand \@href[1]{\@@startlink{#1}\@@href}%
\providecommand \@@href[1]{\endgroup#1\@@endlink}%
\providecommand \@sanitize@url [0]{\catcode `\\12\catcode `\$12\catcode
  `\&12\catcode `\#12\catcode `\^12\catcode `\_12\catcode `\%12\relax}%
\providecommand \@@startlink[1]{}%
\providecommand \@@endlink[0]{}%
\providecommand \url  [0]{\begingroup\@sanitize@url \@url }%
\providecommand \@url [1]{\endgroup\@href {#1}{\urlprefix }}%
\providecommand \urlprefix  [0]{URL }%
\providecommand \Eprint [0]{\href }%
\providecommand \doibase [0]{http://dx.doi.org/}%
\providecommand \selectlanguage [0]{\@gobble}%
\providecommand \bibinfo  [0]{\@secondoftwo}%
\providecommand \bibfield  [0]{\@secondoftwo}%
\providecommand \translation [1]{[#1]}%
\providecommand \BibitemOpen [0]{}%
\providecommand \bibitemStop [0]{}%
\providecommand \bibitemNoStop [0]{.\EOS\space}%
\providecommand \EOS [0]{\spacefactor3000\relax}%
\providecommand \BibitemShut  [1]{\csname bibitem#1\endcsname}%
\let\auto@bib@innerbib\@empty
\bibitem [{\citenamefont {Sparks}\ \emph {et~al.}(2019)\citenamefont {Sparks},
  \citenamefont {Germer},\ and\ \citenamefont {Sparks}}]{sparks2019classical}%
  \BibitemOpen
  \bibfield  {author} {\bibinfo {author} {\bibfnamefont {W.~B.}\ \bibnamefont
  {Sparks}}, \bibinfo {author} {\bibfnamefont {T.~A.}\ \bibnamefont {Germer}},
  \ and\ \bibinfo {author} {\bibfnamefont {R.~M.}\ \bibnamefont {Sparks}},\
  }\href@noop {} {\bibfield  {journal} {\bibinfo  {journal} {Publications of
  the Astronomical Society of the Pacific}\ }\textbf {\bibinfo {volume}
  {131}},\ \bibinfo {pages} {075002} (\bibinfo {year} {2019})}\BibitemShut
  {NoStop}%
\bibitem [{\citenamefont {Whittaker}\ \emph {et~al.}(1994)\citenamefont
  {Whittaker}, \citenamefont {Kloner}, \citenamefont {Boughner},\ and\
  \citenamefont {Pickering}}]{whittaker1994quantitative}%
  \BibitemOpen
  \bibfield  {author} {\bibinfo {author} {\bibfnamefont {P.}~\bibnamefont
  {Whittaker}}, \bibinfo {author} {\bibfnamefont {R.}~\bibnamefont {Kloner}},
  \bibinfo {author} {\bibfnamefont {D.}~\bibnamefont {Boughner}}, \ and\
  \bibinfo {author} {\bibfnamefont {J.}~\bibnamefont {Pickering}},\ }\href@noop
  {} {\bibfield  {journal} {\bibinfo  {journal} {Basic research in cardiology}\
  }\textbf {\bibinfo {volume} {89}},\ \bibinfo {pages} {397} (\bibinfo {year}
  {1994})}\BibitemShut {NoStop}%
\bibitem [{\citenamefont {Louie}\ \emph {et~al.}(2018)\citenamefont {Louie},
  \citenamefont {Tchvialeva}, \citenamefont {Kalia}, \citenamefont {Lui},\ and\
  \citenamefont {Lee}}]{10.1117/12.2288761}%
  \BibitemOpen
  \bibfield  {author} {\bibinfo {author} {\bibfnamefont {D.~C.}\ \bibnamefont
  {Louie}}, \bibinfo {author} {\bibfnamefont {L.}~\bibnamefont {Tchvialeva}},
  \bibinfo {author} {\bibfnamefont {S.}~\bibnamefont {Kalia}}, \bibinfo
  {author} {\bibfnamefont {H.}~\bibnamefont {Lui}}, \ and\ \bibinfo {author}
  {\bibfnamefont {T.~K.}\ \bibnamefont {Lee}},\ }in\ \href {\doibase
  10.1117/12.2288761} {\emph {\bibinfo {booktitle} {Photonics in Dermatology
  and Plastic Surgery 2018}}},\ Vol.\ \bibinfo {volume} {10467},\ \bibinfo
  {editor} {edited by\ \bibinfo {editor} {\bibfnamefont {B.}~\bibnamefont
  {Choi}}\ and\ \bibinfo {editor} {\bibfnamefont {H.}~\bibnamefont {Zeng}}},\
  \bibinfo {organization} {International Society for Optics and Photonics}\
  (\bibinfo  {publisher} {SPIE},\ \bibinfo {year} {2018})\ pp.\ \bibinfo
  {pages} {44 -- 52}\BibitemShut {NoStop}%
\bibitem [{\citenamefont {Chang}\ \emph {et~al.}(2016)\citenamefont {Chang},
  \citenamefont {He}, \citenamefont {Wang}, \citenamefont {Huang},
  \citenamefont {Li}, \citenamefont {He}, \citenamefont {Liao}, \citenamefont
  {Zeng}, \citenamefont {Liu},\ and\ \citenamefont
  {Ma}}]{10.1117/1.JBO.21.5.056002}%
  \BibitemOpen
  \bibfield  {author} {\bibinfo {author} {\bibfnamefont {J.}~\bibnamefont
  {Chang}}, \bibinfo {author} {\bibfnamefont {H.}~\bibnamefont {He}}, \bibinfo
  {author} {\bibfnamefont {Y.}~\bibnamefont {Wang}}, \bibinfo {author}
  {\bibfnamefont {Y.}~\bibnamefont {Huang}}, \bibinfo {author} {\bibfnamefont
  {X.}~\bibnamefont {Li}}, \bibinfo {author} {\bibfnamefont {C.}~\bibnamefont
  {He}}, \bibinfo {author} {\bibfnamefont {R.}~\bibnamefont {Liao}}, \bibinfo
  {author} {\bibfnamefont {N.}~\bibnamefont {Zeng}}, \bibinfo {author}
  {\bibfnamefont {S.}~\bibnamefont {Liu}}, \ and\ \bibinfo {author}
  {\bibfnamefont {H.}~\bibnamefont {Ma}},\ }\href {\doibase
  10.1117/1.JBO.21.5.056002} {\bibfield  {journal} {\bibinfo  {journal}
  {Journal of Biomedical Optics}\ }\textbf {\bibinfo {volume} {21}},\ \bibinfo
  {pages} {1 } (\bibinfo {year} {2016})}\BibitemShut {NoStop}%
\bibitem [{\citenamefont {Gao}\ \emph {et~al.}(2012)\citenamefont {Gao},
  \citenamefont {Fallahi}, \citenamefont {Togan}, \citenamefont
  {Miguel-S{\'a}nchez},\ and\ \citenamefont {Imamoglu}}]{gao2012observation}%
  \BibitemOpen
  \bibfield  {author} {\bibinfo {author} {\bibfnamefont {W.}~\bibnamefont
  {Gao}}, \bibinfo {author} {\bibfnamefont {P.}~\bibnamefont {Fallahi}},
  \bibinfo {author} {\bibfnamefont {E.}~\bibnamefont {Togan}}, \bibinfo
  {author} {\bibfnamefont {J.}~\bibnamefont {Miguel-S{\'a}nchez}}, \ and\
  \bibinfo {author} {\bibfnamefont {A.}~\bibnamefont {Imamoglu}},\ }\href@noop
  {} {\bibfield  {journal} {\bibinfo  {journal} {Nature}\ }\textbf {\bibinfo
  {volume} {491}},\ \bibinfo {pages} {426} (\bibinfo {year}
  {2012})}\BibitemShut {NoStop}%
\bibitem [{\citenamefont {Bhaskar}\ \emph {et~al.}(2020)\citenamefont
  {Bhaskar}, \citenamefont {Riedinger}, \citenamefont {Machielse},
  \citenamefont {Levonian}, \citenamefont {Nguyen}, \citenamefont {Knall},
  \citenamefont {Park}, \citenamefont {Englund}, \citenamefont {Lon{\v{c}}ar},
  \citenamefont {Sukachev} \emph {et~al.}}]{bhaskar2020experimental}%
  \BibitemOpen
  \bibfield  {author} {\bibinfo {author} {\bibfnamefont {M.~K.}\ \bibnamefont
  {Bhaskar}}, \bibinfo {author} {\bibfnamefont {R.}~\bibnamefont {Riedinger}},
  \bibinfo {author} {\bibfnamefont {B.}~\bibnamefont {Machielse}}, \bibinfo
  {author} {\bibfnamefont {D.~S.}\ \bibnamefont {Levonian}}, \bibinfo {author}
  {\bibfnamefont {C.~T.}\ \bibnamefont {Nguyen}}, \bibinfo {author}
  {\bibfnamefont {E.~N.}\ \bibnamefont {Knall}}, \bibinfo {author}
  {\bibfnamefont {H.}~\bibnamefont {Park}}, \bibinfo {author} {\bibfnamefont
  {D.}~\bibnamefont {Englund}}, \bibinfo {author} {\bibfnamefont
  {M.}~\bibnamefont {Lon{\v{c}}ar}}, \bibinfo {author} {\bibfnamefont {D.~D.}\
  \bibnamefont {Sukachev}},  \emph {et~al.},\ }\href@noop {} {\bibfield
  {journal} {\bibinfo  {journal} {Nature}\ }\textbf {\bibinfo {volume} {580}},\
  \bibinfo {pages} {60} (\bibinfo {year} {2020})}\BibitemShut {NoStop}%
\bibitem [{\citenamefont {Togan}\ \emph {et~al.}(2010)\citenamefont {Togan},
  \citenamefont {Chu}, \citenamefont {Trifonov}, \citenamefont {Jiang},
  \citenamefont {Maze}, \citenamefont {Childress}, \citenamefont {Dutt},
  \citenamefont {S{\o}rensen}, \citenamefont {Hemmer}, \citenamefont {Zibrov}
  \emph {et~al.}}]{togan2010quantum}%
  \BibitemOpen
  \bibfield  {author} {\bibinfo {author} {\bibfnamefont {E.}~\bibnamefont
  {Togan}}, \bibinfo {author} {\bibfnamefont {Y.}~\bibnamefont {Chu}}, \bibinfo
  {author} {\bibfnamefont {A.~S.}\ \bibnamefont {Trifonov}}, \bibinfo {author}
  {\bibfnamefont {L.}~\bibnamefont {Jiang}}, \bibinfo {author} {\bibfnamefont
  {J.}~\bibnamefont {Maze}}, \bibinfo {author} {\bibfnamefont {L.}~\bibnamefont
  {Childress}}, \bibinfo {author} {\bibfnamefont {M.~G.}\ \bibnamefont {Dutt}},
  \bibinfo {author} {\bibfnamefont {A.~S.}\ \bibnamefont {S{\o}rensen}},
  \bibinfo {author} {\bibfnamefont {P.}~\bibnamefont {Hemmer}}, \bibinfo
  {author} {\bibfnamefont {A.~S.}\ \bibnamefont {Zibrov}},  \emph {et~al.},\
  }\href@noop {} {\bibfield  {journal} {\bibinfo  {journal} {Nature}\ }\textbf
  {\bibinfo {volume} {466}},\ \bibinfo {pages} {730} (\bibinfo {year}
  {2010})}\BibitemShut {NoStop}%
\bibitem [{\citenamefont {Rao}\ \emph {et~al.}(2015)\citenamefont {Rao},
  \citenamefont {Yang},\ and\ \citenamefont {Wrachtrup}}]{PhysRevB.92.081301}%
  \BibitemOpen
  \bibfield  {author} {\bibinfo {author} {\bibfnamefont {D.~D.~B}\
  \bibnamefont {Rao}}, \bibinfo {author} {\bibfnamefont {S.}~\bibnamefont
  {Yang}}, \ and\ \bibinfo {author} {\bibfnamefont {J.}~\bibnamefont
  {Wrachtrup}},\ }\href {\doibase 10.1103/PhysRevB.92.081301} {\bibfield
  {journal} {\bibinfo  {journal} {Phys. Rev. B}\ }\textbf {\bibinfo {volume}
  {92}},\ \bibinfo {pages} {081301(R)} (\bibinfo {year} {2015})}\BibitemShut
  {NoStop}%
\bibitem [{\citenamefont {Berry}\ \emph {et~al.}(1977)\citenamefont {Berry},
  \citenamefont {Gabrielse},\ and\ \citenamefont {Livingston}}]{Berry:s}%
  \BibitemOpen
  \bibfield  {author} {\bibinfo {author} {\bibfnamefont {H.~G.}\ \bibnamefont
  {Berry}}, \bibinfo {author} {\bibfnamefont {G.}~\bibnamefont {Gabrielse}}, \
  and\ \bibinfo {author} {\bibfnamefont {A.~E.}\ \bibnamefont {Livingston}},\
  }\href {\doibase 10.1364/AO.16.003200} {\bibfield  {journal} {\bibinfo
  {journal} {Appl. Opt.}\ }\textbf {\bibinfo {volume} {16}},\ \bibinfo {pages}
  {3200} (\bibinfo {year} {1977})}\BibitemShut {NoStop}%
\bibitem [{\citenamefont {Basiri}\ \emph {et~al.}(2019)\citenamefont {Basiri},
  \citenamefont {Chen}, \citenamefont {Bai}, \citenamefont {Amrollahi},
  \citenamefont {Carpenter}, \citenamefont {Holman}, \citenamefont {Wang},\
  and\ \citenamefont {Yao}}]{basiri2019nature}%
  \BibitemOpen
  \bibfield  {author} {\bibinfo {author} {\bibfnamefont {A.}~\bibnamefont
  {Basiri}}, \bibinfo {author} {\bibfnamefont {X.}~\bibnamefont {Chen}},
  \bibinfo {author} {\bibfnamefont {J.}~\bibnamefont {Bai}}, \bibinfo {author}
  {\bibfnamefont {P.}~\bibnamefont {Amrollahi}}, \bibinfo {author}
  {\bibfnamefont {J.}~\bibnamefont {Carpenter}}, \bibinfo {author}
  {\bibfnamefont {Z.}~\bibnamefont {Holman}}, \bibinfo {author} {\bibfnamefont
  {C.}~\bibnamefont {Wang}}, \ and\ \bibinfo {author} {\bibfnamefont
  {Y.}~\bibnamefont {Yao}},\ }\href@noop {} {\bibfield  {journal} {\bibinfo
  {journal} {Light: Science \& Applications}\ }\textbf {\bibinfo {volume}
  {8}},\ \bibinfo {pages} {1} (\bibinfo {year} {2019})}\BibitemShut {NoStop}%
\bibitem [{\citenamefont {Awartani}\ \emph {et~al.}(2014)\citenamefont
  {Awartani}, \citenamefont {Kudenov},\ and\ \citenamefont
  {O'Connor}}]{doi:10.1063/1.4868041}%
  \BibitemOpen
  \bibfield  {author} {\bibinfo {author} {\bibfnamefont {O.}~\bibnamefont
  {Awartani}}, \bibinfo {author} {\bibfnamefont {M.~W.}\ \bibnamefont
  {Kudenov}}, \ and\ \bibinfo {author} {\bibfnamefont {B.~T.}\ \bibnamefont
  {O'Connor}},\ }\href {\doibase 10.1063/1.4868041} {\bibfield  {journal}
  {\bibinfo  {journal} {Applied Physics Letters}\ }\textbf {\bibinfo {volume}
  {104}},\ \bibinfo {pages} {093306} (\bibinfo {year} {2014})},\ \Eprint
  {http://arxiv.org/abs/https://doi.org/10.1063/1.4868041}
  {https://doi.org/10.1063/1.4868041} \BibitemShut {NoStop}%
\bibitem [{\citenamefont {Roy}\ \emph {et~al.}(2016)\citenamefont {Roy},
  \citenamefont {Awartani}, \citenamefont {Sen}, \citenamefont {O'Connor},\
  and\ \citenamefont {Kudenov}}]{Roy:16}%
  \BibitemOpen
  \bibfield  {author} {\bibinfo {author} {\bibfnamefont {S.~G.}\ \bibnamefont
  {Roy}}, \bibinfo {author} {\bibfnamefont {O.~M.}\ \bibnamefont {Awartani}},
  \bibinfo {author} {\bibfnamefont {P.}~\bibnamefont {Sen}}, \bibinfo {author}
  {\bibfnamefont {B.}~\bibnamefont {O'Connor}}, \ and\ \bibinfo {author}
  {\bibfnamefont {M.~W.}\ \bibnamefont {Kudenov}},\ }\href {\doibase
  10.1364/OE.24.014737} {\bibfield  {journal} {\bibinfo  {journal} {Opt.
  Express}\ }\textbf {\bibinfo {volume} {24}},\ \bibinfo {pages} {14737}
  (\bibinfo {year} {2016})}\BibitemShut {NoStop}%
\bibitem [{\citenamefont {Yang}\ \emph {et~al.}(2017)\citenamefont {Yang},
  \citenamefont {Sen}, \citenamefont {O'Connor},\ and\ \citenamefont
  {Kudenov}}]{Yang:17}%
  \BibitemOpen
  \bibfield  {author} {\bibinfo {author} {\bibfnamefont {R.}~\bibnamefont
  {Yang}}, \bibinfo {author} {\bibfnamefont {P.}~\bibnamefont {Sen}}, \bibinfo
  {author} {\bibfnamefont {B.~T.}\ \bibnamefont {O'Connor}}, \ and\ \bibinfo
  {author} {\bibfnamefont {M.~W.}\ \bibnamefont {Kudenov}},\ }\href {\doibase
  10.1364/AO.56.001768} {\bibfield  {journal} {\bibinfo  {journal} {Appl.
  Opt.}\ }\textbf {\bibinfo {volume} {56}},\ \bibinfo {pages} {1768} (\bibinfo
  {year} {2017})}\BibitemShut {NoStop}%
\bibitem [{\citenamefont {Bai}\ \emph {et~al.}(1051)\citenamefont {Bai},
  \citenamefont {Wang}, \citenamefont {Chen}, \citenamefont {Basiri},
  \citenamefont {Wang},\ and\ \citenamefont {Yao}}]{Bai:s}%
  \BibitemOpen
  \bibfield  {author} {\bibinfo {author} {\bibfnamefont {J.}~\bibnamefont
  {Bai}}, \bibinfo {author} {\bibfnamefont {C.}~\bibnamefont {Wang}}, \bibinfo
  {author} {\bibfnamefont {X.}~\bibnamefont {Chen}}, \bibinfo {author}
  {\bibfnamefont {A.}~\bibnamefont {Basiri}}, \bibinfo {author} {\bibfnamefont
  {C.}~\bibnamefont {Wang}}, \ and\ \bibinfo {author} {\bibfnamefont
  {Y.}~\bibnamefont {Yao}},\ }\href@noop {} {\bibfield  {journal} {\bibinfo
  {journal} {Photon. Res.}\ }\textbf {\bibinfo {volume} {7}},\ \bibinfo {pages}
  {1051} (\bibinfo {year} {{ts }, doi = {10.1364/PRJ.7.001051}})}\BibitemShut
  {NoStop}%
\bibitem [{\citenamefont {Akbari}\ \emph {et~al.}(2018)\citenamefont {Akbari},
  \citenamefont {Gao},\ and\ \citenamefont {Yang}}]{Akbari:18}%
  \BibitemOpen
  \bibfield  {author} {\bibinfo {author} {\bibfnamefont {M.}~\bibnamefont
  {Akbari}}, \bibinfo {author} {\bibfnamefont {J.}~\bibnamefont {Gao}}, \ and\
  \bibinfo {author} {\bibfnamefont {X.}~\bibnamefont {Yang}},\ }\href {\doibase
  10.1364/OE.26.021194} {\bibfield  {journal} {\bibinfo  {journal} {Opt.
  Express}\ }\textbf {\bibinfo {volume} {26}},\ \bibinfo {pages} {21194}
  (\bibinfo {year} {2018})}\BibitemShut {NoStop}%
\bibitem [{\citenamefont {Hu}\ \emph {et~al.}(2017)\citenamefont {Hu},
  \citenamefont {Zhao}, \citenamefont {Lin}, \citenamefont {Zhu}, \citenamefont
  {Zhu}, \citenamefont {Guo}, \citenamefont {Cao},\ and\ \citenamefont
  {Wang}}]{hu2017all}%
  \BibitemOpen
  \bibfield  {author} {\bibinfo {author} {\bibfnamefont {J.}~\bibnamefont
  {Hu}}, \bibinfo {author} {\bibfnamefont {X.}~\bibnamefont {Zhao}}, \bibinfo
  {author} {\bibfnamefont {Y.}~\bibnamefont {Lin}}, \bibinfo {author}
  {\bibfnamefont {A.}~\bibnamefont {Zhu}}, \bibinfo {author} {\bibfnamefont
  {X.}~\bibnamefont {Zhu}}, \bibinfo {author} {\bibfnamefont {P.}~\bibnamefont
  {Guo}}, \bibinfo {author} {\bibfnamefont {B.}~\bibnamefont {Cao}}, \ and\
  \bibinfo {author} {\bibfnamefont {C.}~\bibnamefont {Wang}},\ }\href@noop {}
  {\bibfield  {journal} {\bibinfo  {journal} {Scientific Reports}\ }\textbf
  {\bibinfo {volume} {7}},\ \bibinfo {pages} {41893} (\bibinfo {year}
  {2017})}\BibitemShut {NoStop}%
\bibitem [{\citenamefont {Zhao}\ \emph {et~al.}(2012)\citenamefont {Zhao},
  \citenamefont {Belkin},\ and\ \citenamefont {Al{\`u}}}]{zhao2012twisted}%
  \BibitemOpen
  \bibfield  {author} {\bibinfo {author} {\bibfnamefont {Y.}~\bibnamefont
  {Zhao}}, \bibinfo {author} {\bibfnamefont {M.}~\bibnamefont {Belkin}}, \ and\
  \bibinfo {author} {\bibfnamefont {A.}~\bibnamefont {Al{\`u}}},\ }\href@noop
  {} {\bibfield  {journal} {\bibinfo  {journal} {Nature communications}\
  }\textbf {\bibinfo {volume} {3}},\ \bibinfo {pages} {1} (\bibinfo {year}
  {2012})}\BibitemShut {NoStop}%
\bibitem [{\citenamefont {Lin}\ \emph {et~al.}(2019)\citenamefont {Lin},
  \citenamefont {Rusch}, \citenamefont {Chen},\ and\ \citenamefont
  {Shi}}]{Lin:19}%
  \BibitemOpen
  \bibfield  {author} {\bibinfo {author} {\bibfnamefont {Z.}~\bibnamefont
  {Lin}}, \bibinfo {author} {\bibfnamefont {L.}~\bibnamefont {Rusch}}, \bibinfo
  {author} {\bibfnamefont {Y.}~\bibnamefont {Chen}}, \ and\ \bibinfo {author}
  {\bibfnamefont {W.}~\bibnamefont {Shi}},\ }\href {\doibase
  10.1364/OE.27.004867} {\bibfield  {journal} {\bibinfo  {journal} {Opt.
  Express}\ }\textbf {\bibinfo {volume} {27}},\ \bibinfo {pages} {4867}
  (\bibinfo {year} {2019})}\BibitemShut {NoStop}%
\bibitem [{\citenamefont {Jiang}\ \emph {et~al.}(2020)\citenamefont {Jiang},
  \citenamefont {Du}, \citenamefont {Jiang}, \citenamefont {Liu}, \citenamefont
  {Liu}, \citenamefont {Li}, \citenamefont {Liu}, \citenamefont {Lin},
  \citenamefont {Zhu},\ and\ \citenamefont {Fang}}]{C9NR10768A}%
  \BibitemOpen
  \bibfield  {author} {\bibinfo {author} {\bibfnamefont {Q.}~\bibnamefont
  {Jiang}}, \bibinfo {author} {\bibfnamefont {B.}~\bibnamefont {Du}}, \bibinfo
  {author} {\bibfnamefont {M.}~\bibnamefont {Jiang}}, \bibinfo {author}
  {\bibfnamefont {D.}~\bibnamefont {Liu}}, \bibinfo {author} {\bibfnamefont
  {Z.}~\bibnamefont {Liu}}, \bibinfo {author} {\bibfnamefont {B.}~\bibnamefont
  {Li}}, \bibinfo {author} {\bibfnamefont {Z.}~\bibnamefont {Liu}}, \bibinfo
  {author} {\bibfnamefont {F.}~\bibnamefont {Lin}}, \bibinfo {author}
  {\bibfnamefont {X.}~\bibnamefont {Zhu}}, \ and\ \bibinfo {author}
  {\bibfnamefont {Z.}~\bibnamefont {Fang}},\ }\href {\doibase
  10.1039/C9NR10768A} {\bibfield  {journal} {\bibinfo  {journal} {Nanoscale}\
  }\textbf {\bibinfo {volume} {12}},\ \bibinfo {pages} {5906} (\bibinfo {year}
  {2020})}\BibitemShut {NoStop}%
\bibitem [{\citenamefont {Stoevelaar}\ \emph {et~al.}(2020)\citenamefont
  {Stoevelaar}, \citenamefont {Berzin\v{s}}, \citenamefont {Silvestri},
  \citenamefont {Fasold}, \citenamefont {Kamali}, \citenamefont {Knopf},
  \citenamefont {Eilenberger}, \citenamefont {Setzpfandt}, \citenamefont
  {Pertsch}, \citenamefont {B\"{a}umer},\ and\ \citenamefont
  {Gerini}}]{PjotrStoevelaar:20}%
  \BibitemOpen
  \bibfield  {author} {\bibinfo {author} {\bibfnamefont {L.~P.}\ \bibnamefont
  {Stoevelaar}}, \bibinfo {author} {\bibfnamefont {J.}~\bibnamefont
  {Berzin\v{s}}}, \bibinfo {author} {\bibfnamefont {F.}~\bibnamefont
  {Silvestri}}, \bibinfo {author} {\bibfnamefont {S.}~\bibnamefont {Fasold}},
  \bibinfo {author} {\bibfnamefont {K.~Z.}\ \bibnamefont {Kamali}}, \bibinfo
  {author} {\bibfnamefont {H.}~\bibnamefont {Knopf}}, \bibinfo {author}
  {\bibfnamefont {F.}~\bibnamefont {Eilenberger}}, \bibinfo {author}
  {\bibfnamefont {F.}~\bibnamefont {Setzpfandt}}, \bibinfo {author}
  {\bibfnamefont {T.}~\bibnamefont {Pertsch}}, \bibinfo {author} {\bibfnamefont
  {S.~M.~B.}\ \bibnamefont {B\"{a}umer}}, \ and\ \bibinfo {author}
  {\bibfnamefont {G.}~\bibnamefont {Gerini}},\ }\href {\doibase
  10.1364/OE.392536} {\bibfield  {journal} {\bibinfo  {journal} {Opt. Express}\
  }\textbf {\bibinfo {volume} {28}},\ \bibinfo {pages} {19818} (\bibinfo {year}
  {2020})}\BibitemShut {NoStop}%
\bibitem [{\citenamefont {Wu}\ \emph {et~al.}(2019)\citenamefont {Wu},
  \citenamefont {Yu}, \citenamefont {Liu},\ and\ \citenamefont
  {Zhang}}]{FullyintegratedCMOScompatiblepolarizationanalyzer}%
  \BibitemOpen
  \bibfield  {author} {\bibinfo {author} {\bibfnamefont {W.}~\bibnamefont
  {Wu}}, \bibinfo {author} {\bibfnamefont {Y.}~\bibnamefont {Yu}}, \bibinfo
  {author} {\bibfnamefont {W.}~\bibnamefont {Liu}}, \ and\ \bibinfo {author}
  {\bibfnamefont {X.}~\bibnamefont {Zhang}},\ }\href {\doibase
  https://doi.org/10.1515/nanoph-2018-0205} {\bibfield  {journal} {\bibinfo
  {journal} {Nanophotonics}\ }\textbf {\bibinfo {volume} {8}},\ \bibinfo
  {pages} {467 } (\bibinfo {year} {01 Mar. 2019})}\BibitemShut {NoStop}%
\bibitem [{\citenamefont {Dong}\ and\ \citenamefont
  {Zhou}(2020)}]{DONG2020125598}%
  \BibitemOpen
  \bibfield  {author} {\bibinfo {author} {\bibfnamefont {J.}~\bibnamefont
  {Dong}}\ and\ \bibinfo {author} {\bibfnamefont {H.}~\bibnamefont {Zhou}},\
  }\href {\doibase https://doi.org/10.1016/j.optcom.2020.125598} {\bibfield
  {journal} {\bibinfo  {journal} {Optics Communications}\ }\textbf {\bibinfo
  {volume} {465}},\ \bibinfo {pages} {125598} (\bibinfo {year}
  {2020})}\BibitemShut {NoStop}%
\bibitem [{\citenamefont {Ando}\ \emph {et~al.}(2010)\citenamefont {Ando},
  \citenamefont {Morikawa}, \citenamefont {Trypiniotis}, \citenamefont
  {Fujikawa}, \citenamefont {Barnes},\ and\ \citenamefont
  {Saitoh}}]{doi:10.1063/1.3327809}%
  \BibitemOpen
  \bibfield  {author} {\bibinfo {author} {\bibfnamefont {K.}~\bibnamefont
  {Ando}}, \bibinfo {author} {\bibfnamefont {M.}~\bibnamefont {Morikawa}},
  \bibinfo {author} {\bibfnamefont {T.}~\bibnamefont {Trypiniotis}}, \bibinfo
  {author} {\bibfnamefont {Y.}~\bibnamefont {Fujikawa}}, \bibinfo {author}
  {\bibfnamefont {C.~H.~W.}\ \bibnamefont {Barnes}}, \ and\ \bibinfo {author}
  {\bibfnamefont {E.}~\bibnamefont {Saitoh}},\ }\href {\doibase
  10.1063/1.3327809} {\bibfield  {journal} {\bibinfo  {journal} {Applied
  Physics Letters}\ }\textbf {\bibinfo {volume} {96}},\ \bibinfo {pages}
  {082502} (\bibinfo {year} {2010})},\ \Eprint
  {http://arxiv.org/abs/https://doi.org/10.1063/1.3327809}
  {https://doi.org/10.1063/1.3327809} \BibitemShut {NoStop}%
\bibitem [{\citenamefont {Khamari}\ \emph {et~al.}(2015)\citenamefont
  {Khamari}, \citenamefont {Porwal}, \citenamefont {Oak},\ and\ \citenamefont
  {Sharma}}]{doi:10.1063/1.4929326}%
  \BibitemOpen
  \bibfield  {author} {\bibinfo {author} {\bibfnamefont {S.~K.}\ \bibnamefont
  {Khamari}}, \bibinfo {author} {\bibfnamefont {S.}~\bibnamefont {Porwal}},
  \bibinfo {author} {\bibfnamefont {S.~M.}\ \bibnamefont {Oak}}, \ and\
  \bibinfo {author} {\bibfnamefont {T.~K.}\ \bibnamefont {Sharma}},\ }\href
  {\doibase 10.1063/1.4929326} {\bibfield  {journal} {\bibinfo  {journal}
  {Applied Physics Letters}\ }\textbf {\bibinfo {volume} {107}},\ \bibinfo
  {pages} {072108} (\bibinfo {year} {2015})},\ \Eprint
  {http://arxiv.org/abs/https://doi.org/10.1063/1.4929326}
  {https://doi.org/10.1063/1.4929326} \BibitemShut {NoStop}%
\bibitem [{\citenamefont {Ivchenko}\ and\ \citenamefont
  {Ganichev}(2017)}]{ivchenko2017spindependent}%
  \BibitemOpen
  \bibfield  {author} {\bibinfo {author} {\bibfnamefont {E.~L.}\ \bibnamefont
  {Ivchenko}}\ and\ \bibinfo {author} {\bibfnamefont {S.~D.}\ \bibnamefont
  {Ganichev}},\ }\href@noop {} {\enquote {\bibinfo {title} {Spin-dependent
  photogalvanic effects (a review)},}\ } (\bibinfo {year} {2017}),\ \Eprint
  {http://arxiv.org/abs/1710.09223} {arXiv:1710.09223 [cond-mat.mes-hall]}
  \BibitemShut {NoStop}%
\bibitem [{\citenamefont {Saitoh}\ \emph {et~al.}(2006)\citenamefont {Saitoh},
  \citenamefont {Ueda}, \citenamefont {Miyajima},\ and\ \citenamefont
  {Tatara}}]{doi:10.1063/1.2199473}%
  \BibitemOpen
  \bibfield  {author} {\bibinfo {author} {\bibfnamefont {E.}~\bibnamefont
  {Saitoh}}, \bibinfo {author} {\bibfnamefont {M.}~\bibnamefont {Ueda}},
  \bibinfo {author} {\bibfnamefont {H.}~\bibnamefont {Miyajima}}, \ and\
  \bibinfo {author} {\bibfnamefont {G.}~\bibnamefont {Tatara}},\ }\href
  {\doibase 10.1063/1.2199473} {\bibfield  {journal} {\bibinfo  {journal}
  {Applied Physics Letters}\ }\textbf {\bibinfo {volume} {88}},\ \bibinfo
  {pages} {182509} (\bibinfo {year} {2006})},\ \Eprint
  {http://arxiv.org/abs/https://doi.org/10.1063/1.2199473}
  {https://doi.org/10.1063/1.2199473} \BibitemShut {NoStop}%
\bibitem [{\citenamefont {Lepine}(1972)}]{PhysRevB.6.436}%
  \BibitemOpen
  \bibfield  {author} {\bibinfo {author} {\bibfnamefont {D.~J.}\ \bibnamefont
  {Lepine}},\ }\href {\doibase 10.1103/PhysRevB.6.436} {\bibfield  {journal}
  {\bibinfo  {journal} {Phys. Rev. B}\ }\textbf {\bibinfo {volume} {6}},\
  \bibinfo {pages} {436} (\bibinfo {year} {1972})}\BibitemShut {NoStop}%
\bibitem [{\citenamefont {Weisbuch}\ and\ \citenamefont
  {Lampel}(1974)}]{WEISBUCH1974141}%
  \BibitemOpen
  \bibfield  {author} {\bibinfo {author} {\bibfnamefont {C.}~\bibnamefont
  {Weisbuch}}\ and\ \bibinfo {author} {\bibfnamefont {G.}~\bibnamefont
  {Lampel}},\ }\href {\doibase https://doi.org/10.1016/0038-1098(74)90202-6}
  {\bibfield  {journal} {\bibinfo  {journal} {Solid State Communications}\
  }\textbf {\bibinfo {volume} {14}},\ \bibinfo {pages} {141 } (\bibinfo {year}
  {1974})}\BibitemShut {NoStop}%
\bibitem [{\citenamefont {Paget}(1984)}]{PhysRevB.30.931}%
  \BibitemOpen
  \bibfield  {author} {\bibinfo {author} {\bibfnamefont {D.}~\bibnamefont
  {Paget}},\ }\href {\doibase 10.1103/PhysRevB.30.931} {\bibfield  {journal}
  {\bibinfo  {journal} {Phys. Rev. B}\ }\textbf {\bibinfo {volume} {30}},\
  \bibinfo {pages} {931} (\bibinfo {year} {1984})}\BibitemShut {NoStop}%
\bibitem [{\citenamefont {Kalevich}\ \emph {et~al.}(2005)\citenamefont
  {Kalevich}, \citenamefont {Ivchenko}, \citenamefont {Afanasiev},
  \citenamefont {Shiryaev}, \citenamefont {Egorov}, \citenamefont {Ustinov},
  \citenamefont {Pal},\ and\ \citenamefont {Masumoto}}]{Kalevich2005}%
  \BibitemOpen
  \bibfield  {author} {\bibinfo {author} {\bibfnamefont {V.~K.}\ \bibnamefont
  {Kalevich}}, \bibinfo {author} {\bibfnamefont {E.~L.}\ \bibnamefont
  {Ivchenko}}, \bibinfo {author} {\bibfnamefont {M.~M.}\ \bibnamefont
  {Afanasiev}}, \bibinfo {author} {\bibfnamefont {A.~Y.}\ \bibnamefont
  {Shiryaev}}, \bibinfo {author} {\bibfnamefont {A.~Y.}\ \bibnamefont
  {Egorov}}, \bibinfo {author} {\bibfnamefont {V.~M.}\ \bibnamefont {Ustinov}},
  \bibinfo {author} {\bibfnamefont {B.}~\bibnamefont {Pal}}, \ and\ \bibinfo
  {author} {\bibfnamefont {Y.}~\bibnamefont {Masumoto}},\ }\href {\doibase
  10.1134/1.2142877} {\bibfield  {journal} {\bibinfo  {journal} {Journal of
  Experimental and Theoretical Physics Letters}\ }\textbf {\bibinfo {volume}
  {82}},\ \bibinfo {pages} {455} (\bibinfo {year} {2005})}\BibitemShut
  {NoStop}%
\bibitem [{\citenamefont {Lombez}\ \emph {et~al.}(2005)\citenamefont {Lombez},
  \citenamefont {Braun}, \citenamefont {Carr\'ere}, \citenamefont {Urbaszek},
  \citenamefont {Renucci}, \citenamefont {Amand}, \citenamefont {Marie},
  \citenamefont {Harmand},\ and\ \citenamefont
  {Kalevich}}]{doi:10.1063/1.2150252}%
  \BibitemOpen
  \bibfield  {author} {\bibinfo {author} {\bibfnamefont {L.}~\bibnamefont
  {Lombez}}, \bibinfo {author} {\bibfnamefont {P.-F.}\ \bibnamefont {Braun}},
  \bibinfo {author} {\bibfnamefont {H.}~\bibnamefont {Carr\'ere}}, \bibinfo
  {author} {\bibfnamefont {B.}~\bibnamefont {Urbaszek}}, \bibinfo {author}
  {\bibfnamefont {P.}~\bibnamefont {Renucci}}, \bibinfo {author} {\bibfnamefont
  {T.}~\bibnamefont {Amand}}, \bibinfo {author} {\bibfnamefont
  {X.}~\bibnamefont {Marie}}, \bibinfo {author} {\bibfnamefont {J.~C.}\
  \bibnamefont {Harmand}}, \ and\ \bibinfo {author} {\bibfnamefont {V.~K.}\
  \bibnamefont {Kalevich}},\ }\href {\doibase 10.1063/1.2150252} {\bibfield
  {journal} {\bibinfo  {journal} {Applied Physics Letters}\ }\textbf {\bibinfo
  {volume} {87}},\ \bibinfo {pages} {252115} (\bibinfo {year} {2005})},\
  \Eprint {http://arxiv.org/abs/https://doi.org/10.1063/1.2150252}
  {https://doi.org/10.1063/1.2150252} \BibitemShut {NoStop}%
\bibitem [{\citenamefont {Kalevich}\ \emph {et~al.}(2006)\citenamefont
  {Kalevich}, \citenamefont {Shiryaev}, \citenamefont {Ivchenko}, \citenamefont
  {Egorov}, \citenamefont {Lombez}, \citenamefont {Lagarde}, \citenamefont
  {Marie},\ and\ \citenamefont {Amand}}]{Kalevich2007}%
  \BibitemOpen
  \bibfield  {author} {\bibinfo {author} {\bibfnamefont {V.~K.}\ \bibnamefont
  {Kalevich}}, \bibinfo {author} {\bibfnamefont {A.~Y.}\ \bibnamefont
  {Shiryaev}}, \bibinfo {author} {\bibfnamefont {E.~L.}\ \bibnamefont
  {Ivchenko}}, \bibinfo {author} {\bibfnamefont {A.~Y.}\ \bibnamefont
  {Egorov}}, \bibinfo {author} {\bibfnamefont {L.}~\bibnamefont {Lombez}},
  \bibinfo {author} {\bibfnamefont {D.}~\bibnamefont {Lagarde}}, \bibinfo
  {author} {\bibfnamefont {X.}~\bibnamefont {Marie}}, \ and\ \bibinfo {author}
  {\bibfnamefont {T.}~\bibnamefont {Amand}},\ }\href {\doibase
  10.1134/S0021364007030095} {\bibfield  {journal} {\bibinfo  {journal} {JETP
  Letters}\ }\textbf {\bibinfo {volume} {85}},\ \bibinfo {pages} {174}
  (\bibinfo {year} {2006})}\BibitemShut {NoStop}%
\bibitem [{\citenamefont {Lagarde}\ \emph {et~al.}(2007)\citenamefont
  {Lagarde}, \citenamefont {Lombez}, \citenamefont {Marie}, \citenamefont
  {Balocchi}, \citenamefont {Amand}, \citenamefont {Kalevich}, \citenamefont
  {Shiryaev}, \citenamefont {Ivchenko},\ and\ \citenamefont
  {Egorov}}]{doi:10.1002/pssa.200673009}%
  \BibitemOpen
  \bibfield  {author} {\bibinfo {author} {\bibfnamefont {D.}~\bibnamefont
  {Lagarde}}, \bibinfo {author} {\bibfnamefont {L.}~\bibnamefont {Lombez}},
  \bibinfo {author} {\bibfnamefont {X.}~\bibnamefont {Marie}}, \bibinfo
  {author} {\bibfnamefont {A.}~\bibnamefont {Balocchi}}, \bibinfo {author}
  {\bibfnamefont {T.}~\bibnamefont {Amand}}, \bibinfo {author} {\bibfnamefont
  {V.~K.}\ \bibnamefont {Kalevich}}, \bibinfo {author} {\bibfnamefont
  {A.}~\bibnamefont {Shiryaev}}, \bibinfo {author} {\bibfnamefont
  {E.}~\bibnamefont {Ivchenko}}, \ and\ \bibinfo {author} {\bibfnamefont
  {A.}~\bibnamefont {Egorov}},\ }\href {\doibase 10.1002/pssa.200673009}
  {\bibfield  {journal} {\bibinfo  {journal} {physica status solidi (a)}\
  }\textbf {\bibinfo {volume} {204}},\ \bibinfo {pages} {208} (\bibinfo {year}
  {2007})},\ \Eprint
  {http://arxiv.org/abs/https://onlinelibrary.wiley.com/doi/pdf/10.1002/pssa.200673009}
  {https://onlinelibrary.wiley.com/doi/pdf/10.1002/pssa.200673009} \BibitemShut
  {NoStop}%
\bibitem [{\citenamefont {Zhao}\ \emph
  {et~al.}(2009{\natexlab{a}})\citenamefont {Zhao}, \citenamefont {Balocchi},
  \citenamefont {Truong}, \citenamefont {Amand}, \citenamefont {Marie},
  \citenamefont {Wang}, \citenamefont {Buyanova}, \citenamefont {Chen},\ and\
  \citenamefont {Harmand}}]{Zhao_2009}%
  \BibitemOpen
  \bibfield  {author} {\bibinfo {author} {\bibfnamefont {F.}~\bibnamefont
  {Zhao}}, \bibinfo {author} {\bibfnamefont {A.}~\bibnamefont {Balocchi}},
  \bibinfo {author} {\bibfnamefont {G.}~\bibnamefont {Truong}}, \bibinfo
  {author} {\bibfnamefont {T.}~\bibnamefont {Amand}}, \bibinfo {author}
  {\bibfnamefont {X.}~\bibnamefont {Marie}}, \bibinfo {author} {\bibfnamefont
  {X.~J.}\ \bibnamefont {Wang}}, \bibinfo {author} {\bibfnamefont {I.~A.}\
  \bibnamefont {Buyanova}}, \bibinfo {author} {\bibfnamefont {W.~M.}\
  \bibnamefont {Chen}}, \ and\ \bibinfo {author} {\bibfnamefont {J.~C.}\
  \bibnamefont {Harmand}},\ }\href {\doibase 10.1088/0953-8984/21/17/174211}
  {\bibfield  {journal} {\bibinfo  {journal} {Journal of Physics: Condensed
  Matter}\ }\textbf {\bibinfo {volume} {21}},\ \bibinfo {pages} {174211}
  (\bibinfo {year} {2009}{\natexlab{a}})}\BibitemShut {NoStop}%
\bibitem [{\citenamefont {Wang}\ \emph
  {et~al.}(2009{\natexlab{a}})\citenamefont {Wang}, \citenamefont {Buyanova},
  \citenamefont {Zhao}, \citenamefont {Lagarde}, \citenamefont {Balocchi},
  \citenamefont {Marie}, \citenamefont {Tu}, \citenamefont {Harmand},\ and\
  \citenamefont {Chen}}]{wang2009room}%
  \BibitemOpen
  \bibfield  {author} {\bibinfo {author} {\bibfnamefont {X.}~\bibnamefont
  {Wang}}, \bibinfo {author} {\bibfnamefont {I.~A.}\ \bibnamefont {Buyanova}},
  \bibinfo {author} {\bibfnamefont {F.}~\bibnamefont {Zhao}}, \bibinfo {author}
  {\bibfnamefont {D.}~\bibnamefont {Lagarde}}, \bibinfo {author} {\bibfnamefont
  {A.}~\bibnamefont {Balocchi}}, \bibinfo {author} {\bibfnamefont
  {X.}~\bibnamefont {Marie}}, \bibinfo {author} {\bibfnamefont
  {C.}~\bibnamefont {Tu}}, \bibinfo {author} {\bibfnamefont {J.}~\bibnamefont
  {Harmand}}, \ and\ \bibinfo {author} {\bibfnamefont {W.}~\bibnamefont
  {Chen}},\ }\href@noop {} {\bibfield  {journal} {\bibinfo  {journal} {Nature
  materials}\ }\textbf {\bibinfo {volume} {8}},\ \bibinfo {pages} {198}
  (\bibinfo {year} {2009}{\natexlab{a}})}\BibitemShut {NoStop}%
\bibitem [{\citenamefont {Kalevich}\ \emph {et~al.}(2009)\citenamefont
  {Kalevich}, \citenamefont {Shiryaev}, \citenamefont {Ivchenko}, \citenamefont
  {Afanasiev}, \citenamefont {Egorov}, \citenamefont {Ustinov},\ and\
  \citenamefont {Masumoto}}]{KALEVICH20094929}%
  \BibitemOpen
  \bibfield  {author} {\bibinfo {author} {\bibfnamefont {V.}~\bibnamefont
  {Kalevich}}, \bibinfo {author} {\bibfnamefont {A.}~\bibnamefont {Shiryaev}},
  \bibinfo {author} {\bibfnamefont {E.}~\bibnamefont {Ivchenko}}, \bibinfo
  {author} {\bibfnamefont {M.}~\bibnamefont {Afanasiev}}, \bibinfo {author}
  {\bibfnamefont {A.}~\bibnamefont {Egorov}}, \bibinfo {author} {\bibfnamefont
  {V.}~\bibnamefont {Ustinov}}, \ and\ \bibinfo {author} {\bibfnamefont
  {Y.}~\bibnamefont {Masumoto}},\ }\href {\doibase
  https://doi.org/10.1016/j.physb.2009.08.234} {\bibfield  {journal} {\bibinfo
  {journal} {Physica B: Condensed Matter}\ }\textbf {\bibinfo {volume} {404}},\
  \bibinfo {pages} {4929 } (\bibinfo {year} {2009})}\BibitemShut {NoStop}%
\bibitem [{\citenamefont {Zhao}\ \emph
  {et~al.}(2009{\natexlab{b}})\citenamefont {Zhao}, \citenamefont {Lombez},
  \citenamefont {Liu}, \citenamefont {Sun}, \citenamefont {Xue}, \citenamefont
  {Chen},\ and\ \citenamefont {Marie}}]{doi:10.1063/1.3186076}%
  \BibitemOpen
  \bibfield  {author} {\bibinfo {author} {\bibfnamefont {H.~M.}\ \bibnamefont
  {Zhao}}, \bibinfo {author} {\bibfnamefont {L.}~\bibnamefont {Lombez}},
  \bibinfo {author} {\bibfnamefont {B.~L.}\ \bibnamefont {Liu}}, \bibinfo
  {author} {\bibfnamefont {B.~Q.}\ \bibnamefont {Sun}}, \bibinfo {author}
  {\bibfnamefont {Q.~K.}\ \bibnamefont {Xue}}, \bibinfo {author} {\bibfnamefont
  {D.~M.}\ \bibnamefont {Chen}}, \ and\ \bibinfo {author} {\bibfnamefont
  {X.}~\bibnamefont {Marie}},\ }\href {\doibase 10.1063/1.3186076} {\bibfield
  {journal} {\bibinfo  {journal} {Applied Physics Letters}\ }\textbf {\bibinfo
  {volume} {95}},\ \bibinfo {pages} {041911} (\bibinfo {year}
  {2009}{\natexlab{b}})},\ \Eprint
  {http://arxiv.org/abs/https://doi.org/10.1063/1.3186076}
  {https://doi.org/10.1063/1.3186076} \BibitemShut {NoStop}%
\bibitem [{\citenamefont {Zhao}\ \emph
  {et~al.}(2009{\natexlab{c}})\citenamefont {Zhao}, \citenamefont {Balocchi},
  \citenamefont {Kunold}, \citenamefont {Carrey}, \citenamefont {Car\'e},
  \citenamefont {Amand}, \citenamefont {Ben~Abdallah}, \citenamefont
  {Harmand},\ and\ \citenamefont {Marie}}]{doi:10.1063/1.3273393}%
  \BibitemOpen
  \bibfield  {author} {\bibinfo {author} {\bibfnamefont {F.}~\bibnamefont
  {Zhao}}, \bibinfo {author} {\bibfnamefont {A.}~\bibnamefont {Balocchi}},
  \bibinfo {author} {\bibfnamefont {A.}~\bibnamefont {Kunold}}, \bibinfo
  {author} {\bibfnamefont {J.}~\bibnamefont {Carrey}}, \bibinfo {author}
  {\bibfnamefont {H.}~\bibnamefont {Car\'e}}, \bibinfo {author} {\bibfnamefont
  {T.}~\bibnamefont {Amand}}, \bibinfo {author} {\bibfnamefont
  {N.}~\bibnamefont {Ben~Abdallah}}, \bibinfo {author} {\bibfnamefont {J.~C.}\
  \bibnamefont {Harmand}}, \ and\ \bibinfo {author} {\bibfnamefont
  {X.}~\bibnamefont {Marie}},\ }\href {\doibase 10.1063/1.3273393} {\bibfield
  {journal} {\bibinfo  {journal} {Applied Physics Letters}\ }\textbf {\bibinfo
  {volume} {95}},\ \bibinfo {pages} {241104} (\bibinfo {year}
  {2009}{\natexlab{c}})},\ \Eprint
  {http://arxiv.org/abs/https://doi.org/10.1063/1.3273393}
  {https://doi.org/10.1063/1.3273393} \BibitemShut {NoStop}%
\bibitem [{\citenamefont {Wang}\ \emph
  {et~al.}(2009{\natexlab{b}})\citenamefont {Wang}, \citenamefont {Puttisong},
  \citenamefont {Tu}, \citenamefont {Ptak}, \citenamefont {Kalevich},
  \citenamefont {Egorov}, \citenamefont {Geelhaar}, \citenamefont {Riechert},
  \citenamefont {Chen},\ and\ \citenamefont
  {Buyanova}}]{doi:10.1063/1.3275703}%
  \BibitemOpen
  \bibfield  {author} {\bibinfo {author} {\bibfnamefont {X.~J.}\ \bibnamefont
  {Wang}}, \bibinfo {author} {\bibfnamefont {Y.}~\bibnamefont {Puttisong}},
  \bibinfo {author} {\bibfnamefont {C.~W.}\ \bibnamefont {Tu}}, \bibinfo
  {author} {\bibfnamefont {A.~J.}\ \bibnamefont {Ptak}}, \bibinfo {author}
  {\bibfnamefont {V.~K.}\ \bibnamefont {Kalevich}}, \bibinfo {author}
  {\bibfnamefont {A.~Y.}\ \bibnamefont {Egorov}}, \bibinfo {author}
  {\bibfnamefont {L.}~\bibnamefont {Geelhaar}}, \bibinfo {author}
  {\bibfnamefont {H.}~\bibnamefont {Riechert}}, \bibinfo {author}
  {\bibfnamefont {W.~M.}\ \bibnamefont {Chen}}, \ and\ \bibinfo {author}
  {\bibfnamefont {I.~A.}\ \bibnamefont {Buyanova}},\ }\href {\doibase
  10.1063/1.3275703} {\bibfield  {journal} {\bibinfo  {journal} {Applied
  Physics Letters}\ }\textbf {\bibinfo {volume} {95}},\ \bibinfo {pages}
  {241904} (\bibinfo {year} {2009}{\natexlab{b}})},\ \Eprint
  {http://arxiv.org/abs/https://doi.org/10.1063/1.3275703}
  {https://doi.org/10.1063/1.3275703} \BibitemShut {NoStop}%
\bibitem [{\citenamefont {Puttisong}\ \emph {et~al.}(2010)\citenamefont
  {Puttisong}, \citenamefont {Wang}, \citenamefont {Buyanova}, \citenamefont
  {Carr{\`e}re}, \citenamefont {Zhao}, \citenamefont {Balocchi}, \citenamefont
  {Marie}, \citenamefont {Tu},\ and\ \citenamefont
  {Chen}}]{doi:10.1063/1.3299015}%
  \BibitemOpen
  \bibfield  {author} {\bibinfo {author} {\bibfnamefont {Y.}~\bibnamefont
  {Puttisong}}, \bibinfo {author} {\bibfnamefont {X.~J.}\ \bibnamefont {Wang}},
  \bibinfo {author} {\bibfnamefont {I.~A.}\ \bibnamefont {Buyanova}}, \bibinfo
  {author} {\bibfnamefont {H.}~\bibnamefont {Carr{\`e}re}}, \bibinfo {author}
  {\bibfnamefont {F.}~\bibnamefont {Zhao}}, \bibinfo {author} {\bibfnamefont
  {A.}~\bibnamefont {Balocchi}}, \bibinfo {author} {\bibfnamefont
  {X.}~\bibnamefont {Marie}}, \bibinfo {author} {\bibfnamefont {C.~W.}\
  \bibnamefont {Tu}}, \ and\ \bibinfo {author} {\bibfnamefont {W.~M.}\
  \bibnamefont {Chen}},\ }\href {\doibase 10.1063/1.3299015} {\bibfield
  {journal} {\bibinfo  {journal} {Applied Physics Letters}\ }\textbf {\bibinfo
  {volume} {96}},\ \bibinfo {pages} {052104} (\bibinfo {year} {2010})},\
  \Eprint {http://arxiv.org/abs/https://doi.org/10.1063/1.3299015}
  {https://doi.org/10.1063/1.3299015} \BibitemShut {NoStop}%
\bibitem [{\citenamefont {Ivchenko}\ \emph {et~al.}(2010)\citenamefont
  {Ivchenko}, \citenamefont {Kalevich}, \citenamefont {Shiryaev}, \citenamefont
  {Afanasiev},\ and\ \citenamefont {Masumoto}}]{Ivchenko_2010}%
  \BibitemOpen
  \bibfield  {author} {\bibinfo {author} {\bibfnamefont {E.~L.}\ \bibnamefont
  {Ivchenko}}, \bibinfo {author} {\bibfnamefont {V.~K.}\ \bibnamefont
  {Kalevich}}, \bibinfo {author} {\bibfnamefont {A.~Y.}\ \bibnamefont
  {Shiryaev}}, \bibinfo {author} {\bibfnamefont {M.~M.}\ \bibnamefont
  {Afanasiev}}, \ and\ \bibinfo {author} {\bibfnamefont {Y.}~\bibnamefont
  {Masumoto}},\ }\href {\doibase 10.1088/0953-8984/22/46/465804} {\bibfield
  {journal} {\bibinfo  {journal} {Journal of Physics: Condensed Matter}\
  }\textbf {\bibinfo {volume} {22}},\ \bibinfo {pages} {465804} (\bibinfo
  {year} {2010})}\BibitemShut {NoStop}%
\bibitem [{\citenamefont {Kunold}\ \emph {et~al.}(2011)\citenamefont {Kunold},
  \citenamefont {Balocchi}, \citenamefont {Zhao}, \citenamefont {Amand},
  \citenamefont {Abdallah}, \citenamefont {Harmand},\ and\ \citenamefont
  {Marie}}]{PhysRevB.83.165202}%
  \BibitemOpen
  \bibfield  {author} {\bibinfo {author} {\bibfnamefont {A.}~\bibnamefont
  {Kunold}}, \bibinfo {author} {\bibfnamefont {A.}~\bibnamefont {Balocchi}},
  \bibinfo {author} {\bibfnamefont {F.}~\bibnamefont {Zhao}}, \bibinfo {author}
  {\bibfnamefont {T.}~\bibnamefont {Amand}}, \bibinfo {author} {\bibfnamefont
  {N.~B.}\ \bibnamefont {Abdallah}}, \bibinfo {author} {\bibfnamefont {J.~C.}\
  \bibnamefont {Harmand}}, \ and\ \bibinfo {author} {\bibfnamefont
  {X.}~\bibnamefont {Marie}},\ }\href {\doibase 10.1103/PhysRevB.83.165202}
  {\bibfield  {journal} {\bibinfo  {journal} {Phys. Rev. B}\ }\textbf {\bibinfo
  {volume} {83}},\ \bibinfo {pages} {165202} (\bibinfo {year}
  {2011})}\BibitemShut {NoStop}%
\bibitem [{\citenamefont {Kalevich}\ \emph {et~al.}(2012)\citenamefont
  {Kalevich}, \citenamefont {Afanasiev}, \citenamefont {Shiryaev},\ and\
  \citenamefont {Egorov}}]{PhysRevB.85.035205}%
  \BibitemOpen
  \bibfield  {author} {\bibinfo {author} {\bibfnamefont {V.~K.}\ \bibnamefont
  {Kalevich}}, \bibinfo {author} {\bibfnamefont {M.~M.}\ \bibnamefont
  {Afanasiev}}, \bibinfo {author} {\bibfnamefont {A.~Y.}\ \bibnamefont
  {Shiryaev}}, \ and\ \bibinfo {author} {\bibfnamefont {A.~Y.}\ \bibnamefont
  {Egorov}},\ }\href {\doibase 10.1103/PhysRevB.85.035205} {\bibfield
  {journal} {\bibinfo  {journal} {Phys. Rev. B}\ }\textbf {\bibinfo {volume}
  {85}},\ \bibinfo {pages} {035205} (\bibinfo {year} {2012})}\BibitemShut
  {NoStop}%
\bibitem [{\citenamefont {Nguyen}\ \emph {et~al.}(2013)\citenamefont {Nguyen},
  \citenamefont {Balocchi}, \citenamefont {Lagarde}, \citenamefont {Zhang},
  \citenamefont {Carr{\`e}re}, \citenamefont {Mazzucato}, \citenamefont
  {Barate}, \citenamefont {Galopin}, \citenamefont {Gierak}, \citenamefont
  {Bourhis}, \citenamefont {Harmand}, \citenamefont {Amand},\ and\
  \citenamefont {Marie}}]{doi:10.1063/1.4816970}%
  \BibitemOpen
  \bibfield  {author} {\bibinfo {author} {\bibfnamefont {C.~T.}\ \bibnamefont
  {Nguyen}}, \bibinfo {author} {\bibfnamefont {A.}~\bibnamefont {Balocchi}},
  \bibinfo {author} {\bibfnamefont {D.}~\bibnamefont {Lagarde}}, \bibinfo
  {author} {\bibfnamefont {T.~T.}\ \bibnamefont {Zhang}}, \bibinfo {author}
  {\bibfnamefont {H.}~\bibnamefont {Carr{\`e}re}}, \bibinfo {author}
  {\bibfnamefont {S.}~\bibnamefont {Mazzucato}}, \bibinfo {author}
  {\bibfnamefont {P.}~\bibnamefont {Barate}}, \bibinfo {author} {\bibfnamefont
  {E.}~\bibnamefont {Galopin}}, \bibinfo {author} {\bibfnamefont
  {J.}~\bibnamefont {Gierak}}, \bibinfo {author} {\bibfnamefont
  {E.}~\bibnamefont {Bourhis}}, \bibinfo {author} {\bibfnamefont {J.~C.}\
  \bibnamefont {Harmand}}, \bibinfo {author} {\bibfnamefont {T.}~\bibnamefont
  {Amand}}, \ and\ \bibinfo {author} {\bibfnamefont {X.}~\bibnamefont
  {Marie}},\ }\href {\doibase 10.1063/1.4816970} {\bibfield  {journal}
  {\bibinfo  {journal} {Applied Physics Letters}\ }\textbf {\bibinfo {volume}
  {103}},\ \bibinfo {pages} {052403} (\bibinfo {year} {2013})},\ \Eprint
  {http://arxiv.org/abs/https://doi.org/10.1063/1.4816970}
  {https://doi.org/10.1063/1.4816970} \BibitemShut {NoStop}%
\bibitem [{\citenamefont {Kalevich}\ \emph {et~al.}(2013)\citenamefont
  {Kalevich}, \citenamefont {Afanasiev}, \citenamefont {Shiryaev},\ and\
  \citenamefont {Egorov}}]{Kalevich2013}%
  \BibitemOpen
  \bibfield  {author} {\bibinfo {author} {\bibfnamefont {V.~K.}\ \bibnamefont
  {Kalevich}}, \bibinfo {author} {\bibfnamefont {M.~M.}\ \bibnamefont
  {Afanasiev}}, \bibinfo {author} {\bibfnamefont {A.~Y.}\ \bibnamefont
  {Shiryaev}}, \ and\ \bibinfo {author} {\bibfnamefont {A.~Y.}\ \bibnamefont
  {Egorov}},\ }\href {\doibase 10.1134/S0021364012210060} {\bibfield  {journal}
  {\bibinfo  {journal} {JETP Letters}\ }\textbf {\bibinfo {volume} {96}},\
  \bibinfo {pages} {567} (\bibinfo {year} {2013})}\BibitemShut {NoStop}%
\bibitem [{\citenamefont {Puttisong}\ \emph
  {et~al.}(2013{\natexlab{a}})\citenamefont {Puttisong}, \citenamefont {Wang},
  \citenamefont {Buyanova}, \citenamefont {Geelhaar}, \citenamefont {Riechert},
  \citenamefont {Ptak}, \citenamefont {Tu},\ and\ \citenamefont
  {Chen}}]{puttisong2013efficient}%
  \BibitemOpen
  \bibfield  {author} {\bibinfo {author} {\bibfnamefont {Y.}~\bibnamefont
  {Puttisong}}, \bibinfo {author} {\bibfnamefont {X.}~\bibnamefont {Wang}},
  \bibinfo {author} {\bibfnamefont {I.}~\bibnamefont {Buyanova}}, \bibinfo
  {author} {\bibfnamefont {L.}~\bibnamefont {Geelhaar}}, \bibinfo {author}
  {\bibfnamefont {H.}~\bibnamefont {Riechert}}, \bibinfo {author}
  {\bibfnamefont {A.}~\bibnamefont {Ptak}}, \bibinfo {author} {\bibfnamefont
  {C.}~\bibnamefont {Tu}}, \ and\ \bibinfo {author} {\bibfnamefont
  {W.}~\bibnamefont {Chen}},\ }\href@noop {} {\bibfield  {journal} {\bibinfo
  {journal} {Nature communications}\ }\textbf {\bibinfo {volume} {4}},\
  \bibinfo {pages} {1751} (\bibinfo {year} {2013}{\natexlab{a}})}\BibitemShut
  {NoStop}%
\bibitem [{\citenamefont {Puttisong}\ \emph
  {et~al.}(2013{\natexlab{b}})\citenamefont {Puttisong}, \citenamefont {Wang},
  \citenamefont {Buyanova},\ and\ \citenamefont {Chen}}]{PhysRevB.87.125202}%
  \BibitemOpen
  \bibfield  {author} {\bibinfo {author} {\bibfnamefont {Y.}~\bibnamefont
  {Puttisong}}, \bibinfo {author} {\bibfnamefont {X.~J.}\ \bibnamefont {Wang}},
  \bibinfo {author} {\bibfnamefont {I.~A.}\ \bibnamefont {Buyanova}}, \ and\
  \bibinfo {author} {\bibfnamefont {W.~M.}\ \bibnamefont {Chen}},\ }\href
  {\doibase 10.1103/PhysRevB.87.125202} {\bibfield  {journal} {\bibinfo
  {journal} {Phys. Rev. B}\ }\textbf {\bibinfo {volume} {87}},\ \bibinfo
  {pages} {125202} (\bibinfo {year} {2013}{\natexlab{b}})}\BibitemShut
  {NoStop}%
\bibitem [{\citenamefont {Sandoval-Santana}\ \emph {et~al.}(2014)\citenamefont
  {Sandoval-Santana}, \citenamefont {Balocchi}, \citenamefont {Amand},
  \citenamefont {Harmand}, \citenamefont {Kunold},\ and\ \citenamefont
  {Marie}}]{PhysRevB.90.115205}%
  \BibitemOpen
  \bibfield  {author} {\bibinfo {author} {\bibfnamefont {C.}~\bibnamefont
  {Sandoval-Santana}}, \bibinfo {author} {\bibfnamefont {A.}~\bibnamefont
  {Balocchi}}, \bibinfo {author} {\bibfnamefont {T.}~\bibnamefont {Amand}},
  \bibinfo {author} {\bibfnamefont {J.~C.}\ \bibnamefont {Harmand}}, \bibinfo
  {author} {\bibfnamefont {A.}~\bibnamefont {Kunold}}, \ and\ \bibinfo {author}
  {\bibfnamefont {X.}~\bibnamefont {Marie}},\ }\href {\doibase
  10.1103/PhysRevB.90.115205} {\bibfield  {journal} {\bibinfo  {journal} {Phys.
  Rev. B}\ }\textbf {\bibinfo {volume} {90}},\ \bibinfo {pages} {115205}
  (\bibinfo {year} {2014})}\BibitemShut {NoStop}%
\bibitem [{\citenamefont {Ivchenko}\ \emph {et~al.}(2015)\citenamefont
  {Ivchenko}, \citenamefont {Bakaleinikov},\ and\ \citenamefont
  {Kalevich}}]{PhysRevB.91.205202}%
  \BibitemOpen
  \bibfield  {author} {\bibinfo {author} {\bibfnamefont {E.~L.}\ \bibnamefont
  {Ivchenko}}, \bibinfo {author} {\bibfnamefont {L.~A.}\ \bibnamefont
  {Bakaleinikov}}, \ and\ \bibinfo {author} {\bibfnamefont {V.~K.}\
  \bibnamefont {Kalevich}},\ }\href {\doibase 10.1103/PhysRevB.91.205202}
  {\bibfield  {journal} {\bibinfo  {journal} {Phys. Rev. B}\ }\textbf {\bibinfo
  {volume} {91}},\ \bibinfo {pages} {205202} (\bibinfo {year}
  {2015})}\BibitemShut {NoStop}%
\bibitem [{\citenamefont {Ivchenko}\ \emph {et~al.}(2016)\citenamefont
  {Ivchenko}, \citenamefont {Bakaleinikov}, \citenamefont {Afanasiev},\ and\
  \citenamefont {Kalevich}}]{Ivchenko2016}%
  \BibitemOpen
  \bibfield  {author} {\bibinfo {author} {\bibfnamefont {E.~L.}\ \bibnamefont
  {Ivchenko}}, \bibinfo {author} {\bibfnamefont {L.~A.}\ \bibnamefont
  {Bakaleinikov}}, \bibinfo {author} {\bibfnamefont {M.~M.}\ \bibnamefont
  {Afanasiev}}, \ and\ \bibinfo {author} {\bibfnamefont {V.~K.}\ \bibnamefont
  {Kalevich}},\ }\href {\doibase 10.1134/S106378341608014X} {\bibfield
  {journal} {\bibinfo  {journal} {Physics of the Solid State}\ }\textbf
  {\bibinfo {volume} {58}},\ \bibinfo {pages} {1539} (\bibinfo {year}
  {2016})}\BibitemShut {NoStop}%
\bibitem [{\citenamefont {Ibarra-Sierra}\ \emph {et~al.}(2017)\citenamefont
  {Ibarra-Sierra}, \citenamefont {Sandoval-Santana}, \citenamefont {Azaizia},
  \citenamefont {Carr\`ere}, \citenamefont {Bakaleinikov}, \citenamefont
  {Kalevich}, \citenamefont {Ivchenko}, \citenamefont {Marie}, \citenamefont
  {Amand}, \citenamefont {Balocchi},\ and\ \citenamefont
  {Kunold}}]{PhysRevB.95.195204}%
  \BibitemOpen
  \bibfield  {author} {\bibinfo {author} {\bibfnamefont {V.~G.}\ \bibnamefont
  {Ibarra-Sierra}}, \bibinfo {author} {\bibfnamefont {J.~C.}\ \bibnamefont
  {Sandoval-Santana}}, \bibinfo {author} {\bibfnamefont {S.}~\bibnamefont
  {Azaizia}}, \bibinfo {author} {\bibfnamefont {H.}~\bibnamefont {Carr\`ere}},
  \bibinfo {author} {\bibfnamefont {L.~A.}\ \bibnamefont {Bakaleinikov}},
  \bibinfo {author} {\bibfnamefont {V.~K.}\ \bibnamefont {Kalevich}}, \bibinfo
  {author} {\bibfnamefont {E.~L.}\ \bibnamefont {Ivchenko}}, \bibinfo {author}
  {\bibfnamefont {X.}~\bibnamefont {Marie}}, \bibinfo {author} {\bibfnamefont
  {T.}~\bibnamefont {Amand}}, \bibinfo {author} {\bibfnamefont
  {A.}~\bibnamefont {Balocchi}}, \ and\ \bibinfo {author} {\bibfnamefont
  {A.}~\bibnamefont {Kunold}},\ }\href {\doibase 10.1103/PhysRevB.95.195204}
  {\bibfield  {journal} {\bibinfo  {journal} {Phys. Rev. B}\ }\textbf {\bibinfo
  {volume} {95}},\ \bibinfo {pages} {195204} (\bibinfo {year}
  {2017})}\BibitemShut {NoStop}%
\bibitem [{\citenamefont {Azaizia}\ \emph {et~al.}(2018)\citenamefont
  {Azaizia}, \citenamefont {Carr\`ere}, \citenamefont {Sandoval-Santana},
  \citenamefont {Ibarra-Sierra}, \citenamefont {Kalevich}, \citenamefont
  {Ivchenko}, \citenamefont {Bakaleinikov}, \citenamefont {Marie},
  \citenamefont {Amand}, \citenamefont {Kunold},\ and\ \citenamefont
  {Balocchi}}]{PhysRevB.97.155201}%
  \BibitemOpen
  \bibfield  {author} {\bibinfo {author} {\bibfnamefont {S.}~\bibnamefont
  {Azaizia}}, \bibinfo {author} {\bibfnamefont {H.}~\bibnamefont {Carr\`ere}},
  \bibinfo {author} {\bibfnamefont {J.~C.}\ \bibnamefont {Sandoval-Santana}},
  \bibinfo {author} {\bibfnamefont {V.~G.}\ \bibnamefont {Ibarra-Sierra}},
  \bibinfo {author} {\bibfnamefont {V.~K.}\ \bibnamefont {Kalevich}}, \bibinfo
  {author} {\bibfnamefont {E.~L.}\ \bibnamefont {Ivchenko}}, \bibinfo {author}
  {\bibfnamefont {L.~A.}\ \bibnamefont {Bakaleinikov}}, \bibinfo {author}
  {\bibfnamefont {X.}~\bibnamefont {Marie}}, \bibinfo {author} {\bibfnamefont
  {T.}~\bibnamefont {Amand}}, \bibinfo {author} {\bibfnamefont
  {A.}~\bibnamefont {Kunold}}, \ and\ \bibinfo {author} {\bibfnamefont
  {A.}~\bibnamefont {Balocchi}},\ }\href {\doibase 10.1103/PhysRevB.97.155201}
  {\bibfield  {journal} {\bibinfo  {journal} {Phys. Rev. B}\ }\textbf {\bibinfo
  {volume} {97}},\ \bibinfo {pages} {155201} (\bibinfo {year}
  {2018})}\BibitemShut {NoStop}%
\bibitem [{\citenamefont {Sandoval-Santana}\ \emph {et~al.}(2018)\citenamefont
  {Sandoval-Santana}, \citenamefont {Ibarra-Sierra}, \citenamefont {Azaizia},
  \citenamefont {Carr{\`e}re}, \citenamefont {Bakaleinikov}, \citenamefont
  {Kalevich}, \citenamefont {Ivchenko}, \citenamefont {Marie}, \citenamefont
  {Amand}, \citenamefont {Balocchi},\ and\ \citenamefont
  {Kunold}}]{Sandoval-Santana2018}%
  \BibitemOpen
  \bibfield  {author} {\bibinfo {author} {\bibfnamefont {J.~C.}\ \bibnamefont
  {Sandoval-Santana}}, \bibinfo {author} {\bibfnamefont {V.~G.}\ \bibnamefont
  {Ibarra-Sierra}}, \bibinfo {author} {\bibfnamefont {S.}~\bibnamefont
  {Azaizia}}, \bibinfo {author} {\bibfnamefont {H.}~\bibnamefont
  {Carr{\`e}re}}, \bibinfo {author} {\bibfnamefont {L.~A.}\ \bibnamefont
  {Bakaleinikov}}, \bibinfo {author} {\bibfnamefont {V.~K.}\ \bibnamefont
  {Kalevich}}, \bibinfo {author} {\bibfnamefont {E.~L.}\ \bibnamefont
  {Ivchenko}}, \bibinfo {author} {\bibfnamefont {X.}~\bibnamefont {Marie}},
  \bibinfo {author} {\bibfnamefont {T.}~\bibnamefont {Amand}}, \bibinfo
  {author} {\bibfnamefont {A.}~\bibnamefont {Balocchi}}, \ and\ \bibinfo
  {author} {\bibfnamefont {A.}~\bibnamefont {Kunold}},\ }\href {\doibase
  10.1140/epjp/i2018-11957-4} {\bibfield  {journal} {\bibinfo  {journal} {The
  European Physical Journal Plus}\ }\textbf {\bibinfo {volume} {133}},\
  \bibinfo {pages} {122} (\bibinfo {year} {2018})}\BibitemShut {NoStop}%
\bibitem [{\citenamefont {Ibarra-Sierra}\ \emph {et~al.}(2018)\citenamefont
  {Ibarra-Sierra}, \citenamefont {Sandoval-Santana}, \citenamefont {Azaizia},
  \citenamefont {Carr{\`e}re}, \citenamefont {Bakaleinikov}, \citenamefont
  {Kalevich}, \citenamefont {Ivchenko}, \citenamefont {Marie}, \citenamefont
  {Amand}, \citenamefont {Balocchi} \emph {et~al.}}]{ibarra2018spin}%
  \BibitemOpen
  \bibfield  {author} {\bibinfo {author} {\bibfnamefont {V.}~\bibnamefont
  {Ibarra-Sierra}}, \bibinfo {author} {\bibfnamefont {J.}~\bibnamefont
  {Sandoval-Santana}}, \bibinfo {author} {\bibfnamefont {S.}~\bibnamefont
  {Azaizia}}, \bibinfo {author} {\bibfnamefont {H.}~\bibnamefont
  {Carr{\`e}re}}, \bibinfo {author} {\bibfnamefont {L.}~\bibnamefont
  {Bakaleinikov}}, \bibinfo {author} {\bibfnamefont {V.}~\bibnamefont
  {Kalevich}}, \bibinfo {author} {\bibfnamefont {E.}~\bibnamefont {Ivchenko}},
  \bibinfo {author} {\bibfnamefont {X.}~\bibnamefont {Marie}}, \bibinfo
  {author} {\bibfnamefont {T.}~\bibnamefont {Amand}}, \bibinfo {author}
  {\bibfnamefont {A.}~\bibnamefont {Balocchi}},  \emph {et~al.},\ }\href@noop
  {} {\bibfield  {journal} {\bibinfo  {journal} {Journal of Materials Science:
  Materials in Electronics}\ }\textbf {\bibinfo {volume} {29}},\ \bibinfo
  {pages} {15307} (\bibinfo {year} {2018})}\BibitemShut {NoStop}%
\bibitem [{\citenamefont {Chen}\ \emph {et~al.}(2018)\citenamefont {Chen},
  \citenamefont {Huang}, \citenamefont {Visser}, \citenamefont {Anand},
  \citenamefont {Buyanova},\ and\ \citenamefont {Chen}}]{chen2018room}%
  \BibitemOpen
  \bibfield  {author} {\bibinfo {author} {\bibfnamefont {S.}~\bibnamefont
  {Chen}}, \bibinfo {author} {\bibfnamefont {Y.}~\bibnamefont {Huang}},
  \bibinfo {author} {\bibfnamefont {D.}~\bibnamefont {Visser}}, \bibinfo
  {author} {\bibfnamefont {S.}~\bibnamefont {Anand}}, \bibinfo {author}
  {\bibfnamefont {I.~A.}\ \bibnamefont {Buyanova}}, \ and\ \bibinfo {author}
  {\bibfnamefont {W.~M.}\ \bibnamefont {Chen}},\ }\href@noop {} {\bibfield
  {journal} {\bibinfo  {journal} {Nature communications}\ }\textbf {\bibinfo
  {volume} {9}},\ \bibinfo {pages} {3575} (\bibinfo {year} {2018})}\BibitemShut
  {NoStop}%
\bibitem [{\citenamefont {Meier}\ and\ \citenamefont
  {Zakharchenya}(2012)}]{meier2012optical}%
  \BibitemOpen
  \bibfield  {author} {\bibinfo {author} {\bibfnamefont {F.}~\bibnamefont
  {Meier}}\ and\ \bibinfo {author} {\bibfnamefont {B.~P.}\ \bibnamefont
  {Zakharchenya}},\ }\href@noop {} {\emph {\bibinfo {title} {Optical
  orientation}}}\ (\bibinfo  {publisher} {Elsevier},\ \bibinfo {year}
  {2012})\BibitemShut {NoStop}%
\bibitem [{\citenamefont {Reason}\ \emph {et~al.}(2007)\citenamefont {Reason},
  \citenamefont {Jin}, \citenamefont {McKay}, \citenamefont {Mangan},
  \citenamefont {Mao}, \citenamefont {Goldman}, \citenamefont {Bai},\ and\
  \citenamefont {Kurdak}}]{doi:10.1063/1.2798629}%
  \BibitemOpen
  \bibfield  {author} {\bibinfo {author} {\bibfnamefont {M.}~\bibnamefont
  {Reason}}, \bibinfo {author} {\bibfnamefont {Y.}~\bibnamefont {Jin}},
  \bibinfo {author} {\bibfnamefont {H.~A.}\ \bibnamefont {McKay}}, \bibinfo
  {author} {\bibfnamefont {N.}~\bibnamefont {Mangan}}, \bibinfo {author}
  {\bibfnamefont {D.}~\bibnamefont {Mao}}, \bibinfo {author} {\bibfnamefont
  {R.~S.}\ \bibnamefont {Goldman}}, \bibinfo {author} {\bibfnamefont
  {X.}~\bibnamefont {Bai}}, \ and\ \bibinfo {author} {\bibfnamefont
  {C.}~\bibnamefont {Kurdak}},\ }\href {\doibase 10.1063/1.2798629} {\bibfield
  {journal} {\bibinfo  {journal} {Journal of Applied Physics}\ }\textbf
  {\bibinfo {volume} {102}},\ \bibinfo {pages} {103710} (\bibinfo {year}
  {2007})},\ \Eprint {http://arxiv.org/abs/https://doi.org/10.1063/1.2798629}
  {https://doi.org/10.1063/1.2798629} \BibitemShut {NoStop}%
\bibitem [{\citenamefont {Dhar}\ \emph {et~al.}(2007)\citenamefont {Dhar},
  \citenamefont {Mondal},\ and\ \citenamefont {Das}}]{Dhar_2007}%
  \BibitemOpen
  \bibfield  {author} {\bibinfo {author} {\bibfnamefont {S.}~\bibnamefont
  {Dhar}}, \bibinfo {author} {\bibfnamefont {A.}~\bibnamefont {Mondal}}, \ and\
  \bibinfo {author} {\bibfnamefont {T.~D.}\ \bibnamefont {Das}},\ }\href
  {\doibase 10.1088/0268-1242/23/1/015007} {\bibfield  {journal} {\bibinfo
  {journal} {Semiconductor Science and Technology}\ }\textbf {\bibinfo {volume}
  {23}},\ \bibinfo {pages} {015007} (\bibinfo {year} {2007})}\BibitemShut
  {NoStop}%
\bibitem [{\citenamefont {Ibáñez}\ \emph {et~al.}(2008)\citenamefont
  {Ibáñez}, \citenamefont {Cuscó}, \citenamefont {Alarcón-Lladó},
  \citenamefont {Artús}, \citenamefont {Patanè}, \citenamefont {Fowler},
  \citenamefont {Eaves}, \citenamefont {Uesugi},\ and\ \citenamefont
  {Suemune}}]{doi:10.1063/1.2927387}%
  \BibitemOpen
  \bibfield  {author} {\bibinfo {author} {\bibfnamefont {J.}~\bibnamefont
  {Ibáñez}}, \bibinfo {author} {\bibfnamefont {R.}~\bibnamefont {Cuscó}},
  \bibinfo {author} {\bibfnamefont {E.}~\bibnamefont {Alarcón-Lladó}},
  \bibinfo {author} {\bibfnamefont {L.}~\bibnamefont {Artús}}, \bibinfo
  {author} {\bibfnamefont {A.}~\bibnamefont {Patanè}}, \bibinfo {author}
  {\bibfnamefont {D.}~\bibnamefont {Fowler}}, \bibinfo {author} {\bibfnamefont
  {L.}~\bibnamefont {Eaves}}, \bibinfo {author} {\bibfnamefont
  {K.}~\bibnamefont {Uesugi}}, \ and\ \bibinfo {author} {\bibfnamefont
  {I.}~\bibnamefont {Suemune}},\ }\href {\doibase 10.1063/1.2927387} {\bibfield
   {journal} {\bibinfo  {journal} {Journal of Applied Physics}\ }\textbf
  {\bibinfo {volume} {103}},\ \bibinfo {pages} {103528} (\bibinfo {year}
  {2008})},\ \Eprint {http://arxiv.org/abs/https://doi.org/10.1063/1.2927387}
  {https://doi.org/10.1063/1.2927387} \BibitemShut {NoStop}%
\bibitem [{\citenamefont {{Suzuki}}\ \emph {et~al.}(2009)\citenamefont
  {{Suzuki}}, \citenamefont {{Hashiguchi}}, \citenamefont {{Kojima}},
  \citenamefont {{Ohshita}},\ and\ \citenamefont {{Yamaguchi}}}]{5411156}%
  \BibitemOpen
  \bibfield  {author} {\bibinfo {author} {\bibfnamefont {H.}~\bibnamefont
  {{Suzuki}}}, \bibinfo {author} {\bibfnamefont {T.}~\bibnamefont
  {{Hashiguchi}}}, \bibinfo {author} {\bibfnamefont {N.}~\bibnamefont
  {{Kojima}}}, \bibinfo {author} {\bibfnamefont {Y.}~\bibnamefont {{Ohshita}}},
  \ and\ \bibinfo {author} {\bibfnamefont {M.}~\bibnamefont {{Yamaguchi}}},\
  }in\ \href {\doibase 10.1109/PVSC.2009.5411156} {\emph {\bibinfo {booktitle}
  {2009 34th IEEE Photovoltaic Specialists Conference (PVSC)}}}\ (\bibinfo
  {year} {2009})\ pp.\ \bibinfo {pages} {000848--000851}\BibitemShut {NoStop}%
\bibitem [{\citenamefont {Inagaki}\ \emph {et~al.}(2013)\citenamefont
  {Inagaki}, \citenamefont {Ikeda}, \citenamefont {Kowaki}, \citenamefont
  {Ohshita}, \citenamefont {Kojima},\ and\ \citenamefont
  {Yamagichi}}]{https://doi.org/10.1002/pssc.201200383}%
  \BibitemOpen
  \bibfield  {author} {\bibinfo {author} {\bibfnamefont {M.}~\bibnamefont
  {Inagaki}}, \bibinfo {author} {\bibfnamefont {K.}~\bibnamefont {Ikeda}},
  \bibinfo {author} {\bibfnamefont {H.}~\bibnamefont {Kowaki}}, \bibinfo
  {author} {\bibfnamefont {Y.}~\bibnamefont {Ohshita}}, \bibinfo {author}
  {\bibfnamefont {N.}~\bibnamefont {Kojima}}, \ and\ \bibinfo {author}
  {\bibfnamefont {M.}~\bibnamefont {Yamagichi}},\ }\href {\doibase
  https://doi.org/10.1002/pssc.201200383} {\bibfield  {journal} {\bibinfo
  {journal} {physica status solidi c}\ }\textbf {\bibinfo {volume} {10}},\
  \bibinfo {pages} {589} (\bibinfo {year} {2013})},\ \Eprint
  {http://arxiv.org/abs/https://onlinelibrary.wiley.com/doi/pdf/10.1002/pssc.201200383}
  {https://onlinelibrary.wiley.com/doi/pdf/10.1002/pssc.201200383} \BibitemShut
  {NoStop}%
\bibitem [{\citenamefont {Patan{\`{e}}}\ \emph {et~al.}(2009)\citenamefont
  {Patan{\`{e}}}, \citenamefont {Allison}, \citenamefont {Eaves}, \citenamefont
  {Hopkinson}, \citenamefont {Hill},\ and\ \citenamefont
  {Ignatov}}]{Patan_2009}%
  \BibitemOpen
  \bibfield  {author} {\bibinfo {author} {\bibfnamefont {A.}~\bibnamefont
  {Patan{\`{e}}}}, \bibinfo {author} {\bibfnamefont {G.}~\bibnamefont
  {Allison}}, \bibinfo {author} {\bibfnamefont {L.}~\bibnamefont {Eaves}},
  \bibinfo {author} {\bibfnamefont {M.}~\bibnamefont {Hopkinson}}, \bibinfo
  {author} {\bibfnamefont {G.}~\bibnamefont {Hill}}, \ and\ \bibinfo {author}
  {\bibfnamefont {A.}~\bibnamefont {Ignatov}},\ }\href {\doibase
  10.1088/0953-8984/21/17/174209} {\bibfield  {journal} {\bibinfo  {journal}
  {Journal of Physics: Condensed Matter}\ }\textbf {\bibinfo {volume} {21}},\
  \bibinfo {pages} {174209} (\bibinfo {year} {2009})}\BibitemShut {NoStop}%
\bibitem [{\citenamefont {Sandoval-Santana}\ \emph {et~al.}(2020)\citenamefont
  {Sandoval-Santana}, \citenamefont {Ibarra-Sierra}, \citenamefont {Carr\`ere},
  \citenamefont {Afanasiev}, \citenamefont {Bakaleinikov}, \citenamefont
  {Kalevich}, \citenamefont {Ivchenko}, \citenamefont {Marie}, \citenamefont
  {Amand}, \citenamefont {Balocchi},\ and\ \citenamefont
  {Kunold}}]{PhysRevB.101.075201}%
  \BibitemOpen
  \bibfield  {author} {\bibinfo {author} {\bibfnamefont {J.~C.}\ \bibnamefont
  {Sandoval-Santana}}, \bibinfo {author} {\bibfnamefont {V.~G.}\ \bibnamefont
  {Ibarra-Sierra}}, \bibinfo {author} {\bibfnamefont {H.}~\bibnamefont
  {Carr\`ere}}, \bibinfo {author} {\bibfnamefont {M.~M.}\ \bibnamefont
  {Afanasiev}}, \bibinfo {author} {\bibfnamefont {L.~A.}\ \bibnamefont
  {Bakaleinikov}}, \bibinfo {author} {\bibfnamefont {V.~K.}\ \bibnamefont
  {Kalevich}}, \bibinfo {author} {\bibfnamefont {E.~L.}\ \bibnamefont
  {Ivchenko}}, \bibinfo {author} {\bibfnamefont {X.}~\bibnamefont {Marie}},
  \bibinfo {author} {\bibfnamefont {T.}~\bibnamefont {Amand}}, \bibinfo
  {author} {\bibfnamefont {A.}~\bibnamefont {Balocchi}}, \ and\ \bibinfo
  {author} {\bibfnamefont {A.}~\bibnamefont {Kunold}},\ }\href {\doibase
  10.1103/PhysRevB.101.075201} {\bibfield  {journal} {\bibinfo  {journal}
  {Phys. Rev. B}\ }\textbf {\bibinfo {volume} {101}},\ \bibinfo {pages}
  {075201} (\bibinfo {year} {2020})}\BibitemShut {NoStop}%
\bibitem [{\citenamefont {Wangsness}\ and\ \citenamefont
  {Bloch}(1953)}]{PhysRev.89.728}%
  \BibitemOpen
  \bibfield  {author} {\bibinfo {author} {\bibfnamefont {R.~K.}\ \bibnamefont
  {Wangsness}}\ and\ \bibinfo {author} {\bibfnamefont {F.}~\bibnamefont
  {Bloch}},\ }\href {\doibase 10.1103/PhysRev.89.728} {\bibfield  {journal}
  {\bibinfo  {journal} {Phys. Rev.}\ }\textbf {\bibinfo {volume} {89}},\
  \bibinfo {pages} {728} (\bibinfo {year} {1953})}\BibitemShut {NoStop}%
\bibitem [{\citenamefont {Redfield}(1965)}]{REDFIELD19651}%
  \BibitemOpen
  \bibfield  {author} {\bibinfo {author} {\bibfnamefont {A.}~\bibnamefont
  {Redfield}},\ }in\ \href {\doibase
  http://dx.doi.org/10.1016/B978-1-4832-3114-3.50007-6} {\emph {\bibinfo
  {booktitle} {Advances in Magnetic Resonance}}},\ \bibinfo {series} {Advances
  in Magnetic and Optical Resonance}, Vol.~\bibinfo {volume} {1},\ \bibinfo
  {editor} {edited by\ \bibinfo {editor} {\bibfnamefont {J.~S.}\ \bibnamefont
  {Waugh}}}\ (\bibinfo  {publisher} {Academic Press},\ \bibinfo {year} {1965})\
  pp.\ \bibinfo {pages} {1 -- 32}\BibitemShut {NoStop}%
\bibitem [{\citenamefont {Leppelmeier}\ and\ \citenamefont
  {Hahn}(1966)}]{PhysRev.142.179}%
  \BibitemOpen
  \bibfield  {author} {\bibinfo {author} {\bibfnamefont {G.~W.}\ \bibnamefont
  {Leppelmeier}}\ and\ \bibinfo {author} {\bibfnamefont {E.~L.}\ \bibnamefont
  {Hahn}},\ }\href {\doibase 10.1103/PhysRev.142.179} {\bibfield  {journal}
  {\bibinfo  {journal} {Phys. Rev.}\ }\textbf {\bibinfo {volume} {142}},\
  \bibinfo {pages} {179} (\bibinfo {year} {1966})}\BibitemShut {NoStop}%
\bibitem [{\citenamefont {Masud~Chaichian}(1997)}]{book:729155}%
  \BibitemOpen
  \bibfield  {author} {\bibinfo {author} {\bibfnamefont {R.~H.}\ \bibnamefont
  {Masud~Chaichian}},\ }\href
  {http://gen.lib.rus.ec/book/index.php?md5=1ab38df45964db67c309c0b6932798c4}
  {\emph {\bibinfo {title} {Symmetries in Quantum Mechanics: From Angular
  Momentum to Supersymmetry (Graduate Student Series in Physics)}}},\ \bibinfo
  {edition} {1st}\ ed.,\ Graduate Student Series in Physics\ (\bibinfo
  {publisher} {Taylor \& Francis},\ \bibinfo {year} {1997})\BibitemShut
  {NoStop}%
\end{thebibliography}
%

\end{document}